\documentclass[%
    nofootinbib,
    notitlepage,
    preprintnumbers
]{revtex4-1}

\usepackage[caption=false]{subfig}

\usepackage{url}
\usepackage[colorlinks, linkcolor=blue, citecolor=blue, urlcolor=blue]{hyperref}
\usepackage{mathtools}
\usepackage{amsfonts}
\usepackage{amssymb}
\usepackage{bm}
\usepackage{booktabs}
\usepackage{graphicx}
\usepackage[utf8]{inputenc}
\usepackage{natbib}
\usepackage{pifont}
\usepackage{siunitx}
\usepackage[dvipsnames]{xcolor}
\usepackage{xspace}
\usepackage{slashed}
\usepackage{physics}
\usepackage{cleveref}
\usepackage[normalem]{ulem}

\renewcommand{\refeq}[1]{Eq.~(\ref{eq:#1})}
\newcommand{\reffig}[1]{Fig.~\ref{fig:#1}}
\newcommand{\refsec}[1]{Section~\ref{sec:#1}}

\newcommand{\GammaCusp}{\ensuremath{\Gamma_{\text{c}}}}
\newcommand{\gammaE}{\ensuremath{\gamma_{E}}}
\newcommand{\GeV}{\ensuremath{\si{\giga\electronvolt}}\xspace}

\renewcommand{\Im}[1]{\operatorname{Im} #1}
\newcommand\Exp[1]{\mathrm{e}^{#1}}

\allowdisplaybreaks

\begin{document}

\title{Systematic Parametrization of the Leading\texorpdfstring{\\}{ }\texorpdfstring{\boldmath$B$}{B}-meson Light-Cone Distribution Amplitude}

\author{Thorsten Feldmann}
\email{thorsten.feldmann@uni-siegen.de}
\affiliation{Theoretische Physik 1, Universit\"at Siegen, Walter-Flex-Stra\ss{}e 3, D-57068 Siegen, Germany}
\author{Philip L\"ughausen}
\email{philip.lueghausen@tum.de}
\affiliation{
Excellence Cluster ORIGINS, Technische Universität München, D-85748 Garching, Germany
}
\affiliation{
Physik Department T31, Technische Universit\"at M\"unchen, D-85748 Garching, Germany
}
\author{Danny van Dyk}
\email{danny.van.dyk@gmail.com}
\affiliation{
Physik Department T31, Technische Universit\"at M\"unchen, D-85748 Garching, Germany
}

\date{\today}

\preprint{{SI-HEP-2022-05, P3H-22-029, TUM-HEP-1388/22}}

\begin{abstract}
    We propose a parametrization of the leading $B$-meson light-cone distribution amplitude (LCDA)
    in heavy-quark effective theory (HQET).
    In position space, it uses a conformal transformation that yields a systematic
    Taylor expansion and an integral bound, which enables control of the truncation error.
    Our parametrization further produces
    compact analytical expressions for a variety of derived quantities. 
    At a given reference scale, our momentum-space parametrization
    corresponds to an expansion in associated Laguerre polynomials,
    which turn into confluent hypergeometric functions ${}_1F_1$ under renormalization-group evolution at one-loop accuracy.
    Our approach thus allows a straightforward and transparent implementation of a variety of phenomenological constraints,
    regardless of their origin.
    Moreover, we can include theoretical information on the Taylor coefficients by using the local
    operator production expansion.
    We showcase the versatility of the parametrization in a series of phenomenological pseudo-fits.
\end{abstract}

\maketitle

\clearpage 


{\footnotesize 
\tableofcontents 
}


\clearpage 


\section{Introduction}

\label{sec:intro1}

Light-cone distribution amplitudes (LCDA) of the $B$-meson are needed as hadronic input functions for the theoretical descriptions
of exclusive (energetic) $B$-meson decays. These descriptions include factorization theorems in 
Quantum Chromodynamics (QCD), which have first been introduced to tackle charmless non-leptonic $B$-decays \cite{Beneke:1999br,Beneke:2001ev}.
They have later also been applied to other decay modes, including semi-leptonic and radiative decays (see e.g.\ the corresponding chapter in \cite{Belle-II:2018jsg} for a recent overview and an exhaustive list of references).
The descriptions further include light-cone sum rules (LCSR), which are a complementary approach. These sum rules can be used to determine
``soft'' hadronic matrix elements for which the factorization of the initial and final states does not work completely.
A formulation of light-cone sum rules with $B$-meson LCDA has been proposed in Ref.~\cite{Khodjamirian:2005ea,DeFazio:2005dx,Khodjamirian:2006st,DeFazio:2007hw}.
It has the advantage that the very same hadronic input functions appear as in QCD factorization.
This fact has recently been exploited to show that precise theoretical predictions for the benchmark decay mode $B \to \gamma \ell \nu$
can be obtained \cite{Beneke:2011nf,Braun:2012kp,Beneke:2018wjp}, which in turn allows inferring the relevant information on the
$B$-meson LCDA\pl{\sout{s}} from future experimental data, notably from the Belle-2 experiment; see the corresponding paragraph in Ref.~\cite{Belle-II:2018jsg}.\\


The leading $B$-meson LCDA enters the aforementioned theoretical approaches in different ways:
\begin{enumerate}
    \item The leading-power terms in QCD factorization involve logarithmic moments of the $B$-meson LCDA.
    The definition of these logarithmic moments follows later.
    
    \item In LCSR the $B$-meson LCDA enters in the form of integrals where the contributions from large light-cone momenta are parametrically suppressed. We later define appropriate quantities to describe the low-momentum behavior of the $B$-meson LCDA relevant for these sum-rule applications.
\end{enumerate}
Adhoc models of the LCDA introduce non-trivial and potentially unphysical correlations within and between these two sets of quantities.
The modelling itself and together with these correlations give rise to unquantifiable systematic uncertainties in the determination of the
leading-twist LCDA, e.g., from the photoleptonic decay $\bar{B} \to \gamma \ell^- \bar\nu$.
One of the main results of this work is a parametrization of the LCDA that is general enough to avoid unjustified correlations between its
observable features and that includes as much model-independent theoretical information as possible.\\

Parametrizing the soft contribution of the $B$-meson LCDA introduces by definition
a low reference momentum scale (in the following denoted as $\omega_0$), which characterizes hadronic dynamics.
In previously discussed benchmark models, this scale has often been identified with the HQET parameter $\bar \Lambda$
by using theoretical expressions for the positive moments of the LCDA, either at tree level~\cite{Grozin:1996pq}
or including the radiative tail~\cite{Lee:2005gza}.
Our parametrization for the LCDA starts from an infinite series of terms, such that the moment constraints
can be fulfilled at each order of the HQET expansion for any value of $\omega_0$. 
The only constraint on this otherwise free parameter is coming from the requirement that the expansion coefficients
are sufficiently converging, which again forces $\omega_0$ to be \emph{of the same order as} $\bar\Lambda$.
In practice, we can truncate the expansion after a few terms, and the intrinsic uncertainty of the truncation
can be estimated by varying the parameter $\omega_0$ in a reasonable range.\\

The theoretical properties of $B$-meson LCDAs have been studied extensively in the past.
For the scope of this work, two related aspects turn out to be most important:
\begin{itemize}
    \item the behavior of LCDAs under change of the renormalization scale; and 
    \item the behavior of LCDAs at large light-cone momentum of the light quark
    (i.e.\ at short separations of the fields in the defining light-cone operator).
\end{itemize}
In both cases, one has to carefully study the renormalization of light-cone operators in the heavy-quark limit, i.e.\,
the treatment of the $b$-quark as a static source of color in heavy-quark effective theory (HQET).
The resulting renormalization group (RG) equation for the $B$-meson LCDA has been first calculated at the one-loop level
by Lange and Neubert~\cite{Lange:2003ff}.
The eigenfunctions of the one-loop RG kernel have first been identified in Ref.~\cite{Bell:2013tfa}, which shortly thereafter
have been reproduced from conformal symmetry considerations~\cite{Braun:2014owa}.
The latter method has very recently been used to derive the RG kernel for the $B$-meson LCDA at two loops~\cite{Braun:2019wyx},
and the solution of the RG equation and its implementation into QCD factorization theorems have been 
discussed in Ref.~\cite{Liu:2020eqe,Galda:2020epp}.
Here we will restrict ourselves to one-loop accuracy. However, our formalism is general enough to allow the
implementation of two-loop effects.\\


This article is structured as follows.
We summarize the properties of the leading $B$-meson LCDA $\phi_+$ and define our notations in \refsec{intro}. This includes a brief discussion of the relevant analytic properties, the renormalization at one-loop level, the generating function for the logarithmic moments, and the definition of suitable quantities to describe the low-momentum behavior. 
In \refsec{ansatz} we introduce our novel parametrization for the $B$-meson LCDA $\tilde\phi_+(\tau)$ in position space.
Starting from a conformal transformation $\tau \mapsto y$, which maps the real $\tau$ axis onto the unit circle in the complex $y$-plane, we construct a Taylor expansion in the variable $y$, where the Taylor coefficients are constrained by an integral bound.
We translate our parametrization to the so-called ``dual'' space and to momentum space.
In both cases, this results in an expansion in terms of associated Laguerre polynomials. We also provide expressions for the logarithmic moments and discuss different options to implement the effect of the RG evolution.
Moreover, we briefly discuss how to generalize our formalism to higher-twist LCDAs, restricting ourselves to the Wandzura-Wilczek limit.
Our parametrization is generic enough to capture the features of a variety of benchmark models discussed in
the literature. This is illustrated in \refsec{models} where we study the convergence properties of our expansion for four examples of such models.
To set the stage for future phenomenological applications, in \refsec{pseudo-pheno}, we perform numerical fits on the basis of two
pseudo-observables that are expected to be well constrained by future data on the photo-leptonic $B\to \gamma \ell \nu$ decay. 
In addition, we show how including theoretical information from the local operator product expansion (OPE)
yields further constraints of the expansion coefficients in phenomenological fits.
We conclude in \refsec{conclusion} and provide some additional formulas in 
two appendices.
%
\section{Prerequisites}
\label{sec:intro}

The leading-twist\footnote{%
    The notion of twist has to be modified for the discussion of light-cone operators in
    HQET; see Ref.~\cite{Braun:2003wx}.%
} LCDA of the $B$-meson is be defined as the matrix element of a light-cone operator in HQET
normalized to the matrix element of the corresponding local operator~\cite{Grozin:1996pq}:
\begin{equation}
\label{eq:def:phit}
\begin{aligned}
    \tilde\phi_+(\tau;\mu)
        & = \frac{\langle 0| \bar q( \tau n) \, [\tau n,0] \, \slashed n \gamma_5 \, h_v(0)|B(v)\rangle}
                 {\langle 0| \bar q(0) \, \slashed n \gamma_5 \, h_v(0)|B(v)\rangle}\,.
\end{aligned}
\end{equation}
Here $n^\mu$ is a light-like vector with $n^2 = 0$, and the gauge link $[\tau n,0]$ appears as a straight
Wilson line that renders the definition of $\tilde\phi_+(\tau)$ gauge invariant in QCD.
The $B$-meson moves with velocity $v^\mu$. For simplicity we are considering a frame with $v \cdot n =1$.
The limit $m_b \to \infty$ has already been taken in HQET.
Hence, $\tilde\phi_+$ does not depend on the heavy-quark mass $m_b$. The $m_b$-dependence
of physical amplitudes is contained in short-distance coefficient functions that multiply the LCDA,
e.g., in QCD factorization calculations.

\subsection{Mathematical Properties}

In position space, the LCDA fulfills the following three properties. They have previously been discussed, e.g., in Ref.~\cite{Grozin:1994ni}:
\newcommand{\Prop}[1]{\textbf{P#1}}
\begin{itemize}
	\item[\Prop{1}:] $\tilde \phi_+(\tau)$ is analytic in the lower complex half plane ${\rm Im} \, \tau <0$.
	\item[\Prop{2}:] $\tilde \phi_+(\tau)$ is analytic on the real $\tau$ axis, except for a single point $\tau=0$ where it
	has a logarithmic singularity of measure zero, with a branch cut extending along the positive imaginary axis.
	Hence $\tilde \phi_+(\tau)$ is Lebesque-integrable  with 
	\begin{align}
	\lim_{\epsilon \to 0^+} \, \int\limits_{-\infty -i\epsilon}^{\infty -i\epsilon} d\tau \, \tilde\phi_+(\tau,\mu) &= 0
	\end{align}	  
	\item[\Prop{3}:] $\tilde\phi_+(\tau)$ can be analytically continued from the lower complex half plane onto the real $\tau$ axis 
	\emph{almost everywhere} (i.e.\ in all points except for a null set).  
\end{itemize}
In the following we assume that the Fourier transform exists,
\begin{align}
    \phi_+(\omega;\mu)
        & = \int\limits_{-\infty-i\epsilon}^{+\infty-i\epsilon} \frac{d\tau}{2\pi} \, e^{i\omega \tau} \, \tilde \phi_+(\tau;\mu) \,.
\end{align}
It follows from the properties \Prop{1} to \Prop{3} and the
Paley-Wiener theorem \cite[theorem 7.2.4]{Strichartz:2003}
that $\tilde \phi_+(\tau)$ is the \emph{holomorphic Fourier transform} of a function $\phi_+(\omega)$,
\begin{equation}
    \tilde \phi_+(\tau;\mu) = \int_{0}^{\infty} d\omega\, e^{-i\omega\tau} \, \phi_+(\omega;\mu) \,,
\end{equation}
and that $\phi_+(\omega) \in L^2$ on the support $[0, \infty)$.
Plancherel's theorem then provides that both $\tilde \phi_+(\tau)$
and $\phi_+(\omega)$ are square-integrable on the entire real $\tau$ axis and the positive $\omega$ axis, respectively, and their two-norms coincide:
\begin{equation}
    \int_{-\infty}^{+\infty} \frac{\dd{\tau}}{2\pi} \big|\tilde \phi_+(\tau)\big|^2
    = \int_0^\infty \dd{\omega} \big|\phi_+(\omega)\big|^2
    < \infty\,.
\end{equation}
As consequence, the inner product exists in both the $\omega$ space and the $\tau$ space.\\

We further assume that $\phi_+(\omega; \mu) \propto \omega$ for $\omega \to 0$ at large renormalization scales $\mu \gg \Lambda_\text{had}$.
This is supported by the asymptotic behavior due to approximate conformal symmetry within the twist expansion~\cite{Braun:2017liq}.
From this assumed behavior at $\omega = 0$ one further property follows:
\begin{itemize}
    \item[\Prop{4}:] The position space LCDA must asymptotically fall off at least as fast as $1/\tau^2$:
    \begin{align}
        0 \leq\lim_{\tau\to\infty}  \left|\tau^{2}  \, \tilde\phi(\tau)\right| < \infty \,.
    \end{align}
\end{itemize}

In QCD factorization theorems, the momentum-space argument $\omega = n \cdot l$ represents 
the light-cone projection of the light spectator-quark momentum $l^\mu$ in the 
$B$-meson.
We remark that the support of the matrix element 
in \refeq{def:phit} is  different from the corresponding expressions for a light
pseudoscalar meson, due to the different analytic properties of the heavy-quark
propagator in HQET compared to a light-quark propagator in full QCD.
As a consequence, $\omega \in [0, \infty)$.

\subsection{Renormalization and Eigenfunctions}
\label{sec:intro:rge-eigenfunctions}

The $B$-meson LCDA $\phi_+(\omega)$ can be expanded in terms
of a \emph{continuous} set of eigenfunctions of the one-loop renormalization-group (RG) equation,
which can be expressed through Bessel functions of the first kind
\cite{Bell:2013tfa,Braun:2014owa}.
Following the convention of Ref.~\cite{Braun:2014owa} one has\footnote{%
    The transformations in \refeq{phi+-from-eta+} imply that the momentum-space
    LCDA $\phi_+(\omega,\mu)$ grows linearly in $\omega$ for small momenta, and its dual
    $\eta_+(s,\mu)$ goes to a constant at $s \to 0$.
}
\begin{equation}
\label{eq:phi+-from-eta+}
\begin{aligned}
    \phi_+(\omega, \mu)
        & = \int_0^\infty \dd s\, \sqrt{\omega s}\, J_1(2 \sqrt{\omega  s})\,
        \eta_+(s, \mu) \\[0.2em]
    \Leftrightarrow \qquad 
    s \, \eta_+(s;\mu)
        & = \int_0^\infty \frac{\mathrm{d\omega}}{\omega} \,  
            \sqrt{\omega s} \, J_1(2 \sqrt{\omega s})\, \phi_+(\omega;\mu) \,. 
\end{aligned}
\end{equation}
The notation for the function $\eta_+(s)$ is related to the function $\rho_+(\omega')$ as defined
in Ref.~\cite{Bell:2013tfa} via the relation
\begin{align}
    s \, \eta_+(s;\mu)
        & = \rho_+(\omega'=1/s;\mu) \,.
\end{align}
In this work we use the notation of Ref.~\cite{Braun:2014owa}.
For convenience we also quote the relation between the dual-space LCDA and the position-space LCDA,
see also Ref.~\cite{Bell:2013tfa},
\begin{equation}
\label{eq:tildephi+-from-eta+}
\begin{aligned}
    s \, \eta_+(s;\mu)
        & = \int \frac{\dd \tau}{2\pi} \left( 1- e^{-i s/\tau} \right) \tilde \phi_+(\tau; \mu)\\
            \Leftrightarrow \quad \tilde\phi_+(\tau;\mu) &= - \frac{1}{\tau^2} \, 
            \int_0^\infty \dd s \, e^{\frac{i s}{\tau}} \, s \,\eta_+(s; \mu) \,.
\end{aligned}
\end{equation}

The purpose of these integral transformations is to showcase that the function $\eta_+(s)$
obeys a simple multiplicative RG equation at one-loop,\footnote{%
	Recently, the two-loop RG equation has been derived in Ref.~\cite{Braun:2019wyx},
	\begin{align*}
	\left(\frac{d}{d\ln\mu} + \Gamma_{c} \, \ln(\hat \mu \, s ) + \gamma_+ \right) \eta_+(s;\mu) 
	&= 4 C_F \left( \frac{\alpha_s}{4\pi} \right)^2 \int_0^1 du \, \frac{\bar u}{u} \, h(u) \, \eta_+(\bar us;\mu) \,,
	\label{RGE2}
	\end{align*}
	where $\bar{u} \equiv 1 - u$  and the function $h(u)$ is given by
	\begin{align*}
	    h(u) & =\ln \bar u \left[\beta_0 + 2 C_F \left( \ln \bar u - \frac{1+\bar u}{\bar u} \, \ln u - \frac32 \right)\right],
	    \qquad \text{with } h(0) = 0\,.
	\end{align*}
}
\begin{equation}
\begin{aligned}
    \frac{\dd \eta_+(s; \mu)}{\dd \ln\mu}
        & = - \left[\Gamma_{\rm c}(\alpha_s(\mu)) \, 
            \ln \left(\mu \, s \, e^{2 \gammaE}\right) + \gamma_+(\alpha_s(\mu)) \right]
            \eta_+(s; \mu) \,.
\end{aligned}
\end{equation}
Its explicit solution reads
\begin{equation}
\label{eq:eta-evol}
\begin{aligned}
    \eta_+(s;\mu)
        & = e^{V(\mu;\mu_0)} \, \eta_+(s;\mu_0) \,
            \left( \hat \mu_0 \, s \right)^{-g(\mu;\mu_0)} \,.
\end{aligned}
\end{equation}
Here and in the following, we use the short-hand notation 
\begin{equation}
    \hat \mu \equiv \mu \, e^{2\gamma_E} \,,
\end{equation}
and similar for other quantities.
Our definitions of the functions $V(\mu; \mu_0)$ and $g(\mu; \mu_0)$ 
coincide with the conventions used, e.g., in Ref.~\cite{Bell:2013tfa}.
They are given in \refeq{def:V} and \refeq{def:g} in the appendix, respectively. 
For convenience, we quote their RG equations:
\begin{align}
    \label{rge:V+g}
    \frac{\mathrm{d}V(\mu,\mu_0)}{\mathrm{d}\ln\mu}
        & = - \left[ \GammaCusp(\alpha_s(\mu)) \, \ln\left(\frac{\mu}{\mu_0}\right) 
        + \gamma_+(\alpha_s(\mu)) \right] \,, \qquad
    \frac{\mathrm{d}g(\mu,\mu_0)}{\mathrm{d}\ln\mu}
          = \GammaCusp(\alpha_s(\mu)) \,.
\end{align}

\subsection{Logarithmic Moments and Generating Function}
\label{sec:intro:log-moments}

In QCD factorization theorems for exclusive $B$-meson decays \cite{Beneke:1999br,Beneke:2001ev}
the $B$-meson LCDA enters in terms of logarithmic moments.
It is convenient to define these moments directly from the spectral representation
\cite{Bell:2013tfa,Feldmann:2014ika}. In the following, we will use the convention
\begin{align}
  L_n(\mu,\mu_m)
    & =(-1)^n \, \int_0^\infty \dd s \, \ln^n (\hat\mu_m s) \, \eta_+(s;\mu) \,,
\end{align}
where $L_0$ is commonly called $1 / \lambda_B$. 
We emphasize that in the definition of the logarithmic moments $L_n$ with $n\geq 1$, we have considered a fixed reference momentum scale $\mu_m$.
Alternative definitions in the literature have used the renormalization scale $\mu$ itself or the zeroth logarithmic moment $\lambda_B$.
The Mellin transform of $\eta_+(s)$
\begin{equation}
\begin{aligned}
    \label{eq:eta:genfunc}
    F_{[\eta_+]}(t;\mu,\mu_m)
        & \equiv \int_0^\infty \dd s \left( \hat \mu_m s \right)^{-t}\,
            \eta_+(s;\mu)
\end{aligned}
\end{equation}
conveniently generates the moments $L_n$ as the coefficients of its Taylor expansion around $t=0$:
\begin{equation}
    L_n(\mu, \mu_m)
        = \left(\frac{\dd}{\dd t} \right)^n  F_{[\eta_+]}(t;\mu, \mu_m) \big|_{t=0} \,.
\end{equation}
We similarly define the generating function%
\footnote{%
    The function $G_{[\phi_+]}$ 
    has also been used in the first analysis of the 
    RG equation for $\phi_+(\omega;\mu)$ in Ref.~\cite{Lange:2003ff}. 
    As has been shown in Ref.~\cite{Galda:2020epp}, this function also is useful so solve the 2-loop RG equations (referred there to as ''Laplace space'').
}
of the logarithmic moments of $\phi_+(\omega)$
\begin{equation}
\begin{aligned}
    \label{eq:phi:genfunc}
    G_{[\phi_+]}(t;\mu,\mu_m)
        & = \int_0^\infty \frac{\dd{\omega}}{\omega} \left(\frac{\mu_m}{\omega}\right)^{-t} \phi_+(\omega,\mu) \,,
\end{aligned}
\end{equation}
which is related to the previous generating function by
\begin{equation}
\label{eq:Fconv}
\begin{aligned}
    G_{[\phi_+]}(t;\mu,\mu_m)
        & = \frac{\Gamma(1+t)}{\Gamma(1-t)} \, e^{2 \gammaE \, t} \, F_{[\eta_+]}(t;\mu,\mu_m) 
          = F_{[\eta_+]}(t;\mu,\mu_m) \left( 1 + {\cal O}(t^3)\right) \qquad (t<1) \,.
\end{aligned}
\end{equation}
Evidently, the logarithmic moments of $\eta_+$ and $\phi_+$ coincide for $n=0,1,2$.
We regularly omit the argument $\mu_m$ in the logarithmic moments and the generating functionals for brevity.

The logarithmic moments $L_n$ obey simple coupled RG equations at one-loop (see also Ref.~\cite{Bell:2013tfa}),
\begin{align}
\label{rge:Ln}
\frac{\dd L_n(\mu,\mu_m)}{\dd \ln\mu}
    & = \GammaCusp(\mu) \, L_{n+1}(\mu,\mu_m)
      - \GammaCusp(\mu) \, \ln \frac{\mu}{\mu_m}  \, L_{n}(\mu,\mu_m) 
      - \gamma_+(\mu) \,
    L_n(\mu,\mu_m)\,.
\end{align}
For the particular choice $\mu_m=\mu_0$ one obtains the simple solution
\begin{equation}
\begin{aligned}
    L_n(\mu,\mu_0)
        & = e^{V(\mu,\mu_0)} \,  \sum_{k=0}^\infty
            \frac{\left[g(\mu,\mu_0)\right]^k}{k!} \, L_{n+k}(\mu_0,\mu_0) \,.
\end{aligned}
\end{equation}
The result for an arbitrary choice of $\mu_m$ follows from
\begin{equation}
\begin{aligned}
    L_n(\mu,\mu_m)
        & = \sum_{i=0}^n \binom{n}{i} \, L_i(\mu,\mu_0) \, \left(\ln\frac{\mu_0}{\mu_m}\right)^{n-i} \\
        &= \mathrm{e}^{V(\mu,\mu_0)} \, \left(\frac{\mu_0}{\mu_m}\right)^{-g(\mu,\mu_0)}
        \sum_{k=0}^\infty \frac{\left[ g(\mu,\mu_0) \right]^k}{k!} \, L_{n+k}(\mu_0,\mu_m)
        \,.
\end{aligned}
\end{equation}

The generating function $F[\eta_+](t;\mu, \mu_m)$ is particularly useful, because it has a simple scale dependence
that follows from \refeq{eta-evol},
\begin{align}
    F_{[\eta_+]}(t;\mu,\mu_m)
        & = e^{V(\mu;\mu_0)} \, \left(\frac{\mu_0}{\mu_m}\right)^{-g(\mu; \mu_0)}
            F_{[\eta_+]}(t+g(\mu;\mu_0);\mu_0,\mu_m)
\end{align}
This is the solution of the RG equation,
\begin{align}
  \frac{\partial F_{[\eta_+]}(t;\mu)}{\partial\ln\mu} &=
  - \left( \gamma_+(\mu) + \Gamma_c(\mu) \, \ln \frac{\mu}{\mu_m} \right) F_{[\eta_+]}(t;\mu)  + 
  \GammaCusp(\mu) \, \frac{\partial F_{[\eta_+]}(t;\mu)}{\partial t}
\end{align}
The two-loop RG equation for $G_{[\phi_+]}$ and its solution can be found in Ref.~\cite{Galda:2020epp},
which can easily be translated to $F_{[\eta_+]}$ via \refeq{Fconv}.

Finally, we note that the generating function $F_{[\eta_+]}$ can directly be obtained from the 
position-space LCDA via
\begin{align}
    F_{[\eta_+]}(t;\mu, \mu_m)
        & = \frac{\Gamma(1-t)}{t} \, \int_{-\infty}^\infty \frac{d\tau}{2\pi} \,
            \left(\frac{i}{\hat \mu_m\tau} \right)^{t} \, \tilde\phi_+(\tau;\mu) \,.
\end{align}


\subsection{Behavior at Small Momentum}
\label{sec:intro:behaviour-at-small-momentum}

While the theoretical expressions in the 
QCD factorization approach 
probe the logarithmic moments $L_n(\mu)$, typical 
applications of light-cone sum rules (LCSR) with $B$-meson LCDAs 
\cite{DeFazio:2005dx,Khodjamirian:2005ea,Khodjamirian:2006st,DeFazio:2007hw} require 
knowledge of the $B$-meson LCDAs for small momenta $\omega \leq s_0/2E$. Here 
$s_0$ is the effective threshold parameter in the hadronic model for the 
spectral density under consideration, and $E$ is the large recoil energy 
of the physical process. In such applications we may expand the LCDA around $\omega=0$,
in terms of its $n$th derivatives, assuming that the latter exist.
We then obtain
\begin{align}
    \phi_+^{(n)}(0;\mu) &=
    \frac{(-1)^{n+1}}{\Gamma(n)} \, \int_0^\infty ds \, s^{n} \, \eta_+(s;\mu)
     = - \frac{(-\hat\mu_m)^{-n}}{\Gamma(n)}  
      \, F_{[\eta_+]}(-n;\mu)
\end{align}
with the same generating function $F_{[\eta_+]}(t)$.
It is to be stressed here that $\phi_+^{(n)}(0)$ discussed above probe
the function $F_{[\eta_+]}$ at finite (discrete) values $t=-n$ ($n>0$), while
the previously discussed logarithmic moments $L_n$ probe the Taylor coefficients of the function $F_{[\eta_+]}$ around $t=0$.
Thus, LCSR and QCD factorization calculations are sensitive to different features of the underlying LCDA $\phi_+(\omega)$.
In particular, for phenomenological applications beyond the leading factorizable terms, 
it is not sufficient to consider only the behavior at $t = 0$ without also considering the
behavior at $t = -n$. On this point we disagree with the
conclusions drawn in Ref.~\cite{Galda:2020epp} where it has been argued that only the expansion of 
the function $F_{[\phi_+]}(t)$ around $t=0$ is phenomenologically relevant. 

We finally note that in the context of LCSR it has been observed that the strict expansion of the sum rule in $s_0/2E\ll 1$ is numerically not well converging. In this view, we propose another quantity to benchmark parametrizations
of the LCDA, the normalized Laplace transform\footnote{%
    This is not to be confused with what is referred to as the \emph{Laplace transform} in Ref.~\cite{Galda:2020epp},
    which we call the \emph{generating function}; see \refeq{phi:genfunc}.
}
\begin{align}
    \label{eq:laplace}
    \frac{\mathcal{L}[\phi_+](\zeta ,\mu)}{{\mathcal{L}}[\omega](\zeta)}
        & \equiv 
            \frac{\int\limits_0^\infty d\omega \, e^{-\zeta  \omega} \,
            \phi_+(\omega,\mu)}{\int\limits_0^\infty d\omega \, e^{-\zeta \omega} \, \omega}
        = \zeta^2 \,\tilde\phi_+(-i\zeta,\mu)\,.
\end{align}
For $\zeta\to\infty$ this reduces to $\phi_+'(0)$, while for large but finite values of $\zeta$ one is sensitive to the 
low $\omega$-behavior of the LCDA, regardless of whether the derivatives $\phi_+^{(n)}(0)$ exist.

\section{Parametrization of the \texorpdfstring{$B$}{B}-meson LCDA}
\label{sec:ansatz}

\begin{figure}[t!]
    \centering
    \subfloat[Color-marked domain of $\tau$]{
        \noindent\begin{minipage}{0.4\textwidth}
        \hspace{1cm}%
        \includegraphics[]{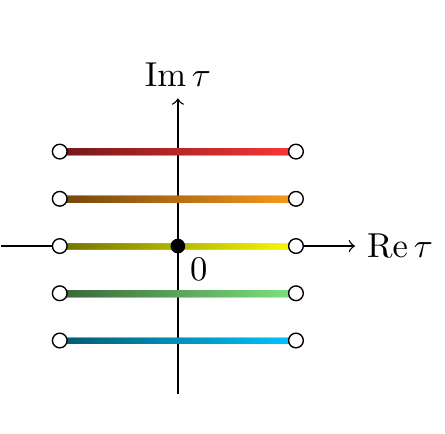}%
        \end{minipage}
    }
    \quad
    \subfloat[Image in the $y(\tau)$-plane.]{
        \noindent\begin{minipage}{0.4\textwidth}
        \hspace{0.55cm}%
        \includegraphics[]{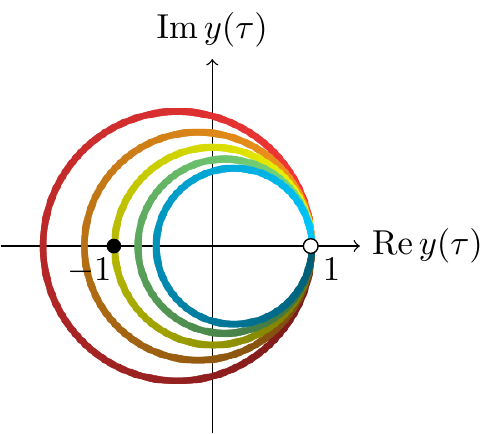}%
        \end{minipage}
    }
    \caption{%
        Illustration of the variable transform $\tau \mapsto y(\tau)$, with $\tau$ in units of $1/\omega_0$.
        Hollow small circles are understood to represent points at $\Re \tau \to \pm \infty$. 
        The small circles and colored lines correspond to each other in the left and right
        sketches.
    }
    \label{fig:yMapping}
\end{figure}

We propose a novel parametrization of the leading-twist $B$-meson LCDA that fulfills the properties discussed in
\refsec{intro}.
We start from the position-space LCDA and study the function $\chi[r]$ defined by the integral
\begin{align}
    \label{eq:chi_r}
    \chi[r](\mu) & \equiv \int\limits_{-\infty}^{\infty} \frac{d\tau}{2\pi} \left| \tilde\phi_+(\tau;\mu) \right|^2 \, |r(\tau;\mu)|^2
\end{align}
with some suitably chosen complex function $r(\tau;\mu)$.
It is instructive to rewrite this integral by means of the variable transform 
\begin{align}
    \label{eq:tauymap}
    \tau \mapsto y(\tau)
        & \equiv \frac{i\omega_0\tau - 1}{i\omega_0 \tau +1} \qquad \Leftrightarrow \qquad i\omega_0\tau(y) = \frac{1+y}{1-y} \,.
\end{align}
This introduces an auxiliary parameter $\omega_0$, which serves as a reference momentum scale.
The variable transform features the following properties, which are visualised in \reffig{yMapping}.
\begin{itemize}
	\item The point $\tau=0$ is mapped onto $y(\tau=0)=-1$.
	\item The points at $|\tau| \to\infty$ are mapped onto $\lim_{|\tau|\to\infty}y(\tau)=+1$.
	\item The real $\tau$ axis is mapped onto the standard unit circle $|y|=1$ in the complex $y$-plane.
	\item The half plane $\Im \tau < 0$ is mapped onto the open unit disk $|y| < 1$.
\end{itemize}

Using the new variable $y$, the integral in \refeq{chi_r} is mapped onto the integral along the boundary $\partial D$
of the unit disc in the complex $y$ plane
\begin{equation}
\begin{aligned}
\chi[r] &= \oint_{\partial D} \frac{\dd{y}}{2\pi}  \left|\tilde\phi_+(\tau(y)) \right|^2 |r(\tau(y))|^2 \, J(\tau(y))
    \cr &= \int\limits_{-\pi}^\pi \frac{d\theta}{2\pi}  \left|\tilde\phi_+(\tau(y)) \right|^2 |r(\tau(y))|^2 \, (-iy) \, J(\tau(y)) \Bigg|_{y=e^{i\theta}}\,.
\end{aligned}
\end{equation}
In the above $\theta = \arg(y)$ and we drop the scale dependence in the arguments for legibility.
The Jacobian $-i y J$ of the chain of variable transforms reads
\begin{align}
-i y \, J(\tau(y)) = -iy \, \frac{d\tau}{dy} = -\frac{2y}{\omega_0 \, (1-y)^2} = \frac{1+\omega_0^2\tau^2}{2\omega_0} \,.
\end{align}
This result inspires us to factorize the LCDA as
\begin{align}
\tilde\phi_+(\tau) & \equiv \frac{f_+(y(\tau))}{r(\tau) \, (1+ i\omega_0\tau)} \,,  \qquad 
\tilde\phi^*_+(\tau)  = \frac{f_+^*(y^*(\tau))}{r^*(\tau) \, (1- i\omega_0\tau)}\,,
\end{align}
This factorization simplifies the expression in \refeq{chi_r}
\begin{align}
    \label{eq:boundConstruction:L2}
    \chi
        & = \frac{1}{2\omega_0} \, 
            \int\limits_{-\pi}^\pi \frac{d\theta}{2\pi} \,  \left|f_+(y) \right|^2 \bigg|_{y=e^{i\theta}}\,,
\end{align}
which is similar in construction to unitarity bounds for hadronic form factors
and is therefore conducive to a systematic parametrization of $\phi_+$ (or equivalently $f_+$)
in terms of orthogonal polynomials on the $y$ unit circle; see Ref.~\cite{Caprini:2019osi}
and references therein. These polynomials coincide with
the monomials $y^n$. Negative powers of $y$ cannot appear in the parametrization of $\phi(\tau)$,
since they would induce singularities on the open unit disk, thereby violating \Prop{1}. The same holds for positive powers of $y^*$.
Therefore, the Taylor expansion of the function $f_+(y)$ corresponds to the Fourier series
\begin{align}
f_+(y) &\equiv \sum_{k=0}^\infty a_k \, y^k \,, \qquad f_+(y)\big|_{y=e^{i\theta}} = \sum_{k=0}^\infty a_k \, e^{i\theta k}
\end{align}
which yields
\begin{align}
\label{eq:akbound}
  \chi[r] &= \frac{1}{2\omega_0} \, \sum_{k=0}^\infty |a_k|^2 \,.
\end{align}
Therefore the sequence $\lbrace a_k \rbrace$ is an element of the $\ell^2$ space of sequences and must fall off faster than $\sqrt{1/k}$
as $k\to \infty$.
In this way we have constructed a converging expansion for the LCDA in position space.
The expansion can be truncated at some value $k=K$, and the truncation error is controlled by the value of the integral $2\omega_0\chi$.
From a different point of view,
as the partial series is monotonously growing with $K$,
a higher saturation due to the truncated parameters implies a better approximation by the truncated parametrisation.
In contrast to the unitarity bounds for hadronic form factors, however, the value of the bound
$\chi[r]$ for the leading-twist LCDA is presently not known.
We find that $\chi[r]$ is finite as long as
\begin{itemize}
    \item $|\lim_{\tau\to \infty} r(\tau;\mu) / \tau| < \infty$, by \Prop{4}; and
    \item $r(\tau; \mu)$ is regular as $\tau \to 0$, by \Prop{2}.
\end{itemize}

As our default choice for the weight function $r$
we take the simplest form that is consistent with the analyticity requirements of $\tilde\phi_+(\tau)$ and that leads to at least a $1/\tau^2$ suppression of $\tilde\phi_+(\tau)$ for $|\tau| \to \infty$, see \Prop{4},
\begin{align}
    \label{rdef}
    r(\tau;\mu_0) & \equiv 1 + i\omega_0\tau \,,
\end{align}
at a \emph{fixed} reference scale $\mu_0$ for which we require that $\ln \mu_0/\omega_0 \sim {\cal O}(1)$.
Other choices for $r(\tau,\mu_0)$ can  be reduced to Eq.~(\ref{rdef}) by readjusting the parameters $a_n$
in the truncated expansion in $y$.
Note that the choice of the weight function is neither unique nor meaningful for the expansion of the LCDA to infinite order in our basis -- it is critical, however, for the rate of convergence.
We preemptively point out that our choice reproduces the popular exponential model at trivial order, i.e., $a_0 = 1$ and $a_{k>0} = 0$ for some value of $\omega_0$.
Thus one can view our parametrisation as a systematic extension of the exponential model.
Via \refeq{tildephi+-from-eta+} our choice for $r(\tau,\mu_0)$ leads to simple expressions for the dual LCDA, see below.
With this -- as one of the central results of our paper -- 
we obtain the following parametrization of the $B$-meson LCDA in position space,
\begin{equation}
\begin{aligned}
\label{eq:para_default}
    \tilde \phi_+(\tau;\mu_0)
        & = \frac{(1-y(\tau))^2}{4} \, \sum_{k=0}^K a_k(\mu_0) \, (y(\tau))^k \\
        & = \frac{1}{(1+i\omega_0\tau)^2} \, \sum_{k=0}^K a_k(\mu_0) \left( \frac{i\omega_0 \tau -1}{i\omega_0\tau +1}\right)^k\,,
\end{aligned}
\end{equation}
which reflects an expansion in the point $\tau = -i/\omega_0$.
It is to be emphasized that our parametrization does not aim to cover the singular behavior of the LCDA in the local limit $\tau\to 0$. Actually, as can be seen from \refeq{para_default}, the values of $\tilde \phi(\tau)$ and all of its derivatives are finite at $\tau =0$ for any finite value of the truncation $K$, which in turn implies the existence of all non-negative moments $\langle \omega^n\rangle$ in momentum space. Nevertheless -- as we will show in Section~\ref{sec:models} 
-- the parametrization can be used at small but \emph{finite} values
$|\tau_0| \sim 1/\mu_0 \ll 1/\omega_0$ to implement the constraints from the local OPE on $\tilde\phi_+(\tau,\mu_0)$ \cite{Kawamura:2008vq}. In this way, we can also mimic the ``radiative tail'' for intermediate values $\omega \sim \mu_0 \gg \omega_0$ of the $B$-meson LCDA in momentum space \cite{Lee:2005gza}.
Moreover, 
as we show below, we can consistently include the RG evolution within the framework of our parametrization by suitably adjusting the coefficients $a_k(\mu)$ and the function $r(\tau,\mu)$.

\subsection{LCDA in Dual Space and Logarithmic Moments}
\label{sec:dual_space_and_log_mom}

In dual space  our parametrization proposed in \refeq{para_default} translates via \refeq{tildephi+-from-eta+} to
\begin{equation}
\begin{aligned}
    \eta_+(s;\mu_0)
        & = e^{-s\omega_0} \, \sum_{k=0}^K \frac{(-1)^k \, a_k(\mu_0)}{1+k} \, L_k^{(1)}(2\omega_0 s) 
\,,
\end{aligned}
\end{equation}
where $L_k^{(1)}$ are the associated Laguerre polynomials.
The expansion coefficients can be obtained from the orthogonality of the Laguerre polynomials resulting in the projection
\begin{align}
    a_k(\mu_0) &= 4 \, (-1)^k \, \omega_0 \int\limits_0^\infty \dd{s} \,
    (\omega_0 s) \, \Exp{-s\omega_0} \, L_k^{(1)}(2\omega_0 s) \, \eta_+(s; \mu_0) \,.
\end{align}
The expression for the integral $\chi$ reads 
\begin{align}
2\omega_0 \, \chi[r](\mu_0)= \sum_{k=0}^K |a_k|^2 &= 2\omega_0 \,
 \int_0^\infty ds \left( \omega_0^2 \left|s \eta_+(s;\mu_0) \right|^2 
 + \left| \frac{d}{ds} \, (s \eta_+(s;\mu_0)) \right|^2\right) 
 \cr 
 &\equiv 2\omega_0 \, \int_0^\infty ds \, \int_0^\infty ds' 
 \left( s' \eta_+^*(s';\mu_0)\right)  R_{[\eta]}(s',s) \left( s \eta_+(s;\mu_0) \right) \,,
\end{align}
with the corresponding integral transform of our default choice of $|r(\tau; \mu_0)|^2$,
\begin{align}
    \label{eq:Rdual}
    R_{[\eta]}(s',s)
        & = \omega_0^2 \, \delta(s-s') - \delta''(s-s') \,.
\end{align}

The generating function for the logarithmic moments can be expressed as
\begin{align}
    F_{[\eta_+]}(t;\mu_0, \mu_m)
        & = \frac{\Gamma(1-t)}{\omega_0} \, 
            \left(\frac{\hat\mu_m}{\omega_0} \right)^{-t}  \, \sum_{k=0}^K a_k \, {}_2F_1(-k,1+t; 2; 2) \,.
\end{align}
This result can be obtained using Cauchy's residue theorem, where poles of higher order result
in derivatives of the integrand, which can be expressed in terms of binomial coefficients.
The hypergeometric function with negative first argument simplifies to polynomials in $t$
of $n^\text{th}$ order,
\begin{align}
    {}_2F_1(0, 1 + t; 2; 2)
        = 1 \,, & &
    {}_2F_1(-1, 1 + t; 2; 2)
        = -t \,, & &
    {}_2F_1(-2, 1 + t; 2; 2)
        = \frac{1}{3} \left(1 + 2t^2\right) \,, & & \mbox{etc.}\,,
\end{align}
which are even functions of $t$ for even $n$, and odd functions of $t$ for odd $n$.
From this we obtain the expressions for the first few logarithmic moments within our parametrization:
\begin{align}
    \label{eq:ansatz:L0}
    L_0(\mu_0)
        & = \frac{1}{\omega_0} \, \sum_{n=k}^K \frac{1-(-1)^{k+1}}{2} \, \frac{a_k(\mu_0)}{k+1}
          = \frac{a_0 +a_2/3 + \ldots }{\omega_0}\,,\\
    \label{eq:ansatz:L1}
    L_1(\mu_0)
        & = -\left(\ln{\frac{\hat\mu_m}{\omega_0}} - \gamma_{\mathrm{E}}\right)\, L_0(\mu_0)
        + \frac{1}{\omega_0} \, \sum_{k=0}^K a_k \left[\dv{t} \, {}_2F_1(-k, 1+t; 2; 2)\right]_{t = 0}\,,\\
        & = -\left(\ln{\frac{\hat\mu_m}{\omega_0}} - \gamma_{\mathrm{E}}\right)\, L_0(\mu_0)
        + \frac{-a_1 - 2/3 \, a_3 + \dots}{\omega_0} \nonumber \,,\\
    \label{eq:ansatz:L2}
    L_2(\mu_0)
        & = \left[\frac{\pi^2}{6} - \left(\ln{\frac{\hat\mu_m}{\omega_0}} - \gamma_{\mathrm{E}}\right)^2\right]\, L_0(\mu_0)
          - 2 \left(\ln{\frac{\hat\mu_m}{\omega_0}} - \gamma_{\mathrm{E}}\right) L_1(\mu_0)\\
    \nonumber
        & + \frac{1}{\omega_0} \, \sum_{k=0}^K a_k \left[\dv[2]{t} \, {}_2F_1(-k, 1+t; 2; 2)\right]_{t = 0}\,, \\
        & = \left[\frac{\pi^2}{6} - \left(\ln{\frac{\hat\mu_m}{\omega_0}} - \gamma_{\mathrm{E}}\right)^2\right]\, L_0(\mu_0)
          - 2 \left(\ln{\frac{\hat\mu_m}{\omega_0}} - \gamma_{\mathrm{E}}\right) L_1(\mu_0)
          + \frac{4/3 \, a_2 + 4/3 a_4 + 56 / 45 a_6 + \dots}{\omega_0} \nonumber\,.
\end{align}
We emphasize that the properties of the confluent hypergeometric functions appearing in
\refeq{ansatz:L1} and \refeq{ansatz:L2} induce
for $\mu_m = \omega_0 e^{-\gamma_{\mathrm{E}}}$ that the logarithmic moments $L_0$ and $L_2$ only depend on coefficients $a_k$ with even index $k$.
Likewise, the logarithmic moment $L_1$ only depends on coefficients $a_k$ with odd index $k$.
The sequence generated by the hypergeometric functions and their derivatives is a null sequence. This brings along
two important properties:
\begin{enumerate}
    \item convergence of the series representation of the logarithmic moments is possible, even if the series $\sum_{k} a_k$ were
    not convergent; and
    \item at the reference scale $\mu_m = \omega_0 e^{-\gamma_{\mathrm{E}}}$, the logarithmic moments $L_0$ and $L_1$ can be chosen
    independently of each other, i.e., there is no model correlation between the two even for a truncated expansion.
\end{enumerate}

\subsection{Momentum-space LCDA and Behavior at \texorpdfstring{$\omega = 0$}{the Origin}}
\label{sec:mom_space_and_small_omega}

The Fourier transform of our parametrisation in \refeq{para_default} yields the corresponding expansion
of the momentum-space LCDA in terms of generalized Laguerre polynomials:
\begin{align}
    \phi_+(\omega;\mu_0)
        & = \frac{\omega \, \mathrm{e}^{-\omega/\omega_0}}{\omega_0^2} \, \sum_{k=0}^K \frac{a_k(\mu_0)}{1+k} \,
            L_k^{(1)}(2\omega/\omega_0) \,.
            \label{paramom}
\end{align}
The expansion coefficients can be obtained from the orthogonality of the Laguerre polynomials resulting in the projection
\begin{align}
    a_k(\mu_0) &= 4 \int\limits_0^\infty \dd{\omega} \, \Exp{-\omega/\omega_0} \, L_k^{(1)}(2\omega/\omega_0) \, \phi_+(\omega; \mu_0) \,.
    \label{eq:projan}
    \end{align}
Alternatively, they can be obtained as the series coefficients of a single integral expression,
\begin{align}
    a_k(\mu_0)
        & = \eval{
        \frac{1}{k!} \pdv[k]{t} \int\limits_0^\infty \dd{\omega}
        \frac{4}{(1-t)^2} \, \exp{\frac{(t+1) }{(t-1)}\, \frac{\omega}{\omega_0}} \, \phi_+(\omega; \mu_0)
    }_{t=0}
    \,.
\end{align}
We highlight that truncating the parametrization at $K=0$ and fixing $a_0 = 1$ yields the popular exponential model~\cite{Grozin:1996pq}.
We emphasize again that the auxiliary parameter $\omega_0$ in our parametrization has no physical meaning and only serves as a reference scale, which does, however, influence the convergence of the expansion.
The integral $\chi[r]$ can 
be expressed in terms of the momentum-space LCDA as 
\begin{align}
2\omega_0 \, \chi[r](\mu_0)= \sum_{n=0}^N |a_n|^2 &=  2\omega_0 \, 
\int_0^\infty d\omega \left( \left|\phi_+(\omega;\mu_0)\right|^2 + \omega_0^2 \, 
\left| \frac{d\phi_+(\omega;\mu_0)}{d\omega} \right|^2 \right)
\cr & \equiv  2\omega_0 \, 
\int_0^\infty d\omega \, \int_0^\infty d\omega' \, \phi_+^*(\omega';\mu_0)\, R_{[\phi]}(\omega',\omega) \, \phi_+(\omega;\mu_0) 
\label{eq:chi:mom}
\end{align}
with the Fourier transform of our default choice of $|r(\tau; \mu_0)|^2$,
\begin{align}
    R_{[\phi]}(\omega',\omega) &= \delta(\omega-\omega') - \omega_0^2 \, \delta''(\omega-\omega') \,.
\end{align}
We note the similarity with the corresponding expressions in \refeq{Rdual}, which
strengthens the notion of $\eta_+(s)$ being a ``dual space representation'' of $\phi_+(\omega)$.

The Taylor expansion of $\phi_+(\omega;\mu_0)$ around $\omega=0$ is related to our expansion coefficients as follows: 
\begin{equation}
\label{eq:derivexp}
\begin{aligned}
    \phi'_+(0;\mu_0)
        & = \frac{1}{\omega_0^2} \, \sum_{k=0}^\infty  a_k \,, \\
    \phi''_+(0;\mu_0)
        & = - \frac{1}{\omega_0^3} \, \sum_{k=0}^\infty  (2k+2) \, a_k  \,, \quad \mbox{etc.}
\end{aligned}
\end{equation}
where the coefficients $a_k$ in the expressions for the $n^{\rm th}$ derivative are weighted by numbers growing 
power-like with $k^{n-1}$. 
Since $\phi_+'(0)$ exists, the coefficients $a_k$ must either have alternating signs, or they must fall off
faster than $1/k$. However, we cannot constrain the convergence of the series representation for the higher derivatives in \refeq{derivexp}.

As already mentioned above, one should keep in mind that actual applications of $B$-meson LCDA in QCD sum rules
consider \emph{integrals} of $\phi_+(\omega)$ over a finite interval of small $\omega$ values.
Typically, the integrals are computed after Borel transformation, such that the appearing
expressions are Laplace transformations of the the momentum space representation $\phi_+(\omega)$.
For this reason we consider the normalized Laplace transformation of $\phi_+(\omega)$ in \refeq{laplace} at large values $\zeta \equiv n \, t_0$ as an example,
\begin{align}
\ell_n(\mu,t_0) &\equiv
    n^2 t_0^2 \, \tilde\phi_+(-i n t_0; \mu) \\
\ell_n(\mu,1/\lambda_B) &=
    \frac{1}{\lambda_B^2}
    \frac{n^2}{(1+n \, \omega_0/\lambda_B)^2} \, \sum_{k=0}^\infty a_k(\mu)
    \left( \frac{n \, \omega_0/\lambda_B-1}{n \, \omega_0/\lambda_B+1} \right)^k
\end{align}
The expansion of the quantities $\ell_n$ in terms of the coefficients $a_k$ in our parametrization converges for
$0 < n < \infty$.

\subsection{RG Evolution}
\label{sec:ansatz:rge}

At one-loop accuracy, the RG evolution is multiplicative in dual space. Starting from our default parametrization at a fixed scale $\mu_0$
we obtain
\begin{equation}
\label{eq:ansatz:rge:sspace}
\begin{aligned}
    \eta_+(s;\mu)
        & = e^{V(\mu,\mu_0)} \, (\hat \mu_0 s)^{-g(\mu,\mu_0)} \, e^{-s\omega_0}
            \sum_{k=0}^K \frac{(-1)^k \, a_k(\mu_0)}{1 + k} \, L_k^{(1)}(2 \omega_0 s) \,.
\end{aligned}
\end{equation}
We discuss three different ways of implementing the above scale evolution for our parametrisation:
\begin{enumerate}
    \item Use the above equation as is, that is to say, the respective forms in momentum and position space.
    The expansion of $\phi_+(\omega, \mu)$ remains in terms of the coefficients $a_k(\mu_0)$
    while the basis of functions of the parametrisation changes.

    \item Project \refeq{ansatz:rge:sspace} onto our parametrisation \emph{with our default choice of $r(\tau)$}.
    We obtain a matrix
    \begin{equation*}
        a_k'(\mu) \sim \sum_{k=0}^{K} \mathcal{R}_{k'k} \, a_{k}(\mu_0) \,,
        \quad k=0, 1, \dots, \infty \,.
    \end{equation*}
    Put differently, the basis of functions remains the same at all scales while the coefficients evolve.
    In this approach, starting with a truncated set of coefficients $a_{k}(\mu_0)$, $k \leq K$, the evolution generates an \emph{infinite set} of coefficients $a_{k'}(\mu)$.
    For practical applications, we thus need to truncate a second time ($k' \leq K'$). The requirements for the secondary truncation
    parameter $K'$ can be studied numerically.

    \item Project \refeq{ansatz:rge:sspace} onto a modified parametrisation \emph{with a scale-dependent choice of $r(\tau, \mu) = \tilde{r}$}.
    The function $\tilde{r}$ is chosen such that we achieve a coefficient RGE similar to the previous approach,
    with the additional feature that $K'=K$ by construction, i.e., no secondary truncation is necessary:
    \begin{equation*}
        \tilde{a}_{k'}(\mu) \sim \sum_{k=0}^K \tilde{\mathcal{R}}_{k'k} \, a_{k}(\mu_0) \,,
        \quad k'=0, 1, \dots, K\,,
    \end{equation*}
    with $\tilde{a}_{k}(\mu_0) = a_{k}(\mu_0)$.
    This approach guarantees that the coefficients remain bounded, $|\tilde{a}_k(\mu)|< \sqrt{2 \omega_0 \tilde{\chi}(\mu)}$ at any scale $\mu$,
    where $\tilde{\chi} \equiv \chi[\tilde{r}]$.
    
\end{enumerate}
From now on we abbreviate $g \equiv g(\mu, \mu_0)$ and $V \equiv V(\mu, \mu_0)$.\\

First, transforming \refeq{ansatz:rge:sspace} into momentum space, we obtain:
\begin{equation}
\label{eq:ansatz:rge:omegaspace}
\begin{aligned}
    \phi_+(\omega; \mu)
        & = e^{V} \left(\frac{\hat{\mu}_0}{\omega_0}\right)^{-g} \, \frac{\omega}{\omega_0^2}\\
        &   \phantom{=}
            \sum_{k=0}^K
            (-1)^k \frac{a_k(\mu_0)}{1 + k}\,
            \left[\frac{1}{k!}\dv[k]{t}
                \left(\frac{1 - t}{1 + t}\right)^{-g} \frac{\Gamma(2 - g)}{(1 + t)^2}\,
                {}_1F_1\left(2 - g; 2; \frac{t - 1}{t + 1} \frac{\omega}{\omega_0}\right)
            \right]_{t = 0} \,.
\end{aligned}
\end{equation}
The derivatives produce an expansion in ${}_1F_1(n - g; n; -x)$, where  ${}_1F_1(n - g; n; -x) \to e^{-x}$
for $g\to 0$.
Here, the coefficients $a_k(\mu_0)$ fulfill a bound obtained at the initial scale.
Numerical calculations using the LCDA require evaluations of the hypergeometric functions
with non-integer parameters.
Obviously, this procedure is not very convenient for numerical evaluation, especially when taking the necessary
variation of $\omega_0$ into account.\\

For the second case we obtain:
\begingroup
\newcommand{\Mg}{g}
\newcommand{\MV}{V}
\begin{align}
    \label{eq:ansatz:rge:parameter_rge}
    a_{k'}(\mu) & =
    \mathrm{e}^{\MV}
    \left(\frac{\hat\mu_0}{2 \omega_0}\right)^{-\Mg} \,
    \sum_{k=0}^K \mathcal{R}_{k'k}(\mu, \mu_0) \, a_{k}(\mu_0)
    \,,
\end{align}
where the matrix $\mathcal{R}(\mu, \mu_0)$ reads
\begin{align}
    \label{eq:ansatz:rge:r_matrix}
    \mathcal{R}_{k'k}(\mu, \mu_0) &=
    \frac{(-1)^{k'+k}}{1+k} \,
    \int\limits_0^\infty \dd{z} \,
    z^{1-\Mg} \,
    \Exp{-z} \, L_{k'}^{(1)}(z)
     \, L_{k}^{(1)}(z) \\
     &=
     \eval{
        \Gamma(2-\Mg) \,
        \frac{(-1)^{k'+k}}{(1+k)!k'!} \, \dv[k]{u}\dv[k']{v} \, 
        \frac{1}{(1-uv)^2}
        \left(
            \frac{1-uv}{(u-1)(v-1)}
        \right)^{\Mg}
    }_{u, v=0}
    \,,
\end{align}%
\endgroup
with $\mathcal{R}_{k'k}(\mu_0, \mu_0) = \delta_{k'k}$.
This approach is promising for calculation-intensive numerical applications:
for fixed $\mu \neq \mu_0$, the matrix needs to be
calculated only once for any given secondary truncation $K'$; the $\omega_0$-dependence is simply multiplicative; and
observables can fully benefit from the simple and efficient representation.
A closer inspection of \refeq{ansatz:rge:r_matrix} shows that the off-diagonal elements of $\mathcal{R}$ are suppressed by $O(g, 1/|k' - k|)$,
and therefore the secondary truncation $K'<\infty$ is justified.
We quantitatively confirm at hand of a model in \refsec{models:lee_neubert} that stable convergence can be achieved in a realistic scenario,
even when $K'\approx K$.\\

In the third case, we consider the transformation of \refeq{ansatz:rge:sspace} to position space, which yields a
new parametrization:
\begin{equation}
    \label{eq:ansatz:rge:tauspace}
    \tilde \phi_+(\tau;\mu)
        = e^V \, \Gamma(1-g) \, \left(\frac{\omega_0}{\hat \mu_0}\right)^g \left( \frac{1-y}{2} \right)^{2}
            \left(\frac{1+y}{2}\right)^{-g} \sum_{k'=0}^{K'=K} \tilde{a}_{k'}(\mu) \, y^{k'} \,.
\end{equation}
We emphasize the truncation at $K'=K$.
The new coefficients read
\begin{equation}
    \label{eq:coeffEvolution}
    \tilde{a}_{k'}(\mu) = \sum_{k=k'}^{K}
    \tilde{\mathcal{R}}_{k'k}(\mu, \mu_0) \, a_k(\mu_0)\,.
\end{equation}
The transformation is given by an upper triangular matrix
\begin{equation}
    \tilde{\mathcal{R}}_{k'k}(\mu, \mu_0)
        = \begin{dcases}
            \frac{(-1)^{1+k} \, \Gamma(k+g-k')}{(1+k) \, \Gamma(g-1-k') \, \Gamma(1+k-k') \, \Gamma(1+k')} & k' \geq k\\
            0 \phantom{\frac{\Gamma}{\Gamma}} & \text{otherwise}
        \end{dcases}\,. 
\end{equation}
From the prefactor in \refeq{ansatz:rge:tauspace} we can read off the desired function $\tilde{r}$:
\begin{align}
    \tilde{r} \equiv e^{-V} \, \frac{(i\hat\mu_0 \tau)^g \, (1 + i\omega_0\tau)^{1-g} }{\Gamma(1-g)}  \,.
\end{align}
Using this function
\begin{align}
    \tilde\chi
        & = \int\limits_{-\infty}^{\infty} \frac{d\tau}{2\pi} \, \left|\tilde\phi_+(\tau;\mu) \right|^2 
            \left|\tilde{r}(\tau)\right|^2 = \frac{1}{2\omega_0} \, \sum_{k'=0}^{K'=K} |\tilde{a}_{k'}(\mu)|^2 \,.
\end{align}
This modification comes at the expense that the functional basis, especially in momentum space, becomes more complicated as $g$ enters the
parametrisation non-trivially. We remark in closing that this third approach works for the one-loop RG evolution. However, we
do not expect it to work in the two-loop case, where the RG equation in dual space becomes inhomogeneous~\cite{Braun:2019wyx}.

\subsection{Application to Higher Twist}
\label{sec:higher-twist}

At higher twist, further LCDAs contribute to the calculation of exclusive processes.
Given sufficient knowledge about their analytic properties, 
our approach
can and should be applied to these as well. Here we discuss briefly the application to the second
two-particle LCDA of the $B$-meson,
which is denoted as $\tilde\phi_-(\tau)$.
It is commonly split into two terms, $\tilde{\phi}_-(\tau) = \tilde{\phi}_-^{\text{(WW)}} + \tilde{\phi}_-^{\text{(tw3)}}$.
The first term refers to the so-called Wandzura-Wilczek limit and is related to the leading-twist LCDA $\tilde{\phi}_+(\tau)$.
The second term $\tilde{\phi}_-^{\text{(tw3)}}$ is genuinely of twist-three origin and
is related to the three-particle LCDA at twist three~\cite{Kawamura:2001jm,Braun:2017liq}.
Below, we only discuss the Wandzura-Wilczek term, and therefore drop the superscripts
for simplicity.
Its RG equations can be found in Ref.~\cite{Descotes-Genon:2009jif}, see also Ref.~\cite{Bell:2008er}.

In position space, the equation of motion connecting the Wandzura-Wilczek term with the leading-twist
LCDA reads (see e.g.\ \cite{Beneke:2000wa})
\begin{align}
    \tilde\phi_+(\tau) &= \tau \dv{\tau} \tilde\phi_-(\tau) + \tilde\phi_-(\tau) \,.
\end{align}
We can rewrite this equation in terms of the variable $y$, which yields:
\begin{equation}
\begin{aligned}
    \frac{\tilde{\phi}_+(\tau(y))}{(1 - y)^2}
        & = \frac{1}{2} \dv{y} \left[\frac{1 + y}{1 - y} \tilde{\phi}_-(\tau(y)) \right] \,.
\end{aligned}
\end{equation}
The solution to this differential equation can be expressed in terms of $f_+(y)$:
\begin{equation}
    \label{eq:phitildem}
\begin{aligned}
    \tilde{\phi}_-(\tau(y))
        & = 2\,  \frac{1-y}{1+y} \, \int_{-1}^y \dd{x} \, \frac{\tilde{\phi}_+(\tau(x))}{(1 - x)^2}
          = \frac{1}{2} \,  \frac{1-y}{1+y} \, \int_{-1}^y \dd{x} \, f_+(x)\\
        & \equiv
            \frac{1}{2} \,  \frac{1-y}{1+y} \, \left[f_-(y) - f_-(-1)\right]\,.
\end{aligned}
\end{equation}
The integration constant and the lower boundary are fixed by requiring that the local limit $y\to -1$
coincides with the local limit of $\tilde{\phi}_+(\tau)$.\\

Our parametrisation for $\tilde\phi_+$ translates to the following expansion of $f_-$:
\begin{align}
    f_-(y)
        & = \sum_{k=0}^K \frac{a_k}{1 + k} y^{1 + k}
          = \int_0^y \dd x \, f_+(x)\,.
\end{align}
The asymptotic behaviour of $\tilde{\phi}_-(\tau)$ for $|\tau|\to \infty$ is $1/\tau$, as expected.
The coefficients $a_k$ enter with $1/(1+k)$ suppression, yielding a more convergent expansion than
for the leading-twist LCDA. Hence, the truncation error for $\tilde{\phi}_-$ is under the same level of
control as for $\tilde{\phi}_+$.\\

We obtain for the momentum space representation of the Wandzura-Wilczek term in our parametrization:
\begin{equation}
\label{phim}
\begin{aligned}
 \phi_-(\omega; \mu_0)
    & = \int_\omega^\infty \frac{\dd\eta}{\eta} \, \phi_+(\eta;\mu_0)\,, \\
    & = \frac{1}{\omega_0} \, 
        \sum_{k=0}^K \frac{a_k(\mu_0)}{1+k} \, 
        \sum_{i=0}^k \frac{(-2)^i}{i!} \,\binom{k+1}{k-i} \,\Gamma(1+i,\omega/\omega_0) \\
    & = \frac{e^{-\omega/\omega_0}}{\omega_0} 
        \left\{
            a_0 + \frac{a_2}{3} -
            \left( a_1 + \frac{2 a_2}{3}  \right) \frac{\omega}{\omega_0} 
            +\frac{2a_2}{3} \, \frac{\omega^2}{\omega_0^2} + \ldots
        \right\} \,,
\end{aligned}
\end{equation}
where we obtain our result through integration of the explicit representation of the Laguerre polynomials.
Closed solutions can be obtained by Fourier transformation of the basis functions that appear in
\refeq{phitildem}, e.g.,
\begin{equation}
\begin{aligned}
    \frac{1 - y}{1 + y}
        & \mapsto \frac{1}{2\omega_0}\,,  &
    \frac{1 - y}{1 + y} y
        & \mapsto \frac{1}{2\omega_0} \left[-1 + 4 e^{-\omega/\omega_0}\right]\,, &
        & \text{etc.}
\end{aligned}    
\end{equation}

\newpage

\section{Application to Existing Models}
\label{sec:models}

For phenomenological applications, simple models of the $B$-meson LCDAs at a low reference scale $\mu_0$
are commonly used. These models typically feature a small number of parameters. Here, we
study four of these models in regard to how they can be captured by our parametrization.
Our selection of models is chosen to showcase a wide variety of behavior.
For ease of comparison, we discuss each model in terms of the dimensionless ratio 
\begin{equation}
    \xi \equiv \frac{\omega_0}{\lambda_{B}^\text{model}(\mu_0)} > 0\,,
\end{equation}
where $\omega_0$ is the auxiliary scale in our general parametrization, and $\lambda_{B}^\text{model}(\mu_0)$ the prediction for the inverse moment in the specific model.
In each case, the coefficients $a_k$ are matched onto the respective model by means of \refeq{projan}.

For each of the considered models, we study the saturation of four of the relevant quantities as a function of the order of truncation $K$.
The saturation of a quantity $X$ is defined as
\begin{equation}
    \mathrm{Sat}\left[ X \right]_K
        \equiv \frac{\sum_{k=0}^K X\big|_k}{\sum_{k=0}^\infty X\big|_k}
        \,,
\end{equation}
where $X\big|_k$ is the contribution by the coefficient $a_k$ in our parametrization.
In the following we use these quantities:
\begin{itemize}
    \item the result for the integral $\chi$, which provides the bound for the expansion parameters $a_k$ in our parametrization,
    \begin{align}
        \chi\big|_k
            & \equiv \frac{1}{2\omega_0} \abs{a_k}^2
    \end{align}

    \item the derivative of the momentum-space LCDA at the origin,
    \begin{align} 
        \phi_+'(0) \big|_k \equiv 
            \frac{1}{\omega_0^2} \, a_k
    \end{align} 

    \item the normalized Laplace transform at $\zeta=n/\lambda_B$,
    \begin{align} 
        \ell_n \big|_k
            \equiv 
                \frac{1}{\lambda_B^2} \frac{n^2}{(1+n\xi)^2} \, \left(\frac{n\xi-1}{n\xi+1} \right)^k a_k\,;
    \end{align} 

    \item the inverse logarithmic moment,
    \begin{align}
        \label{sat:mom}
        L_0\big|_k = \lambda_{B}^{-1} \big|_k \equiv
        \frac{1}{2\omega_0} \, 
            \frac{1+(-1)^{k}}{1+k} \, a_k \,,
    \end{align}

    \item and the normalized first logarithmic moment,
    \begin{align}
        \label{partial:L1overL0}
        \sigma_B \big|_k
            \equiv - \lambda_B \, L_1(\mu_m = \operatorname{e}^{-\gamma_\text{E}} \lambda_B) \big|_k 
            & =
            \begin{dcases}
                - \ln \xi & k = 0 \\
                - \frac{a_k}{\xi} \left[\dv{t} \, {}_2F_1(-k, 1+t; 2; 2)\right]_{t = 0} & k \geq 1 \text{ and odd} \\
                \phantom{-} 0 & k \geq 1 \text{ and even}
            \end{dcases}
        \,.
    \end{align}
\end{itemize}
All of the above quantities, including the expansion coefficients $a_k$, 
are implicitly understood to be evaluated at a renormalisation scale $\mu = \mu_0$.

We further consider the ``relative growth'' of some of the quantities. Is is defined as
\begin{equation}
    \mathrm{Gr}[ X ]_K \equiv \frac{X\big|_K}{\sum_{k=0}^K X\big|_k}\,.
    \label{growth}
\end{equation}
For each model, we consider the relative growth of the contribution to the bound $\chi$.
We also apply the relative growth to any of the benchmark quantities defined above
if said quantity is ill-defined for a specific model. The relative growth is also
instrumental for model-independent phenomenological studies as a proxy for the corresponding saturation.
Its reliability, however, can only be tested in model studies, as the convergence rate of the
parametrisation is not known a-priori.\\

\subsection{Exponential Model}
\label{sec:models:exponential}

\newcommand{\omegaM}[0]{\omega_{\mathrm{M}}}

\begin{figure}[p!]
        \subfloat[%
            \label{fig:exp:coeff}
            Values for the expansion coefficients $a_k$
        ]{%
            \includegraphics[scale=0.95]{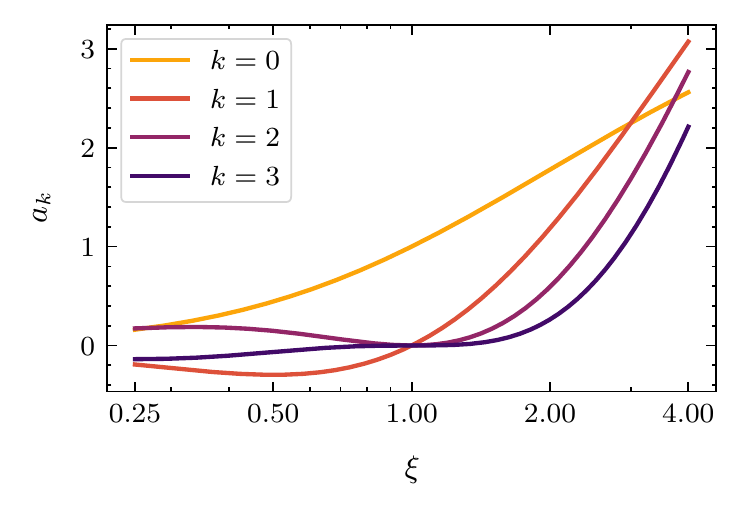}
        }
        \hfill
        \subfloat[%
            \label{fig:exp:window}
            Variability of the momentum-space LCDA for different truncations $K$ in the interval $1/2<\xi<2$.
        ]{%
            \includegraphics[scale=0.95]{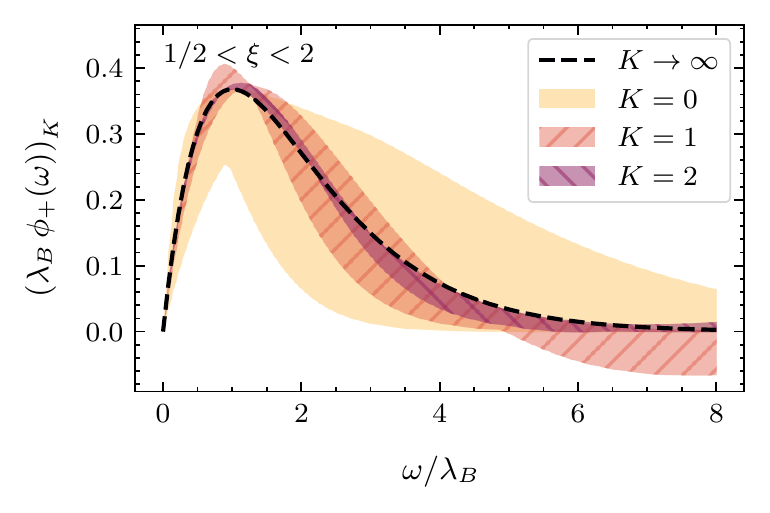}
        }\\
        \subfloat[%
            \label{fig:exp:sat_chi}
            Saturation of the integral bound $\chi$
        ]{%
            \includegraphics[scale=0.95]{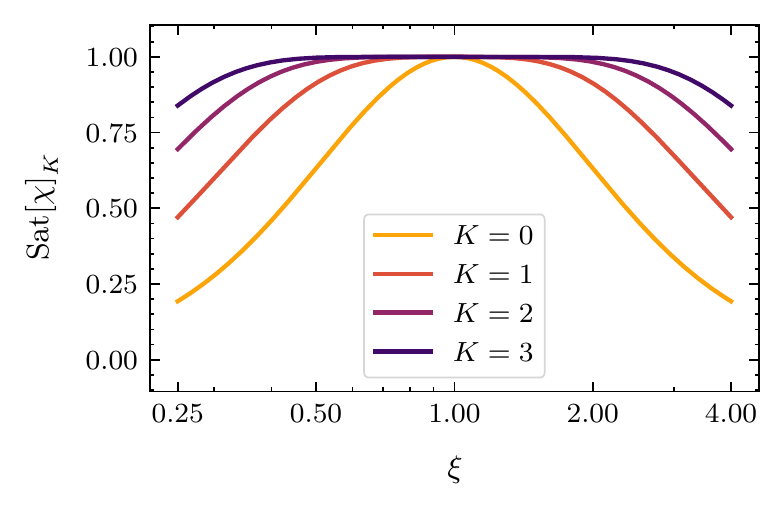}
        }
        \hfill
        \subfloat[%
            \label{fig:exp:growth_chi}
            Relative growth of the integral bound $\chi$
        ]{%
            \includegraphics[scale=0.95]{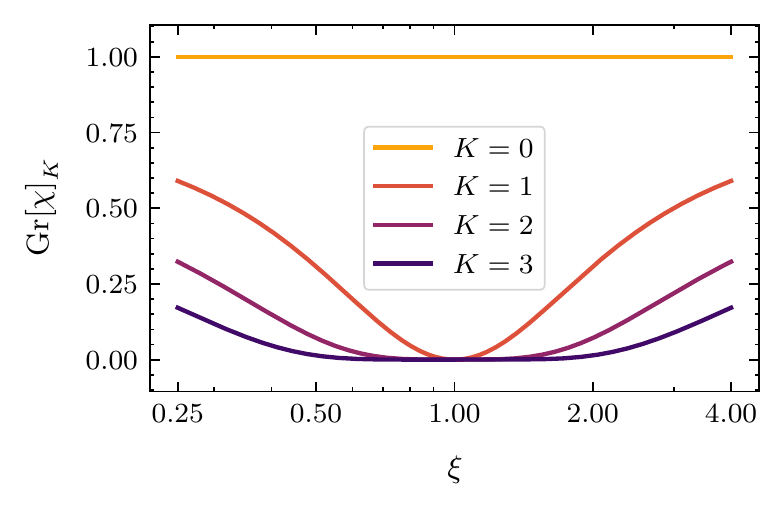}
        }\\
        \subfloat[%
            \label{fig:exp:sat_L0}
            Saturation of the inverse moment $L_0=\lambda_B^{-1}$.
        ]{%
            \includegraphics[scale=0.95]{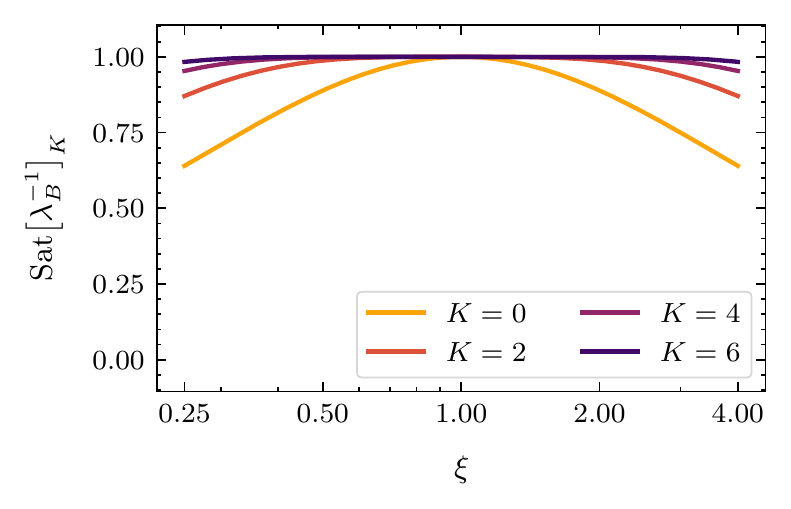}
        }
        \hfill
        \subfloat[%
            \label{fig:exp:sigma}
            Value of $\sigma_B$ for different truncations $K$.
        ]{%
            \includegraphics[scale=0.95]{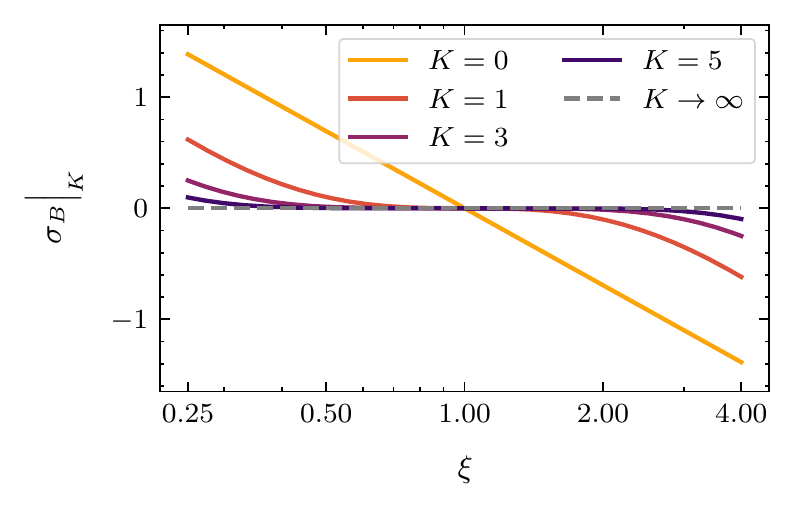}
        }\\
        \subfloat[%
            \label{fig:exp:sat_deriv}
            Saturation of the derivative of $\phi_+(\omega)$ at the origin.
        ]{%
            \includegraphics[scale=0.95]{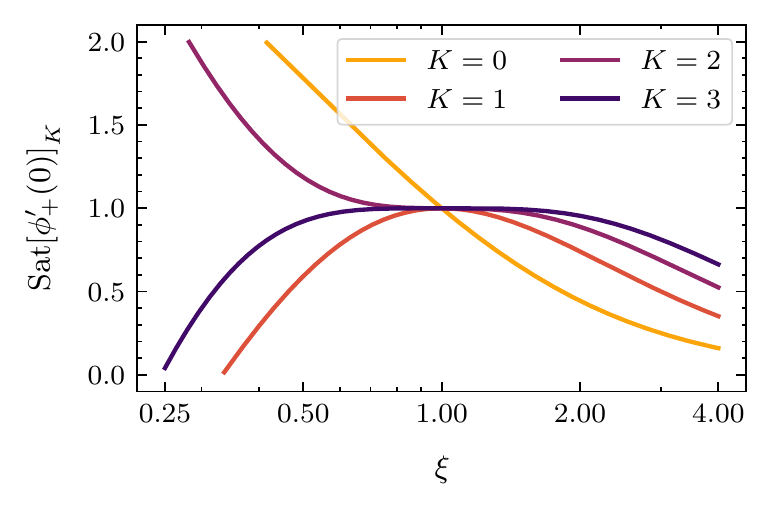}
        }
        \hfill
        \subfloat[%
            \label{fig:exp:sat_l5}
            Saturation of the normalized Laplace transform $\ell_5$
        ]{%
            \includegraphics[scale=0.95]{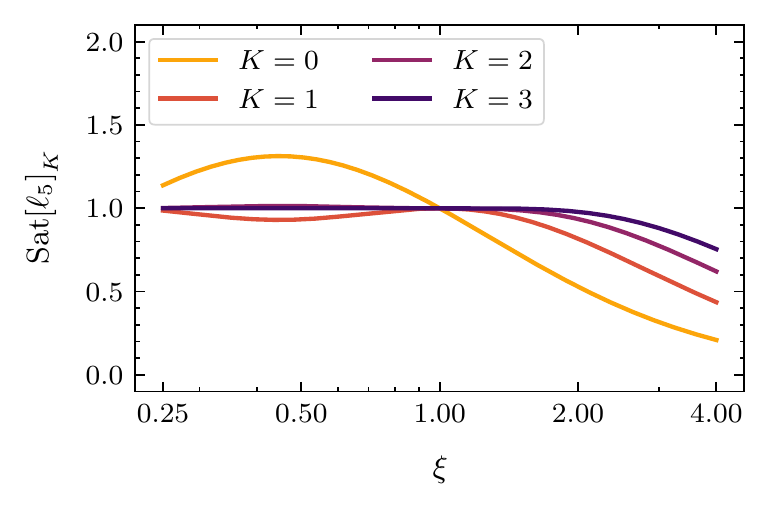}
        }
    \caption{%
        \label{fig:exp}
        Plots illustrating the truncation effects of our parametrization for the exponential model \refeq{exp:model}.
        We often use the ratio $\xi=\omega_0/\lambda_B$, plotted on a logarithmic scale.
    }
\end{figure}

A popular model that is commonly taken as a starting point for a phenomenological analysis is~\cite{Grozin:1996pq}
\begin{align}
    \label{eq:exp:model}
    \phi_+(\omega,\mu_0) = \frac{\omega}{\lambda_B^2} \mathrm{e}^{-\omega/\lambda_B}
    \qquad \mbox{[exp.\ model]} \,.  
\end{align}
Projecting onto our ansatz via \refeq{projan} yields the following result for the expansion coefficients:
\begin{align}
    \label{eq:exp:ak}
    a_k &= (k+1) \left(\frac{2\xi}{1+\xi}\right)^2 \left( \frac{\xi - 1}{\xi + 1} \right)^k 
    \qquad \mbox{[exp.\ model]} \,.
\end{align}
They fall off exponentially for $\xi\neq 1$. For $\xi = 1$, the exponential model trivially matches onto our parametrization with $a_0=1$ and $a_{k>0}=0$.
The result for the first few coefficients as a function of $\xi$ is plotted in \reffig{exp:coeff}.
We observe a rapid fall off of the magnitude of the coefficients $a_k$ for $k>2$ in the entire ``benchmark interval''
\begin{equation}
    \label{eq:benchmark_interval}
    1/2 \lesssim \xi \lesssim 2 \qquad \text{[benchmark interval]}
\end{equation}
This allows us to use the above interval to define an estimator
for the inherent uncertainty of our parametrization, also for other models to be discussed in the following.
The uncertainty estimate is illustrated in \reffig{exp:window}, where we plot the resulting 
variation of the shape of the momentum-space LCDA for different levels of truncation.
Again, already with $K=2$ we find a very narrow envelope for the parametrized function.\\

The integral bound for the exponential model can be 
calculated explicitly, yielding a monotonous function of $\xi$,
\begin{align}
    2\omega_0 \chi = \frac12 \left( \xi + \xi^3 \right) \qquad \mbox{[exp.\ model]}\,.
\end{align}
We plot its saturation and its relative growth as a function of $\xi$ in \reffig{exp:sat_chi}
and \reffig{exp:growth_chi}, respectively,
for different values of the truncation $K$. We observe that both the saturation and the
relative growth give comparable information about the convergence of the parametrization.
As expected, the convergence is very rapid as long as $\xi \simeq 1$: taking $K=2$
and varying in the benchmark interval \refeq{benchmark_interval}, the saturation exceeds
$98\%$ and the relative growth is smaller than $7\%$.\\

We continue to investigate the saturation for the inverse moment $L_0 = \lambda_B^{-1}$, which is plotted
in \reffig{exp:sat_L0}. This quantity also rapidly convergences within our parametrization:
for $K=2$ the saturation within the benchmark interval is better than $98\%$.
It is also instructive to study the normalized first logarithmic moment, which in the model is given by
zero at the given scale $\mu_m$,
\begin{align}
    \sigma_B & = 0 \qquad \mbox{[exp.\ model]} \,.
\end{align}
We show the result as a function of $\xi$ for different truncations $K$
in \reffig{exp:sigma}.
The model result is rapidly reproduced
by the truncated parametrization, and the absolute difference falls below $0.11$ for $K=2$
in the benchmark interval.\\

Finally, we study the saturation of the derivative of the LCDA at the origin $\omega = 0$ and of the normalized
Laplace transform at $n=\zeta \lambda_B$. 
These test how well our parametrization captures the behavior of the LCDA at small light-cone momentum.
In the exponential model, they read:
\begin{equation}
\label{eq:exp:deriv+elln}
\begin{aligned}
    \phi_+'(0,\mu_0)
        & = \lim_{n\to\infty} \ell_n = \frac{1}{\lambda_B^2} \,, \quad
    \ell_n(\mu_0,1/\lambda_B)
        =  \frac{1}{\lambda_B^2} \, \frac{n^2}{(1+n)^2} \,
          \qquad \mbox{[exp.\ model]} \,.
\end{aligned}
\end{equation}
In \reffig{exp:sat_deriv} and \reffig{exp:sat_l5} we show the saturation of $\phi_+'(0)$ and $\ell_5$, respectively, for a number of different truncations $K$.
We find for $K=2$ in the benchmark interval
$0.88 < \mathrm{Sat} \left[ \phi'_+(0) \right]_K < 1.19$ and
$0.93 < \mathrm{Sat} \left[ \ell_5 \right]_K < 1.02$.
Because of the exponential decrease of the individual coefficients $a_k$ in \refeq{exp:ak},
even $\phi_+'(0)$ shows a reasonable convergence.
As discussed in \refsec{mom_space_and_small_omega}, the convergence of $\ell_5$ is expected to be more rapid than
for $\phi_+(0)$, which is confirmed by the plot.\\

We conclude that our parametrisation captures the exponential model with high precision
even for small $K$.
We remark that the parametrisation \emph{to any order} envelopes the model \emph{by construction}; however,
the dependence on the auxiliary parameter $\omega_0$ becomes weaker for growing $K$.
We find that $K=2$ offers sufficient precision for practical applications using the model.\\
Of course, an exemplary behavior of the relatively simple exponential model is expected, since it can be
expressed to trivial order in $K$ for the specific choice of $\omega_0 = \lambda_B$.
Nevertheless, our analysis provides important input for the comparison with other models in the literature.
For that comparison, the exponential model provides a benchmark.

\subsection{Lee-Neubert Model with Radiative Tail}
\label{sec:models:lee_neubert}

\begin{figure}[p]
        \subfloat[%
            \label{fig:LN:coeff}
            Values for the expansion coefficients $a_k$.
        ]{%
            \includegraphics[scale=0.95]{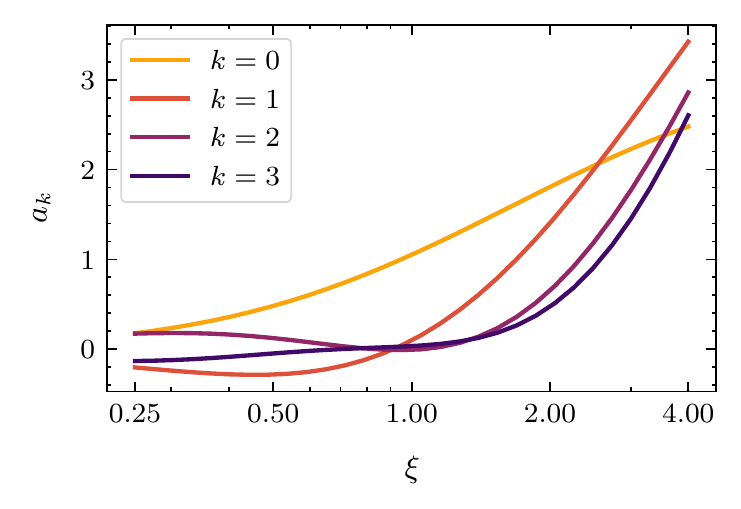}
        }
        \hfill
        \subfloat[%
            \label{fig:LN:phi_window}
            Variability of the momentum-space LCDA for different truncations $K$ in the interval $1/2<\xi<2$.
        ]{%
            \includegraphics[scale=0.95]{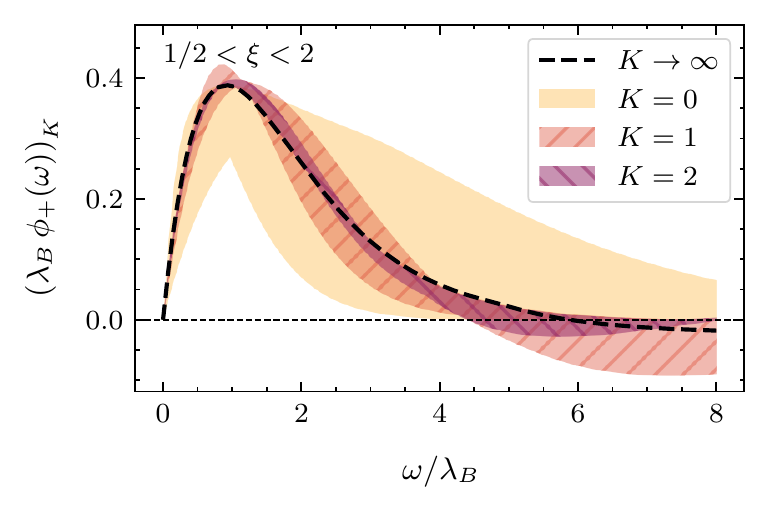}
        }\\
        \subfloat[%
            \label{fig:LN:phi_window_tail}
            Extension of \ref{fig:LN:phi_window}.
        ]{%
            \includegraphics[scale=0.95]{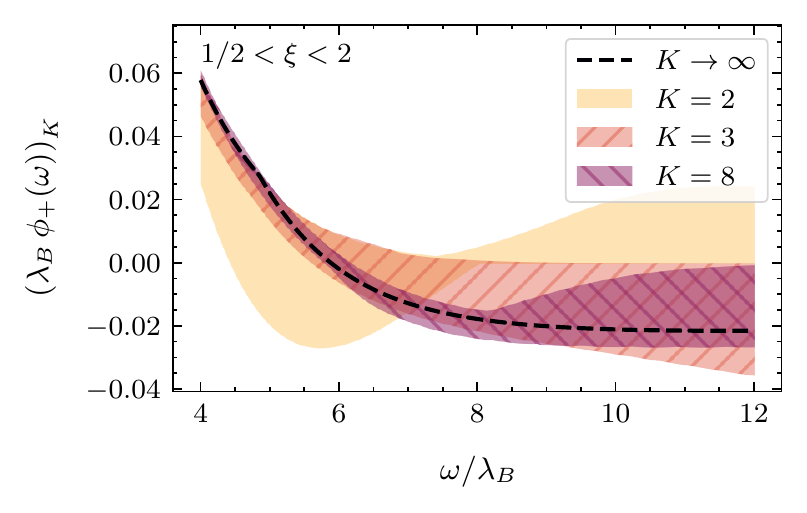}
        }
        \hfill
        \subfloat[%
            \label{fig:LN:sat_chi}
            Saturation of the integral bound $\chi$.
        ]{%
            \includegraphics[scale=0.95]{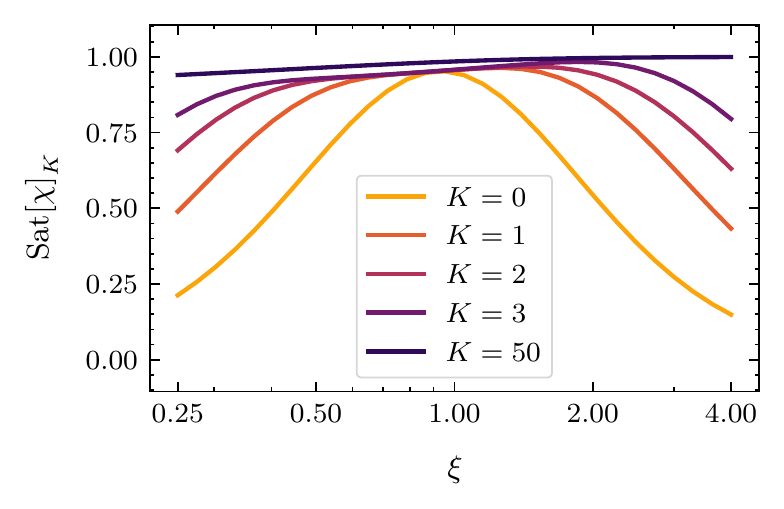}
        }\\
        \subfloat[%
            \label{fig:LN:growth_chi}
            Relative growth of the integral bound $\chi$.
        ]{%
            \includegraphics[scale=0.95]{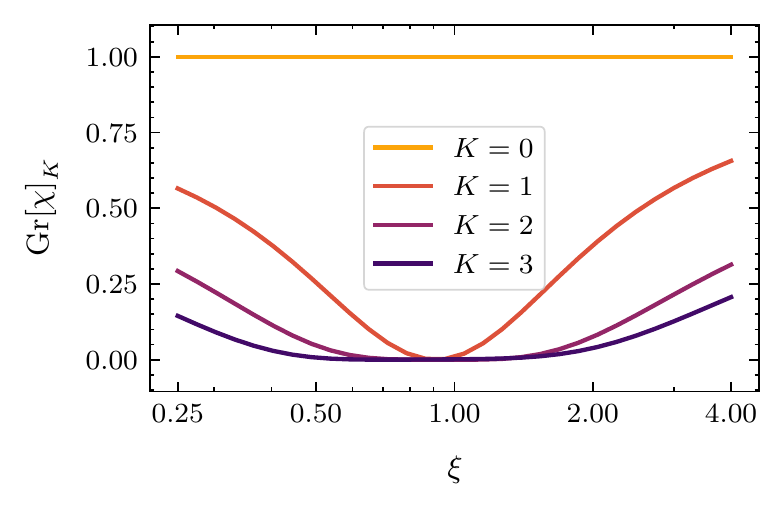}
        }
        \hfill
        \subfloat[%
            \label{fig:LN:sat_L0}
            Saturation of the inverse moment $L_0=\lambda_B^{-1}$.
        ]{%
            \includegraphics[scale=0.95]{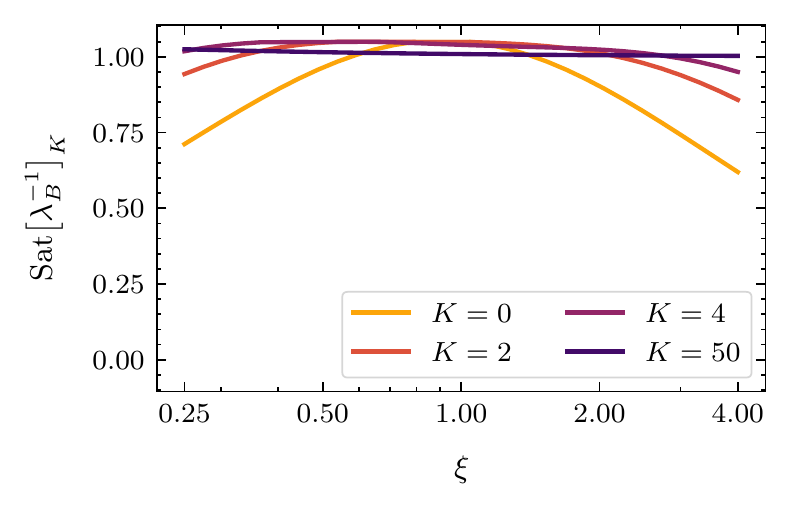}
        }\\
        \subfloat[%
            \label{fig:LN:sigma}
            Value of $\sigma_B$ for different truncations $K$.
        ]{%
            \includegraphics[scale=0.95]{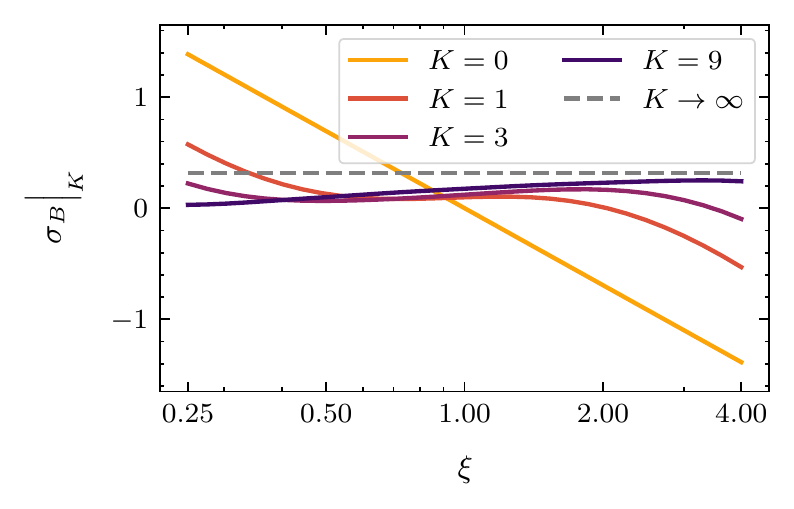}
        }
        \hfill
        \subfloat[%
            \label{fig:LN:RGE}
            Demonstration of the RG evolution.
        ]{%
            \includegraphics[scale=0.95]{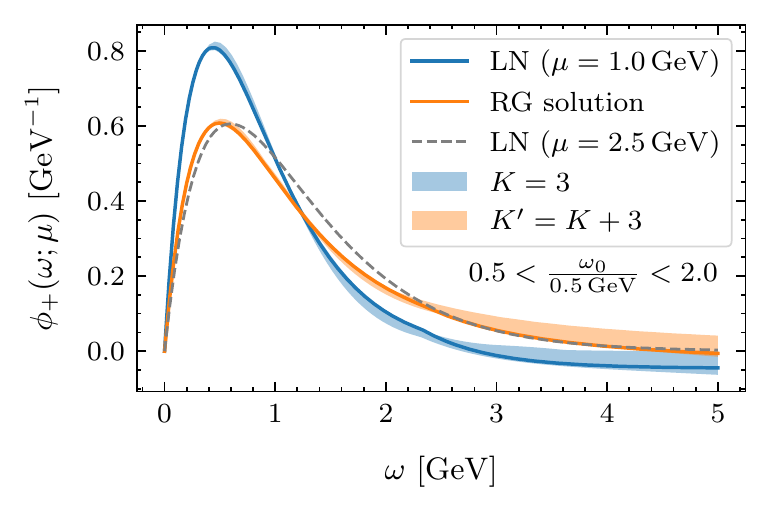}
        }
    \caption{\label{fig:LN}%
        Plots illustrating the truncation effects of our parametrization for the Lee-Neubert model \refeq{model:LN}.
        We use the ratio $\xi=\omega_0/\lambda_B$, plotted on a logarithmic scale.
    }
\end{figure} 

Lee and Neubert \cite{Lee:2005gza} have refined the exponential model by attaching a ``radiative tail'',
which can be deduced from the behavior of the partonic LCDA at large light-cone momenta
$\omega \sim \mu \gg \Lambda_{\rm QCD}$. 
In this model, the LCDA is described by an exponential at low values of $\omega$,
while a radiative tail is added at some intermediate value $\omega_t$,
\begin{align}
    \phi_+(\omega,\mu_0)
        & = \mathcal{N} \, \frac{\omega \, e^{-\omega/\bar \omega}}{\bar\omega^2}
            + \frac{\alpha_s C_F}{\pi} \, \frac{\theta(\omega-\omega_t)}{\omega} 
            \left\{
                \frac12 - \ln \frac{\omega}{\mu} 
                + 4 \, \frac{\bar\Lambda_{\rm DA}}{3\omega} 
                \left(
                    2 - \ln \frac{\omega}{\mu}
                \right)
            \right\}
    \qquad \text{[Lee/Neubert]}\,.
    \label{eq:model:LN}
\end{align}
Here $\bar\Lambda_\text{DA}$ is the HQET mass parameter in a convenient renormalon-free scheme, see Ref.~\cite{Lee:2005gza} for details.
The parameter $\omega_t$ is fixed by requiring the model LCDA to be continuous at $\omega=\omega_t$, while the values for ${\cal N}$ and
$\bar\omega$ are fixed by matching to the partonic calculation. We stress that this model is not supposed to give the correct description at
asymptotically large values $\omega \gg \mu$, which would require to further resum the large logarithms $\ln \omega/\mu$ in the above formula
(see e.g.\ the discussion in Ref.~\cite{Feldmann:2014ika}). With this in mind, we match our parametrization to this model,
and we aim at a reasonable description for small and intermediate values of $\omega$. 
For the following numerical discussion, we adapt the parameter values found in Ref.~\cite{Lee:2005gza} for $\mu_0=1$~GeV,
$$
 \bar\Lambda_{\rm DA} = 519~{\rm MeV}\,, 
 \quad 
 \bar\omega = 438~{\rm MeV} \,, 
 \quad 
 {\cal N} = 0.963 \,, 
 \quad 
 \omega_t = 2.33~{\rm GeV} \,,
$$
with $\alpha_s(\mu_0)=0.5$. For this choice one finds
\begin{align}
    L_0 = \lambda_B^{-1} =1/479~\text{MeV}^{-1}\,, 
    \qquad
    \sigma_B = 0.315
    \qquad \mbox{[Lee/Neubert]}\,,
\end{align}
and the expansion coefficients $a_k$ can easily be calculated numerically. For instance, for $\xi=1$ we find
$$
 a_0 \simeq 1.050 \,, \quad 
 a_1 \simeq 0.096 \,, \quad 
 a_2 \simeq -0.007 \,, \quad 
 a_3 \simeq 0.035 \,, \quad 
 a_4 \simeq -0.051 \,, \quad 
 a_5 \simeq 0.047 \,.
$$
For large values of $k$ the coefficients $a_k$ remain almost constant in magnitude, however, with alternating signs.
The result for $a_{0},\dots,a_{5}$ and the corresponding approximation to the Lee/Neubert model are shown in
\reffig{LN:coeff} and \reffig{LN:phi_window}, respectively. At first glance, the results look qualitatively
very similar to the exponential model. However, the features induced by the radiative tail, namely the cusp
at $\omega=\omega_t$ and the zero at $\omega\simeq 2.82$~GeV, require special attention.
We therefore zoom into the region $4 < \omega/\lambda_B < 12$ in \reffig{LN:phi_window_tail}, where we
also consider larger values for the truncation parameter $K$. 
We find that a reasonably precise description of the radiative tail at intermediate values of $\omega$ requires somewhat higher truncation levels
than the exponential model. Note that -- by construction -- our parametrization is not designed to capture the radiative tail at
values $\omega \gg \mu$. We will revisit this point later in \refsec{pseudo-pheno}.

We continue with the discussion of the integral bound, which in the Lee-Neubert model takes the numerical value
\begin{align}
    2\omega_0 \chi &=0.547 \, \xi + 0.608 \, \xi^3 \qquad \mathrm{[Lee/Neubert]}\,,
\end{align}
which is close to the exponential model.
We emphasize that the bound is finite due to the continuity of the model, despite the derivative in \refeq{chi:mom} acting on the Heaviside distribution.
The saturation and the relative growth of the integral bound are plotted in Fig.~\ref{fig:LN:sat_chi} and Fig.~\ref{fig:LN:growth_chi}, respectively.
First, we observe that the saturation is always smaller than one;
this is clear, as the bound is monotonously increasing with $K$.
Second, we observe that the curves are tilted in comparison to the exponential model:
small values of $\xi$ result in slow convergence, while
best convergence for the integral bound is obtained
for values $\gtrsim 1$.
The peaking structure reflects the fact that we need to include terms of higher order in $k$ to get a reasonable description, such that the curve flattens.
The relative growth of the bound plotted in Fig.~\ref{fig:LN:growth_chi} decreases reasonably within our benchmark interval $1/2<\xi<2$.

In Fig.~\ref{fig:LN:sat_L0} and Fig.~\ref{fig:LN:sigma}, we show the saturation for the inverse moment $L_0 = \lambda_B^{-1}$ and the value of $\sigma_B$ as a function of $\xi$ for different levels of truncation, respectively.
As in the exponential model, we find good convergence of both quantities in our benchmark interval for $\xi$,
with a preference for larger values.

We skip a discussion for $\phi'_+(0)$ and $\ell_5$, which show qualitatively the same behavior as in the exponential model.
This is obvious, since they are are naturally only sensitive to the region of small $\omega$,
where the radiative tail has no effect.

Next, we use the opportunity to demonstrate the RG evolution of the parameters as defined in \refeq{ansatz:rge:parameter_rge}.
Ref.~\cite{Lee:2005gza} provides the model parameters for two different choices of the renormalisation scale
and plots of the momentum-space LCDA, as well as the general RG solution in momentum space.
In Fig.~\ref{fig:LN:RGE},
we show the model at $\mu = 1~\mathrm{GeV}$ and its RG evolution to $\mu=2.5~\mathrm{GeV}$.
They coincide well with our truncated parametrization with $K=3$ at the initial scale and its evolution to $\mu=2.5\,\text{GeV}$ with $K'=K+3$, using
the usual benchmark interval for $\xi = \omega / \lambda_B$ for illustration.
We also plot the model as provided at $\mu = 2.5~\mathrm{GeV}$ purely for reference.
Our plot is visually indistinguishable from the plots shown in Ref.~\cite{Lee:2005gza}.
We further observe that the variation band is consistent for both scales, which verifies the expectation that higher orders in the expansion remain negligible.

\subsection{Na\"ive Parton Model}
\label{sec:models:parton}

\begin{figure}[p]
        \subfloat[%
            \label{fig:parton:coeff}
            Values for the expansion coefficients $a_k$.
        ]{%
            \includegraphics[scale=0.95]{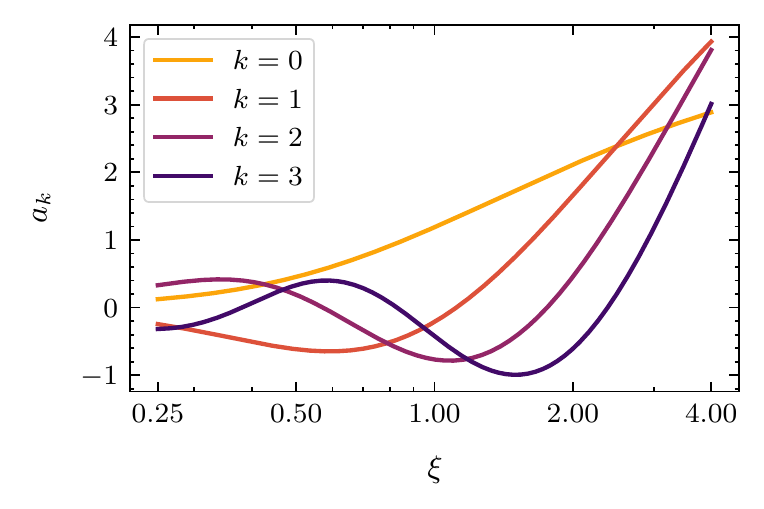}
        }
        \hfill
        \subfloat[%
            \label{fig:parton:window}
            Variability of the momentum-space LCDA for different truncations $K$ in the interval $1/2<\xi<2$.
        ]{%
            \includegraphics[scale=0.95]{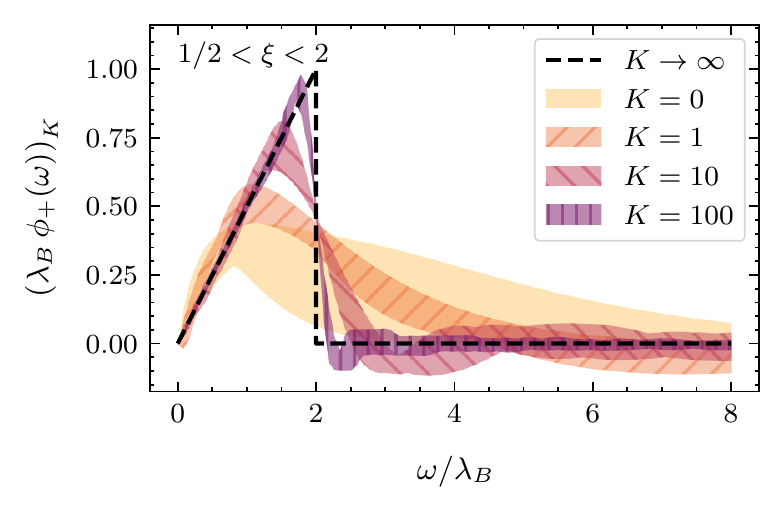}
        }\\
        \subfloat[%
            \label{fig:parton:partial_chi}
            Saturation of the integral bound $\chi$.
        ]{%
            \includegraphics[scale=0.95]{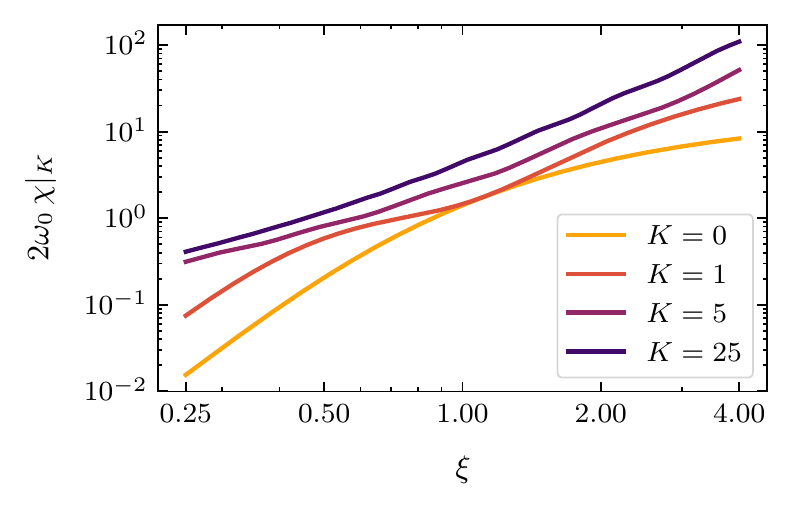}
        }
        \hfill
        \subfloat[%
            \label{fig:parton:growth_chi}
            Relative growth of the integral bound $\chi$.
        ]{%
            \includegraphics[scale=0.95]{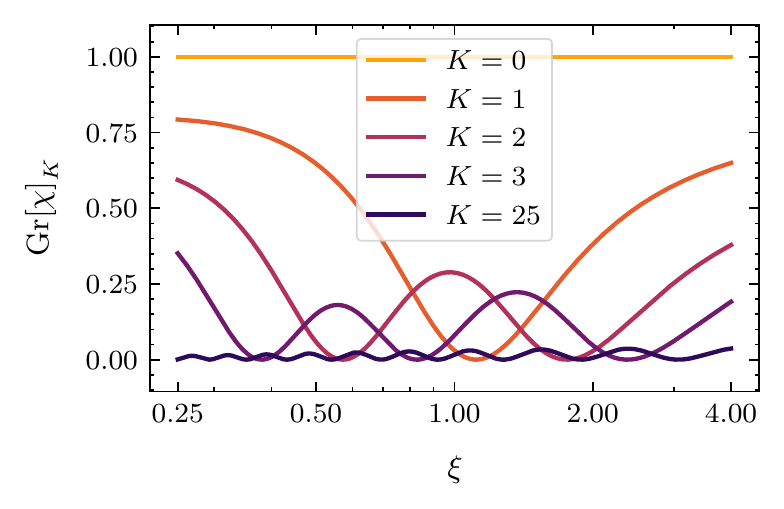}
        }\\
        \subfloat[%
            \label{fig:parton:sat_L0}
            Saturation of the inverse moment $L_0=\lambda_B^{-1}$.
        ]{%
            \includegraphics[scale=0.95]{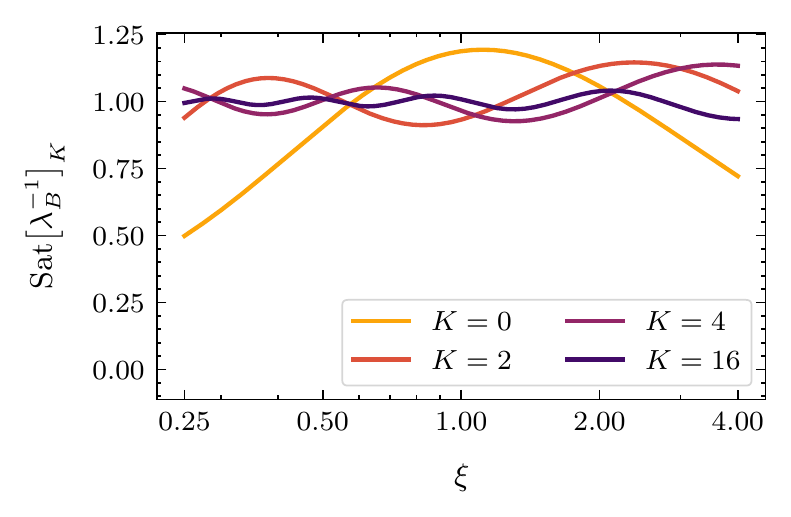}
        }
        \hfill
        \subfloat[%
            \label{fig:parton:sigma}
            Value of $\sigma_B$ for different truncations $K$.
        ]{%
            \includegraphics[scale=0.95]{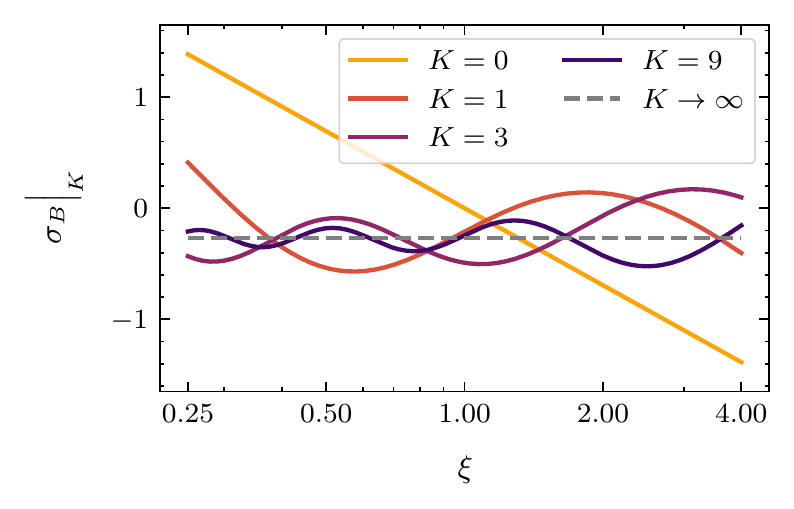}
        }\\
        \subfloat[%
            \label{fig:parton:sat_deriv}
            Saturation of the derivative of $\phi_+(\omega)$ at the origin.
        ]{%
            \includegraphics[scale=0.95]{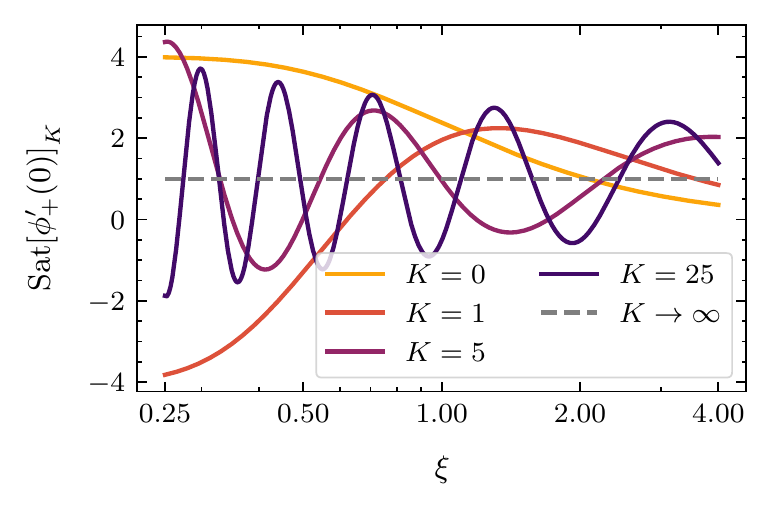}
        }
        \hfill
        \subfloat[%
            \label{fig:parton:sat_l5}
            Saturation of the normalized Laplace transform $\ell_5$.
        ]{%
            \includegraphics[scale=0.95]{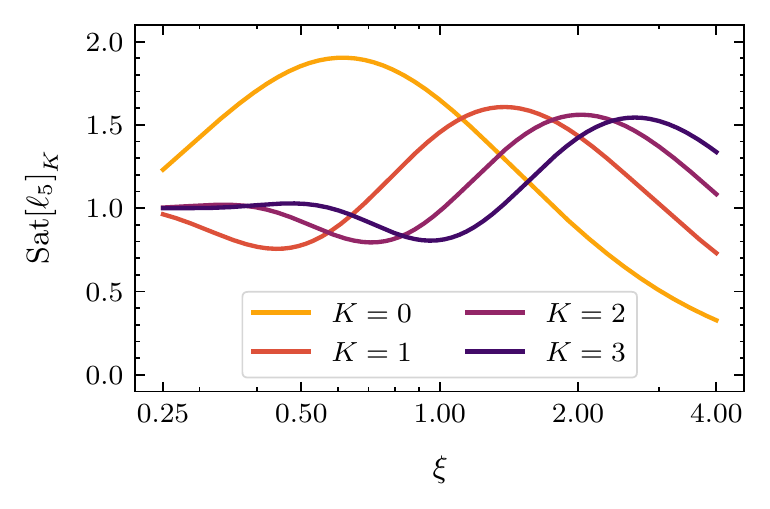}
        }
    \caption{%
        \label{fig:parton}
        Plots illustrating the truncation effects of our parametrization for the parton model \refeq{model:parton}.
        We often use the ratio $\xi=\omega_0/\lambda_B$, plotted on a logarithmic scale.
    }
\end{figure}

In the na\"ive parton model \cite{Kawamura:2001jm} the LCDA takes the form 
\begin{align}
    \label{eq:model:parton}
    \phi_+(\omega)
    = 
    \frac{\omega}{2\lambda_B^2} \, \theta(2\lambda_B - \omega)
    \qquad \mbox{[parton model]} \,,
\end{align}
where $\lambda_B$
is identified with the HQET mass parameter $\bar\Lambda \simeq M_B - m_b$.
In position space this yields
\begin{align}
    \tilde \phi_+(\tau) &= 
    \frac{(1+ 2 i \lambda_B \tau) \, e^{-2i\lambda_B \tau} - 1}{2\lambda_B^2 \tau^2}
    \qquad \mbox{[parton model]} \,,
\end{align}
which only falls off as $1/\tau$ for $|\tau|\to\infty$,
thereby violating \Prop{4}.
The parton-model LCDA is therefore a pathological example of a model.
Nevertheless, it can serve as a toy model to study under which circumstances our parametrization can also capture extreme examples.

To this end, we show the numerical result for the expansion coefficients
and the resulting shapes of the 
momentum-space LCDA for different levels of truncation in Fig.~\ref{fig:parton:coeff} and  Fig.~\ref{fig:parton:window}, respectively.
Indeed, we observe that the expansion coefficients remain sizable even for large values of $k$, without any preferred value for the ratio $\xi$.
This indicates a bad convergence of the expansion.
Similarly, our estimate for the truncation uncertainty in Fig.~\ref{fig:parton:window},
reflected by the variation of $\xi$ in the benchmark interval,
is larger than in the exponential model.
As expected, the triangular shape cannot be reproduced well, even for very high levels of truncation.\\

As $\tilde\phi(\tau)$ only falls off as $1/\tau$, the integral bound $\chi$ in \refeq{chi_r} does not exist for our choice of the function $r(\tau,\mu_0)$.
In Fig.~\ref{fig:parton:partial_chi} we therefore only plot the diverging sum 
$$
2 \omega_0 \chi \big|_K = \sum_{k=0}^K |a_k|^2 \,,
$$
together with its relative growth in Fig.~\ref{fig:parton:growth_chi}. The observed oscillatory behavior of the latter
can be taken as an indicator for the non-convergence of the expansion.\\

We continue with the discussion of $L_0 = \lambda_B^{-1}$ for which we plot the saturation in Fig.~\ref{fig:parton:sat_L0}.
Its saturation oscillates around unity with an amplitude that is only slowly decreasing with increasing $K$.
The normalized first logarithmic moment is given by
\begin{align}
    \sigma_B
        &=1 - \ln 2 - \gamma_{\text{E}} \simeq - 0.270
        \qquad \mbox{[parton model]} \,.
\end{align}
In Fig.~\ref{fig:parton:sigma} we show the result for different truncation $K$.
Again we observe an oscillatory behavior around the true model value.\\

Finally, we study the convergence of our parametrization at low values of $\omega$. 
We obtain
\begin{align}
    \phi_+'(0,\mu_0)=\frac{1}{2\lambda_B^2} \,,
    \qquad
    \ell_n(\mu_0,1/\lambda_B) =
    \frac{1}{2\lambda_B^2} \left(1-(1+2n) \, \mathrm{e}^{-2n} \right)
    \qquad \mbox{[parton model]} \,.
\end{align}
The saturation of the derivative at the origin in Fig.~\ref{fig:parton:sat_deriv} shows oscillatory behavior for the whole range of $\xi$,
while for the normalized Laplace transform we observe that the saturation in Fig.~\ref{fig:parton:sat_l5} approaches unity for sufficiently large values of $K$ and/or small values of $\xi$.

\clearpage 
\subsection{A Model with \texorpdfstring{$\phi_+'(0)\to \infty$}{Diverging Derivative at the Origin}}
\label{sec:models:beneke_braun_2}

\begin{figure}[p]
        \subfloat[%
            \label{fig:betal:coeff}
            Values for the expansion coefficients $a_k$.
        ]{%
            \includegraphics[scale=0.95]{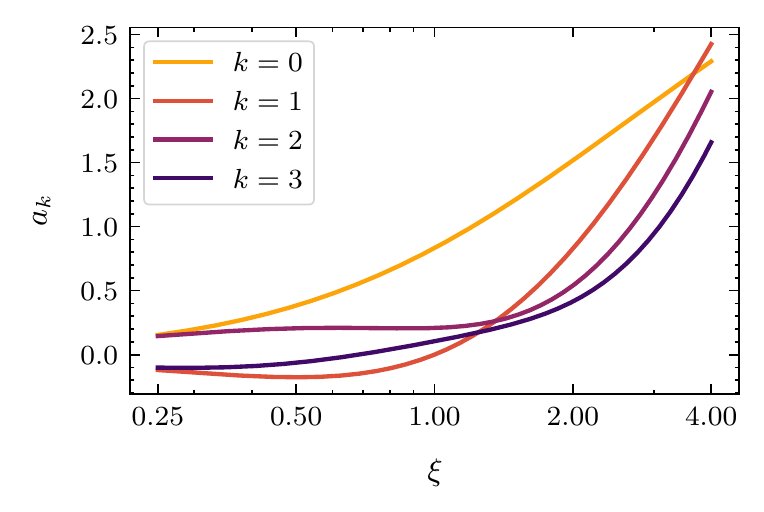}
        }
        \hfill
        \subfloat[%
            \label{fig:betal:window}
            Variability of the momentum-space LCDA for different truncations $K$ in the interval $1/2<\xi<2$.
        ]{%
            \includegraphics[scale=0.95]{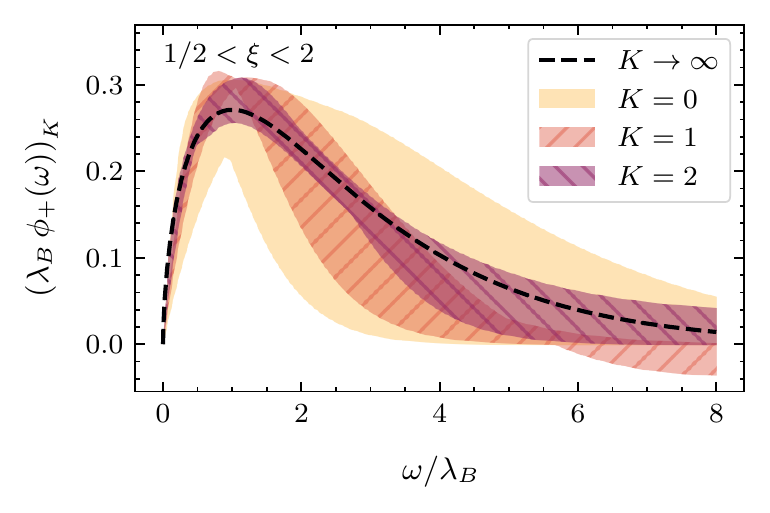}
        }\\
        \subfloat[%
            \label{fig:betal:sat_chi}
            Saturation of the integral bound $\chi$.
        ]{%
            \includegraphics[scale=0.95]{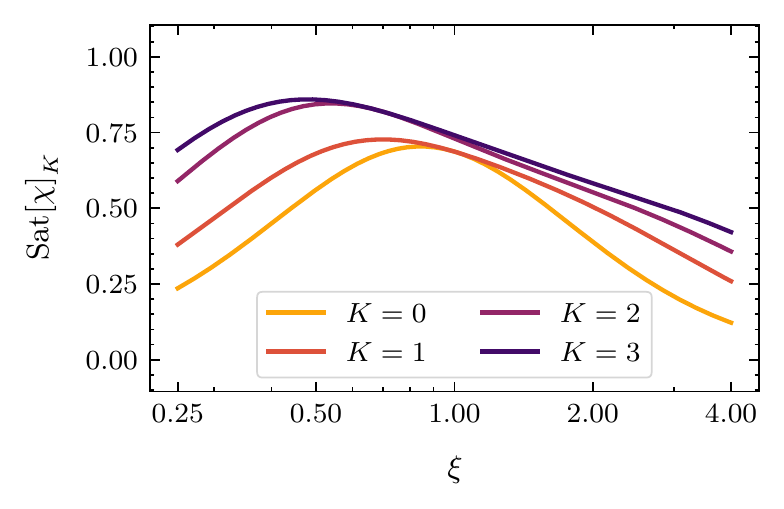}
        }
        \hfill
        \subfloat[%
            \label{fig:betal:growth_chi}
            Relative growth of the integral bound $\chi$.
        ]{%
            \includegraphics[scale=0.95]{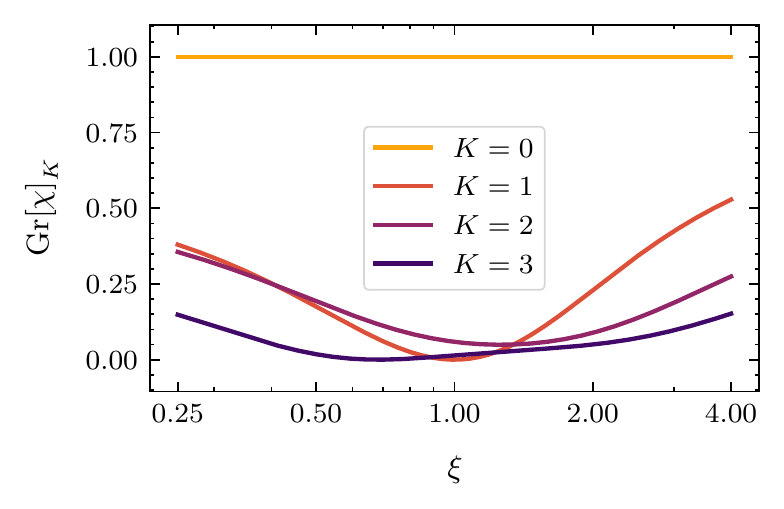}
        }\\
        \subfloat[%
            \label{fig:betal:sat_L0}
            Saturation of the inverse moment $L_0=\lambda_B^{-1}$.
        ]{%
            \includegraphics[scale=0.95]{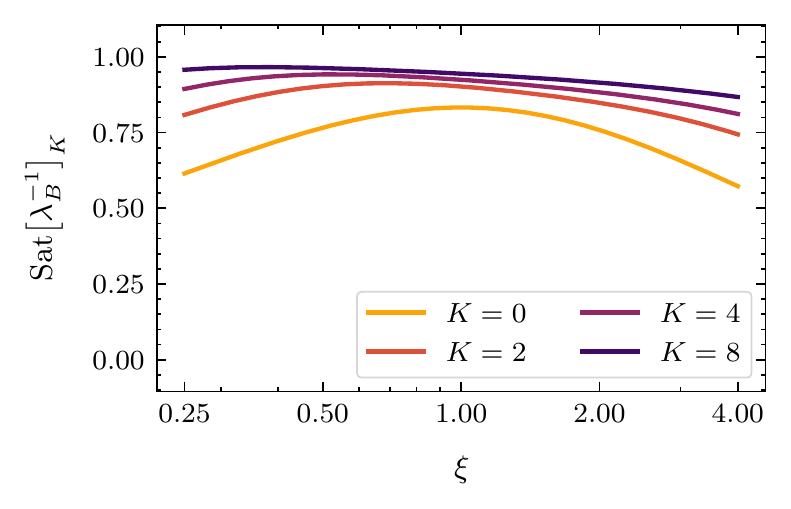}
        }
        \hfill
        \subfloat[%
            \label{fig:betal:sigma}
            Value of $\sigma_B$ for different truncations $K$.
        ]{%
            \includegraphics[scale=0.95]{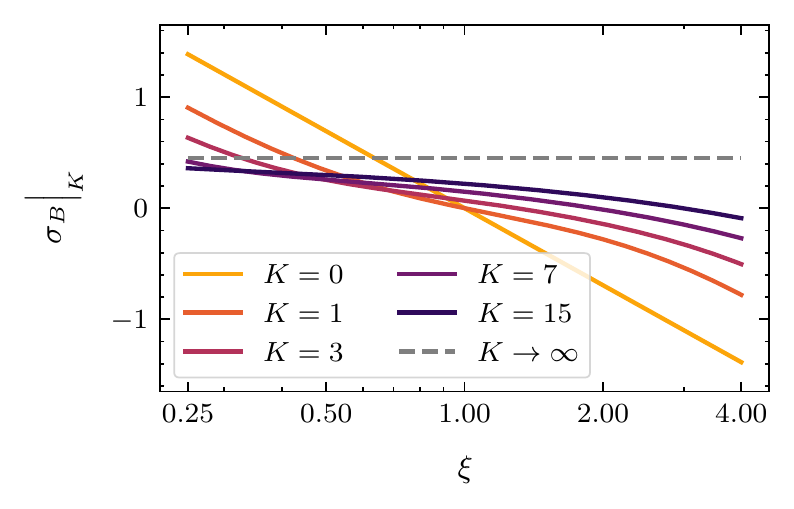}
        }\\
        \subfloat[%
            \label{fig:betal:partial_deriv}
            Partial sum of the derivative at the origin.
        ]{%
            \includegraphics[scale=0.95]{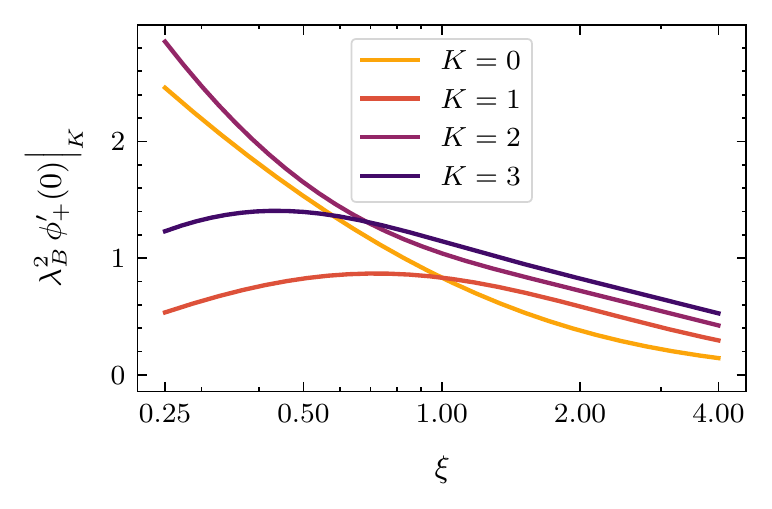}
        }
        \hfill
        \subfloat[%
            \label{fig:betal:sat_l5}
            Saturation of the normalized Laplace transform $\ell_5$.
        ]{%
            \includegraphics[scale=0.95]{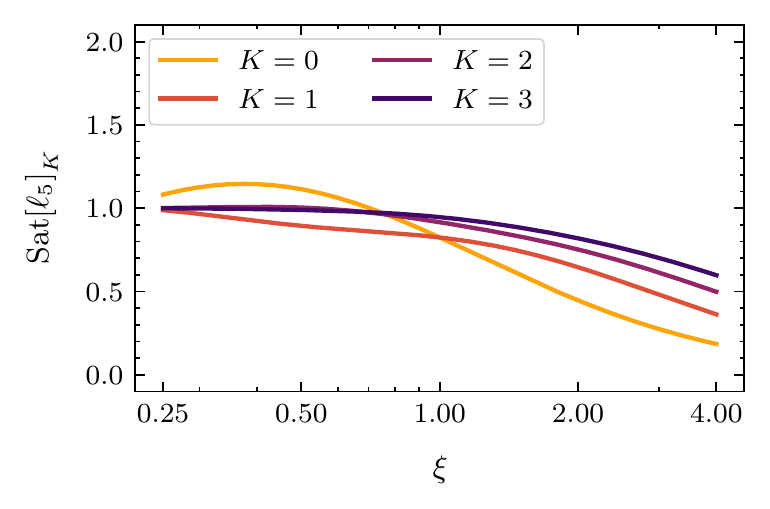}
        }
    \caption{%
        \label{fig:model:betal}
        Plots illustrating the truncation effects of our parametrization for the model \refeq{model:betal} discussed by
        Beneke et al.~\cite{Beneke:2018wjp}.
        We often use the ratio $\xi=\omega_0/\lambda_B$, plotted on a logarithmic scale.
    }
\end{figure}

Beneke et al.~\cite{Beneke:2018wjp} have suggested to consider more general parametrizations for the $B$-meson LCDA, which also include cases where the derivative at the origin $\phi_+'(0)$ does not exist. 
We study their model
\begin{align}
    \label{eq:model:betal}
    \phi_+(\omega,\mu_0) &= 
    \frac{1}{\Gamma(1+a)} \, 
    \left(\frac{(1+a) \, \omega}{\lambda_B} \right)^{1+a} 
    \frac{\mathrm{e}^{-(1+a) \, \omega/\lambda_B}}{\lambda_B} \quad 
    \mbox{[Beneke et al.]} \,.
\end{align}
For $a\to 0$, this function reduces to the simple exponential model.
For $a<0$, the behavior at $\omega \to 0$ is somewhat pathological, since it violates \Prop{4}.
Nevertheless, it is interesting to study the convergence of our parametrization for this behavior.
For concreteness, in the following we only consider the case $a=-0.4$.

We show the coefficients $a_k$ and the resulting shapes to the LCDA for different levels of truncation in Fig.~\ref{fig:betal:coeff} and Fig.~\ref{fig:betal:window}, respectively.
We see that except for the vicinity of $\omega=0$, already $K=2$ yields a reasonable approximation within the estimated uncertainty from the $\xi$-variation.\\

The integral bound for the model follows as 
\begin{align}
    2\omega_0 \chi &= \frac{1}{2^{1+2a}} \, \frac{\Gamma(2+2a)}{\Gamma(1+a)^2} \left( \xi + \frac{(1+a)^2}{1+2a} \, \xi^3 \right) 
    \overset{a\to-0.4}{\simeq}  0.360 \, \xi + 0.649 \, \xi^3
    \qquad 
    \mbox{[Beneke et al.]} \,.
\end{align}
In Fig.~\ref{fig:betal:sat_chi} we observe the opposite behavior as compared to the Lee-Neubert model, i.e.\ the peak of the saturation is tilted to the other side, such that the best convergence is obtained for small values of $\xi$.
Again, the relative growth of the integral bound in Fig.~\ref{fig:betal:growth_chi} remains small in the benchmark interval around $\xi=1$.\\

The saturation of the inverse moment $L_0 = \lambda_B^{-1}$ and the normalized first logarithmic moment,
\begin{equation}
    \sigma_B =
        - \Psi(1+a) + \ln(1+a) - \gamma_{\text{E}}
        \overset{a\to-0.4}{\simeq} 
            + 0.453
    \qquad \mbox{[Beneke et al.]} \,,
\end{equation}
are shown in Fig.~\ref{fig:betal:sat_L0} and Fig.~\ref{fig:betal:sigma}, respectively. 
Compared to the exponential model shown in Fig.~\ref{fig:exp:sat_L0}, the saturation for $L_0$ converges more slowly, and the truncated result for $\sigma_B$
is rather sensitive to the value of $\xi$ (for moderate truncation $K$).
\\
The fact that $\sigma_B$ can be adjusted by an independent parameter $a$ may be viewed as an advantage of the model \refeq{model:betal}.
However the required pathological behavior at $\omega \to 0$ makes it less useful
in phenomenological applications. We therefore advocate our parametrization, which is able to systematically
decorrelate the pseudo observables by including sufficiently many terms.
We illustrate this point in the following section.
\\

Finally, we also show the results for the quantities that characterize the behavior of the LCDA at small values of $\xi$. 
As the derivative at the origin does not exist in the model (\ref{eq:model:betal}), in \reffig{betal:partial_deriv} we only show the truncated sum for a finite value of $K$,
$$
 \phi_+'(0)_K = \frac{1}{\omega_0^2} \,  \sum_{k=0}^K a_k \,.
$$
Indeed, no convergence is apparent.
In Fig.~\ref{fig:betal:sat_l5}, we show the saturation of the normalized Laplace transform,
\begin{equation}
 \ell_n(\mu_0,1/\lambda_B) = 
     \frac{n^2}{\lambda_B^2}
     \left( \frac{1+a}{n+1+a} \right)^{2+a} 
     \qquad \mbox{[Beneke et al.]} \,,
\end{equation}
at $n=5$.
Here, we observe similarly good saturation properties as for the exponential model.

\clearpage
\section{Pseudo-phenomenology}
\label{sec:pseudo-pheno}

In this section we illustrate the feasibility of using our parametrization to describe
the LCDA in its full kinematic range, based on a global analysis of all available phenomenological information in the future.
Here, we do not strive for a rigorous statistical analysis. Instead, we illustrate the
complementarity of the available constraints in a qualitative manner.
Quantitative statements herein should not be mistaken for theoretical predictions.

\subsection{Using \texorpdfstring{$\lambda_B$}{the Inverse Moment} and \texorpdfstring{$\ell_5$}{the normalized Laplace Transform} as Phenomenological Constraints}

In this subsection we study the hypothetical situation that some phenomenological information,
be it experimental or theoretical in nature, constrains the quantities 
(``pseudo observables'')
\begin{equation}
 p_1 \equiv L_0(\mu_0, \mu_m) \qquad \mbox{and} \qquad
 p_2 \equiv \lambda_B^2 \, \ell_5(\mu_0, 1/\lambda_B)
 \label{eq:pseudo}
\end{equation}
at a low reference scale $\mu_0$. 
We select these two pseudo observables, because they emerge in the theoretical description
of the $\bar{B}^-\to \gamma\mu^-\bar\nu$ form factors. An experimental determination of
the these form factors, through measurements of the decays, is foreseen by the Belle II
experiment~\cite{Belle:2015mpp,Belle:2018jqd,Belle-II:2018jsg}.
Moreover, these two pseudo observables probe complementary aspects of the $B$-meson LCDA:
in the following discussion we will neglect the uncertainties on these parameters for simplicity. Of course,
in a realistic fit to experimental data, these uncertainties as well as their correlations have to be taken into account.
With no further theory input at hand, we can use our parametrization to estimate the effect on the $B$-meson LCDA and other derived quantities.\\

Here, we truncate at $K=2$, which yields four independent parameters. They are $\omega_0$ and $a_0$ through $a_2$.
With the two phenomenological constraints above, we can determine two of these parameters.
We choose to determine $a_0$ and $a_1$:
\begin{align}
    a_0
        & = \xi - \frac{a_2}{3} \,, &
    a_1
        & = \frac{(5\xi+1)^3}{25 \,(5\xi-1)} \, p_2
            - \frac{\xi \, (5\xi +1)}{5\xi-1}
            - \frac{2 \, (25\xi^2-20\xi+1)}{3 (5\xi+1)(5\xi-1)} \, a_2\,, &
    \text{where }
    \xi
        & = p_1 \, \omega_0\,.
        \label{eq:pheno1:an}
\end{align}
This leaves two unconstrained parameters: the auxiliary scale ratio $\xi$ 
and the coefficient $a_2$.
In order to constrain the possible ranges for $\xi$ and $a_2$,
we now impose the following conditions, which are motivated by the findings in the previous subsection:
The relative growth of the integral bound $\chi$ is limited to $20\%$ (for $K=1$) and $10\%$ (for $K=2$),
\begin{align}
    \label{eq:growthcond}
    \frac{|a_1|^2}{|a_0|^2+|a_1|^2}
        & \leq 0.2 \,, &
    \frac{|a_2|^2}{|a_0|^2+|a_1|^2+|a_2|^2}
        & \leq 0.1\,.
\end{align}
Combined with \refeq{pheno1:an}
this provides a bounded region for the joint distribution of the parameters $\xi$ and $a_2$.
For any given pseudo observable we can therefore determine its minimal and maximal values for parameter values in that region.

For example, the result for the normalized logarithmic moment
at the reference scale $\mu_m = 1 / p_1 \, e^{-\gamma_E}$
is given by
\begin{equation}
    \sigma_{B}\big|_2
        \equiv
            -\frac{1}{p_1} \, L_1(\mu_0, 1/ p_1 \, \mathrm{e}^{-\gamma_E})\Big|_{K=2}
        = - \ln \xi + \frac{a_1}{\xi} \,.
\end{equation}

In the following, we will determine the resulting variations for the 
normalized logarithmic moment $\sigma_B$ at $K=2$,
as well as for the LCDA $\phi(\omega)$ at different values of $x=\omega \, p_1$, and the normalized Laplace transformation $\ell_n$
at different (real) values of $n$ for each of the
four models discussed in \refsec{models}.
Note that at this stage, the only difference between the four models is given by their predictions for the pseudo-observable $p_2$, while the
pseudo-observable $p_1$ only enters via the auxiliary parameter $\xi$ to set the reference scale in the analysis.

\begin{figure}[t!pb]
\begin{center}
    \subfloat[%
        \label{fig:pseudopheno1:exp}
        Using the exponential model \refeq{exp:model}.
    ]{%
        \includegraphics[scale=0.95]{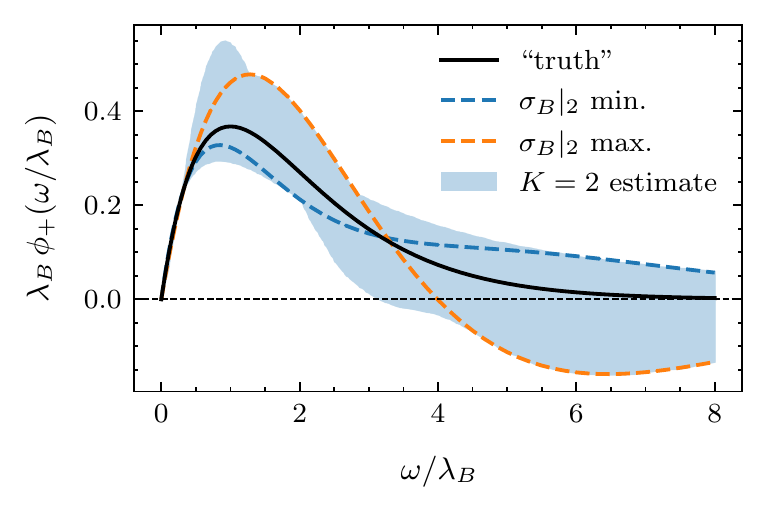}
    }
    \hfill
    \subfloat[%
        \label{fig:pseudopheno1:LN}
        Using the Lee-Neubert model \refeq{model:LN}.
    ]{%
        \includegraphics[scale=0.95]{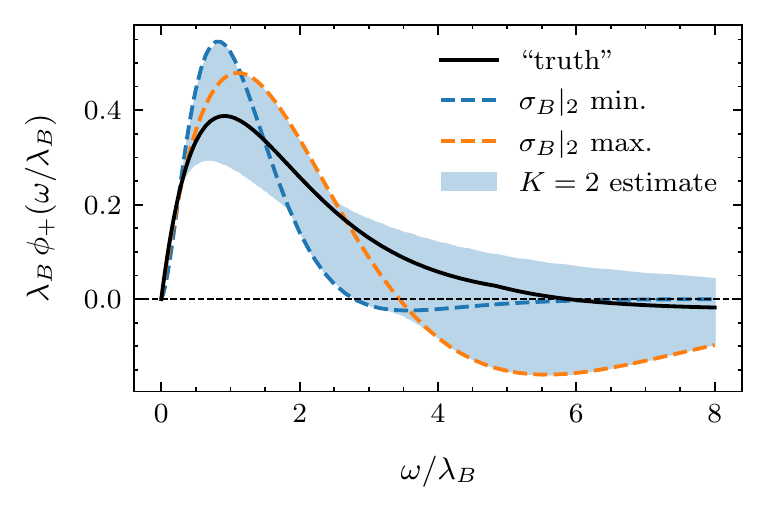}
    }
\end{center}
\caption{\label{fig:pseudopheno1}
    Pseudo-fit to the $B$-meson LCDA in momentum space using the two pseudo-observables $p_{1,2}$ 
    as predicted by two models
    as a function of $x=\omega/\lambda_B$.
    We show the model (``truth''), the curves that lead to extreme values of $\sigma_B$ and the total variation of the parametrized LCDA.
    For the latter, we take the benchmark interval for $\xi = \omega_0/\lambda_B$ and a small $a_2$ into account.
}
\end{figure}

\begin{figure}[t!pb]
\begin{center}
    \subfloat[%
        \label{fig:pseudophenolaplace1:exp}
        Using the exponential model \refeq{exp:model}.
    ]{%
        \includegraphics[scale=0.95]{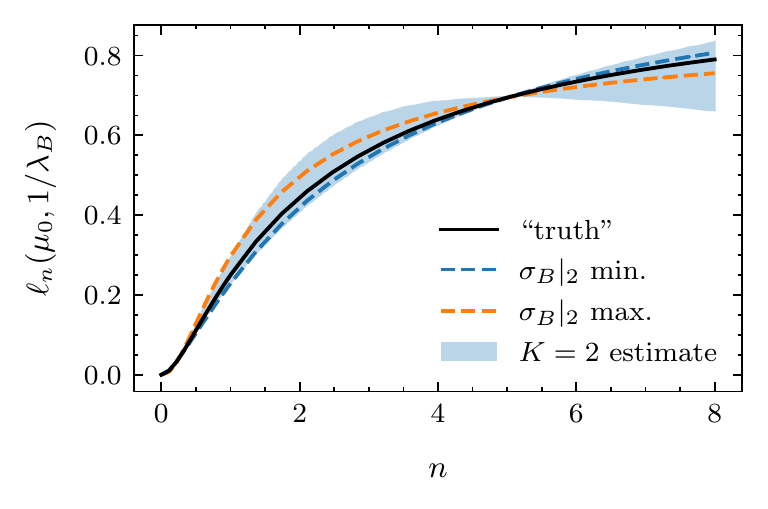}
    }
    \hfill
    \subfloat[%
        \label{fig:pseudophenolaplace1:LN}
        Using the Lee-Neubert model \refeq{model:LN}.
    ]{%
        \includegraphics[scale=0.95]{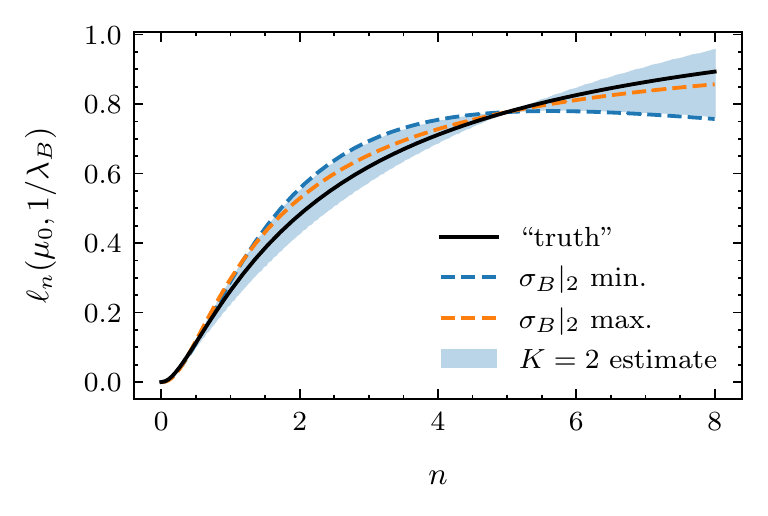}
    }
\end{center}
\caption{\label{fig:pseudophenolaplace1}
    Pseudo-fit to the 
    normalized Laplace transform $\ell_n(\mu_0,1/\lambda_B)$
    using the two pseudo-observables $p_{1,2}$ 
    as predicted by two models
    as a function of $x=\omega/\lambda_B$.
    We show the model (``truth''), the curves that lead to extreme values of $\sigma_B$ and the total variation of the parametrized LCDA.
    For the latter, we take the benchmark interval for $\xi = \omega_0/\lambda_B$ and a small $a_2$ into account.
}
\end{figure}

\subsubsection{Exponential Model}

In the exponential model \refeq{exp:model} the value for
the pseudo-observable $p_2$ is given by
\begin{align}
    p_2 = \frac{25}{36} \simeq 0.694 \,, \qquad \text{[exp.\ model]}
\end{align}
from which we can directly determine the coefficients $a_0$ and $a_1$ in the truncated parametrization.
We may now determine the maximal range of values that $\sigma_B$ can take, when the free parameters $\xi$ and $a_2$ are varied as explained above. This results in
\begin{align}
    \sigma_B\big|_2^{\rm min} &= -0.073 & & \mbox{for $\xi \to 1.332$ and $a_2 \to +0.410$} \,, \\
    \sigma_B\big|_2^{\rm max} &= +0.172 & & \mbox{for $\xi \to 1.489$ and $a_2 \to -0.625$} \,,
\end{align}
while $\sigma_B=0$ in the exponential model.
We further plot the momentum space LCDA $\phi_+(\omega,\mu_0)$ and its corresponding Laplace transform
$\ell_n(\mu_0,1/\lambda_B)$ in Fig.~\ref{fig:pseudopheno1:exp} and Fig.~\ref{fig:pseudophenolaplace1:exp}, respectively,
as well as the corresponding curves for the extreme values of $\sigma_B$.
We find that the parameter values are constrained to the intervals
$0.607 < \xi < 1.499$, $-0.633 < a_2 < 0.477$,
and
$0.607 < \xi < 1.636$, $-0.632 < a_2 < 0.539$%
, respectively.
We observe the following:
\begin{enumerate}
    \item The knowledge of the pseudo-observables $p_1$ and $p_2$ indeed fixes the behavior of $\phi_+(\omega)$ at low momentum, as well as the behavior of its Laplace transform at large values of $\zeta = -i\tau \gg\lambda_B$.
    \item The shape of $\phi_+(\omega)$ at intermediate values of $\omega$ is very sensitive to the variation of $\xi$ and $a_2$, and therefore not very meaningful without additional phenomenological constraints.
    \item Similarly, the behavior of the Laplace transform at small values of $\zeta \lesssim 1$ is sensitive to the variation of $\xi$ and $a_2$, but in contrast to the LCDA in momentum space, the shape of the Laplace transform remains stable, reflecting a monotonous function (at least, as long as $\zeta$ is not too close to zero).
\end{enumerate}
From this we can already see, that a meaningful fit to the LCDA would benefit from additional information on the Laplace transform at small values of $\zeta$ (i.e.\ the LCDA in position space at imaginary light-cone time $-i\tau \lesssim \lambda_B$). This will be further illustrated below.

\subsubsection{Lee-Neubert Model with Radiative Tail}

In the Lee-Neubert model \refeq{model:LN} the value for
the pseudo-observable $p_2$ is given by  
\begin{align}
    \quad p_2  \simeq 0.777 \qquad [\text{Lee/Neubert}] \,,
\end{align}
which is slightly larger than for the exponential model.
The corresponding ranges for the logarithmic moment are obtained as  %
\begin{align}
    \label{range:LN:sig}
    \sigma_B\big|_2^{\rm min} &= 0.045 & & \mbox{for $\xi \to 0.541$ and $a_2 \to -0.225$} \,, \\
    \sigma_B\big|_2^{\rm max} &= 0.321 & & \mbox{for $\xi \to 1.283$ and $a_2 \to -0.543$} \,,
\end{align}
This includes the value $\sigma_B \simeq 0.315$ of the LN model. We observe that values of $p_2$ that are larger than in the exponential model yield larger values of $\sigma_B$.
We will discuss how to consistently implement information on the ``radiative tail'' in a modified fit procedure in \refsec{ppheno:ope}.
We plot the momentum-space LCDA and its Laplace transform in Fig.~\ref{fig:pseudopheno1:LN} and Fig.~\ref{fig:pseudophenolaplace1:LN}.
We find that the parameter values are constrained to the intervals
$0.498 < \xi < 1.342$, $-0.546 < a_2 < 0.439$,
and
$0.540 < \xi < 1.381$, $-0.536 < a_2 < 0.458$%
, respectively.

\subsubsection{Na\"ive Parton Model}

We can repeat the analysis for the na\"ive parton model. Here the pseudo-observable $p_2$ takes the value 
\begin{align}
    \quad p_2 = (1- 11 \, \mathrm{e}^{-10})/2 \simeq 0.5
    \qquad \text{[parton model]} \,,
\end{align}
which now is \emph{smaller} than in the exponential model.
With this, the range of values that $\sigma_B$ can take amounts to
\begin{align}
    \sigma_B\big|_2^{\rm min} &= -0.565 & & \mbox{for $\xi \to 1.810$ and $a_2 \to +0.543$} \,, \\
    \sigma_B\big|_2^{\rm max} &= -0.231 & & \mbox{for $\xi \to 2.228$ and $a_2 \to -0.943$} \,,
    \label{range:part:sig}
\end{align}
which includes the ``true'' value $\sigma_B \simeq - 0.270$ in the na\"ive parton model.
We find that the parameter values are constrained to the intervals
$0.824 < \xi < 2.229$, $-0.948 < a_2 < 0.536$,
and
$0.824 < \xi < 2.242$, $-0.947 < a_2 < 0.697$%
, respectively.
Qualitatively, as before, we observe that now smaller values of $p_2$ tend to yield smaller values of $\sigma_B$ in the pseudo-fit.

\subsubsection{Model with \texorpdfstring{$\phi_+'(0)\to \infty$}{diverging derivative at the origin}}

Repeating the analysis for the model \refeq{model:betal}, we observe that the value of the pseudo-observable
\begin{align}
    p_2 \simeq 0.701 \qquad [\text{Beneke et al.}]
\end{align}
is very close to that of the exponential model. As a consequence -- without further input -- the fit cannot distinguish the two cases.
Note that the range for the first logarithmic moment that results from our fit procedure will not include the actual value
$\sigma_B = 0.453$ in this model. This can be traced back to the pathological behavior of the model at $\omega \to 0$.

\clearpage
\subsection{Adding Theoretical Constraints from the Short-distance OPE}
\label{sec:ppheno:ope}

As has previously been mentioned, the $B$-meson LCDA can be constrained using information obtained
from the short-distance OPE of the light-cone operator. The latter describes the behavior of $\tilde \phi_+(\tau)$
at small values of $|\tau| \sim 1/\mu \ll 1/\Lambda_\text{QCD}$, and it can be obtained from a fixed-order partonic
calculation in HQET~\cite{Kawamura:2008vq},
\begin{align}
    \label{eq:Kawamura}
    \tilde \phi_+(\tau,\mu)
        & = 1 - \frac{\alpha_s C_F}{4\pi} \left( 2 L^2 + 2 L + \frac{5\pi^2}{12} \right)
            - i \tau \, \frac{4\bar\Lambda}{3} 
            \left[
                1 - \frac{\alpha_s C_F}{4\pi} \left(
                    2 L^2 + 4 L - \frac94 + \frac{5\pi^2}{12}
                \right)
            \right]
            + \mathcal{O}(\tau^2 \Lambda_\text{QCD}^2)\,.
\end{align}
In the above, $\bar \Lambda$ is the HQET mass parameter in the on-shell scheme and we abbreviate
\begin{equation*}
 L = \ln(i \tau \mu e^{\gamma_E}) \,.
\end{equation*}
In the Lee-Neubert model \refeq{model:LN} the short-distance information has been added as a radiative tail in momentum space,
by considering cut-off moments 
$M_n(\Lambda_{\rm UV}) = \int_0^{\Lambda_{\rm UV}} d\omega \, \omega^n \, \phi_+(\omega)$.
A more direct and simpler approach is to evaluate the LCDA in position space, using our parametrization, and compare with
\refeq{Kawamura}.
To this end, we have to first expand our parametrization in $\omega_0 / \mu_0 \ll 1$ for a fixed value of
\begin{align}
    x_0 & \equiv i\tau_0 \mu_0 \, e^{\gamma_E} \sim {\cal O}(1)
\end{align}
The analogue of the first two moments $M_0$ and $M_1$ used in the Lee-Neubert model are then taken as the value
and first derivative of $\tilde\phi_+(\tau)$, which defines the two theory inputs
\begin{align}
    t_1 & \equiv \tilde \phi(\tau_0, \mu_0) 
    \,, \qquad 
    t_2 \equiv i \frac{d\tilde\phi(\tau,\mu_0)}{d\tau}\bigg|_{\tau = \tau_0} \,.
\end{align}
Let us first consider a situation where only this theory input is known. A minimal approach would then be to consider our parametrization at
truncation level $K=1$ and fix the parameters $a_0$ and $a_1$ by matching $t_1$ and $t_2$ to \refeq{Kawamura}. This yields
\begin{align}
    a_0 &= 2 - \frac{2\bar\Lambda}{3\omega_0} + \frac{\alpha_s C_F}{4\pi} \left( - \frac{1}{x_0} \, \frac{\mu_0 e^{\gamma_E}}{\omega_0} \, (1+ 2 \ln x_0 ) + \ldots \right) \,, 
    \\
    a_1 &= 1 - \frac{2\bar\Lambda}{3\omega_0} + \frac{\alpha_s C_F}{4\pi} \left(
     - \frac{1}{x_0} \, \frac{\mu_0 e^{\gamma_E}}{\omega_0} \, (1+ 2 \ln x_0 ) + \ldots\right) \,, 
\end{align}
where we only show the $\alpha_s$ corrections that are enhanced by $\mu_0 / \omega_0$. These terms can be absorbed by a redefinition of the HQET mass parameter: For our purpose, a convenient renormalization scheme is%
\footnote{%
    A similar definition has been derived from the analysis of the cut-off moments
    $M_0$ and $M_1$ in Ref.~\cite{Lee:2005gza}, leading to the above-mentioned ``DA-scheme''
    for the HQET mass parameter.
}
\begin{align}
    \bar \Lambda & \equiv 
    \bar \Lambda_a(\mu,x_0)
    \left[ 1 + \frac{\alpha_s C_F}{4\pi} \left( 
    10\, \ln x_0 + \frac{15}{4} \right) \right]
    - \frac{\alpha_s C_F}{4\pi}  \, \frac{3\mu e^{\gamma_E}}{2x_0} \left( 1+ 2 \ln x_0 \right) 
    \label{eq:Lama1}
\end{align}
which yields
\begin{align}
    a_0 &= Z(x_0) \left( 2 - 
    \frac{2 \bar\Lambda_a}{3\omega_0}
    -\frac{8\alpha_s C_F}{3\pi} \, \frac{\bar\Lambda_a x_0}{\mu_0 e^{\gamma_E}}
    \, (1+\ln x_0)
    \right) \,, 
    \\ a_1 &= Z(x_0) \left( 1 - 
    \frac{2 \bar\Lambda_a}{3\omega_0}
    -\frac{4\alpha_s C_F}{3\pi} \, \frac{\bar\Lambda_a x_0}{\mu_0 e^{\gamma_E}}
    \, (1+\ln x_0)
    \right) 
    \,,
\end{align}
with 
\begin{align}
    Z(x_0) &= 1+\frac{\alpha_s C_F}{4\pi} \left( -2\ln^2 x_0+2\ln x_0 + 2 - \frac{5\pi^2}{12} \right) \,.
\end{align}
Our definition of $\bar\Lambda_a$ and $Z(x_0)$ have been chosen such that the result for the position-space LCDA with finite truncation $K$  always satisfies
\begin{align}
 \tilde\phi_+(0)\big|_K &= \sum_{k=1}^K (-1)^k \, a_k = Z(x_0) - \frac{4 \alpha_s C_F}{3\pi} \, \frac{\bar\Lambda_a x_0}{\mu_0 e^{\gamma_E}} \, (1+\ln x_0) \,,
 \cr 
  \tilde\phi_+'(0)\big|_K &= - 2 i \omega_0 \, \sum_{k=1}^K (-1)^k \, (1+k) \, a_k
  = - Z(x_0) \, \frac{4 i \bar \Lambda_a}{3} \,,
  \label{eq:Lama_scheme}
\end{align}
which generalizes the Grozin-Neubert relations in Ref.~\cite{Grozin:1996pq} to one-loop accuracy in our
formalism.\footnote{%
    The conditions \refeq{Lama_scheme} take the same form in dual space, since
    $\eta_+(0)=\tilde \phi_+(0)$ and 
    $\eta_+'(0) = -\frac{i}{2} \, \tilde\phi_+'(0)$ for finite truncation $K$.
}
It is instructive to compare with the approach by Lee and Neubert in Ref.~\cite{Lee:2005gza}, where corresponding expressions are obtained for the zeroth and first moment of the momentum-space LCDA with a UV cut off. The perturbative relation between the parameter $\bar\Lambda_\text{DA}$ defined in that scheme and our scheme reads
\begin{align}
    \bar\Lambda_a(\mu,x_0) &= 
    \bar\Lambda_{\rm DA}(\mu,\mu) \left[1+ \frac{\alpha_s C_F}{4\pi} \left( - 10 \,\ln x_0 -2\right) \right] + \mu \, \frac{\alpha_s C_F}{4\pi} \left( \frac{3 e^{\gamma_E}}{2x_0} \, (1+ 2 \ln x_0) - \frac92\right)\,.
    \label{eq:Lama}
\end{align}
For instance, using $x_0=1$, $\mu_0 = 1\,\GeV$, $\bar\Lambda_{\rm DA}(\mu_0,\mu_0) = 519$~MeV, and $\alpha_s(\mu_0)=0.5$
as in Ref.~\cite{Lee:2005gza}, we obtain
\begin{equation*}
    \bar\Lambda_a(\mu_0, x_0) = 367\,\text{MeV} \,.
\end{equation*}

We are now in the position to include the phenomenological constraints $p_1$ and $p_2$ as defined in the previous subsection. For the sake of legibility, we introduce the quantity 
\begin{align}
 n_0 \equiv i\tau_0 \omega_0 =  \frac{x_0 \omega_0}{\mu_0 e^{\gamma_E} } \,.
\end{align}
For the power expansion defined by the OPE to converge,
we would need $x_0 \sim {\cal O}(1)$ and small values of $n_0$. On the other hand, the (reasonably fast)
convergence of our parametrization requires a finite $n_0 > 0$.
In the following, we use
\begin{equation*}
n_0 =1/3\,, \quad \mu_0 = 1~{\rm GeV}\,,
\end{equation*}
as our default choice,
while the value of $\omega_0$ (and thus of $x_0$) will be varied within the fit, with suitable constraints on the resulting relative growth (see below).
For instance, a value $x_0=1$ corresponds to $\omega_0 \approx 600$~MeV, which appears reasonable.
Note that our choice for $n_0$ corresponds to the value 
$y_0 = -1/2$ in \refeq{tauymap},
which lies exactly halfway between the origin and the local limit ($y=-1$).

With two new constraints we increase the truncation level from $K=2 \to 4$ compared to the previous subsection, leaving $a_4(\mu_0)$ and $\xi = \omega_0/\lambda_B$ as free parameters.
For $\xi=1$ and $x_0=1$ we obtain
\begin{align}
    a_0 &= Z \left( -\frac{28}{25} + \frac{2 \bar\Lambda_a p_1}{15}\right) +
      \frac{112 \alpha_s C_F}{75\pi} \, \bar\Lambda_a n_0 p_1
      + 3 - \frac{972 \, p_2}{625} - \frac{4a_4}{15}
    \,, \\
    a_1 &= Z \left( - 2 + \frac{\bar \Lambda_a p_1}{3} \right)  + 
      \frac{8 \alpha_s C_F}{3\pi} \, \bar\Lambda_a n_0 p_1 + \frac32 - \frac{4a_4}{5}
    \,, \\
    a_2 &= Z \left( \frac {84}{25} - \frac{2\bar\Lambda_a p_1}{5} \right) - 
      \frac{112 \alpha_s C_F}{25\pi}  \, \bar\Lambda_a n_0 p_1
      -6+ \frac{2916 \, p_2}{625} + \frac{a_4}{5}
    \,, \\
    a_3 &= Z \left( \frac{81}{25} - \frac{3\bar\Lambda_a p_1}{5}\right)  - 
      \frac{108 \alpha_s C_F}{25\pi}  \, \bar\Lambda_a n_0 p_1
      - \frac92 + \frac{1944 \, p_2}{625} + \frac{26 a_4}{15} \,,
\end{align}
where $Z\equiv Z(x_0 = 1)$.
The result for arbitrary values of $\xi$ and $x_0$ can be found in Appendix~\ref{app:aksolutions}.
The parameter range for $a_4$ and $\xi$ will be further constrained by analogous conditions on the relative growth as in the previous subsection,
\begin{align}
   {\rm Gr}[\chi]_K \leq \frac{20\%}{K}
\end{align}
which generalizes \refeq{growthcond}.

\begin{figure}[p!t]
\begin{center}
    \subfloat[%
        \label{fig:pheno2:phi}
        Momentum-space LCDA.
    ]{%
        \includegraphics[scale=0.95]{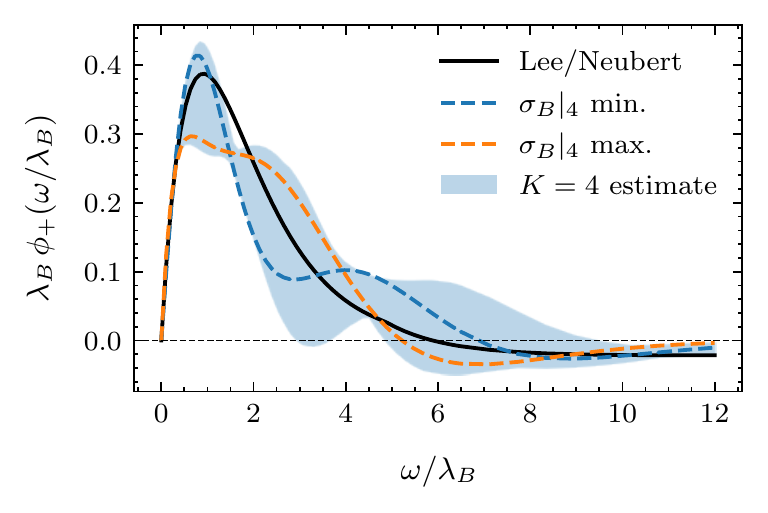}
    }
    \hfill
    \subfloat[%
        \label{fig:pheno2:ell}
        Laplace transform $\ell_n$.
    ]{%
        \includegraphics[scale=0.95]{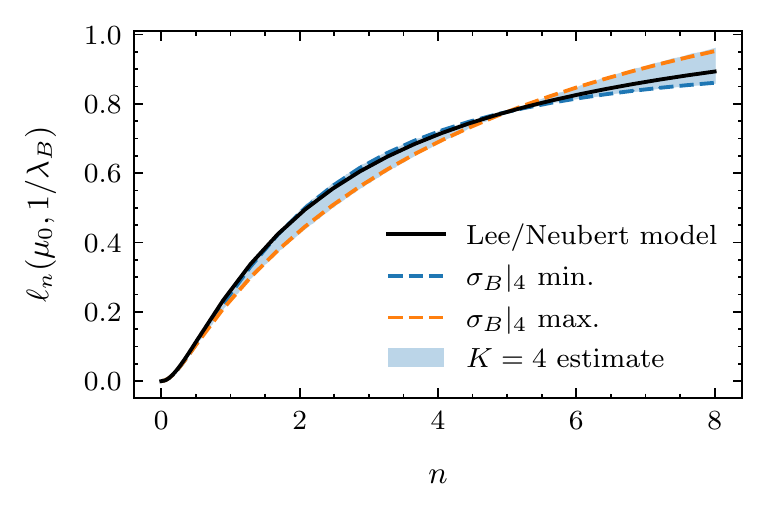}
    }\\
    \subfloat[%
        \label{fig:pheno2:pararms-a-combined}
        Bounded regions for each parameter.
    ]{%
        \includegraphics[scale=0.95]{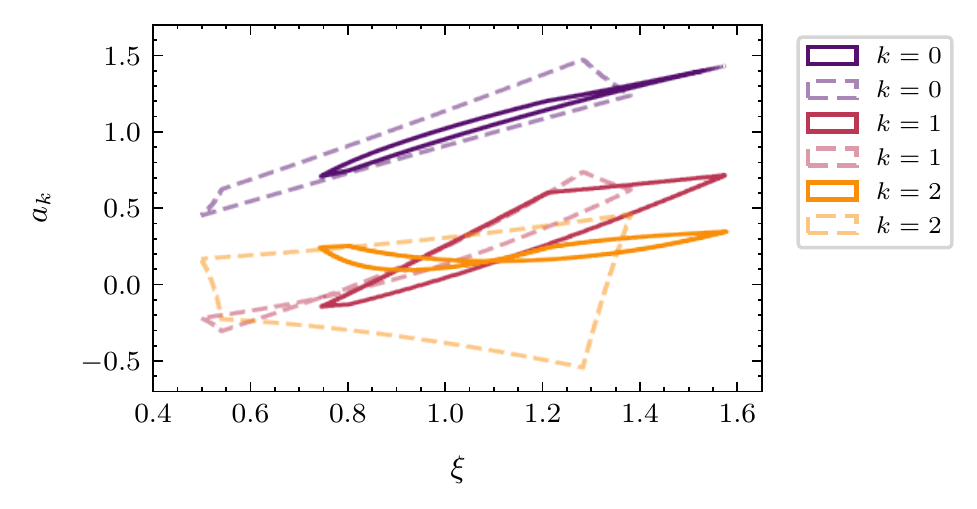}
    }
    \hfill
    \subfloat[%
        \label{fig:pheno2:params-b}
        Bounded regions for each parameter.
    ]{%
        \includegraphics[scale=0.95]{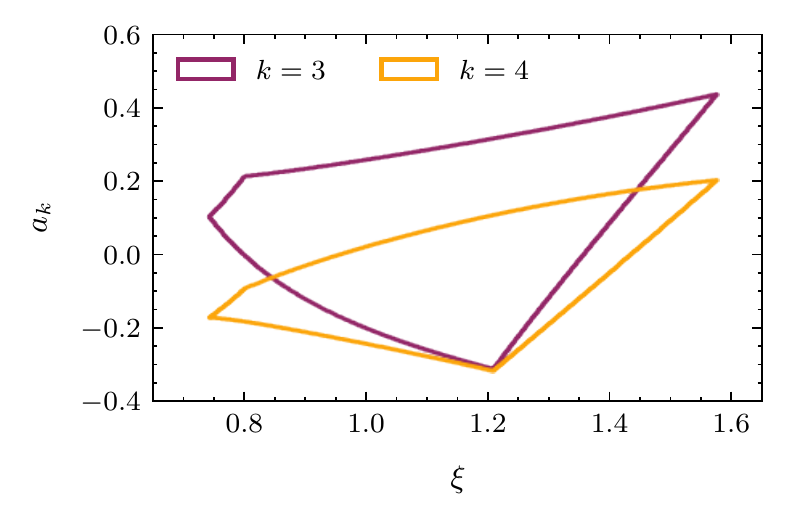}
    }
\end{center}
\caption{\label{fig:pseudopheno2}
    Results from the pseudo-fit employing theoretical constraints in addition to the phenomenological
    inputs $p_1$ and $p_2$.
    The dashed regions in Fig.~\ref{fig:pheno2:pararms-a-combined} correspond to the pseudo-fit \emph{without} theoretical OPE constraints.
}
\end{figure}

Among the four benchmark models discussed in \refsec{models}, only the Lee-Neubert model in
\refeq{model:LN} features a radiative tail that reflects the 
constraints from the local OPE.
We therefore consider this model as a benchmark. We expect that the pseudo-fit should correctly reproduce the main
qualitative and quantitative features of that model. We therefore set 
\begin{align}
    p_1 = 1/\lambda_B = 2.085 \,\text{GeV}^{-1} \,,\quad
    p_2 = 0.777
    \qquad [\text{Lee/Neubert}] \,,
\end{align}
and take the same input for the theory parameters as outlined below \refeq{model:LN}, while $\bar\Lambda_a$ is calculated from \refeq{Lama} as a function of $x_0$. 
As in \refsec{pseudo-pheno}, we consider the normalized first logarithmic moment, now truncated at $K=4$,
\begin{equation}
    \sigma_B\big|_4
        = - \ln \xi + \frac{a_1 + 2/3 \, a_3}{\xi} \,.
\end{equation}
We find the following range of values, constrained by the growth criterion,
\begin{align}
    \sigma_B\big|_4^{\rm min} & = 0.114 & & \mbox{for $\xi \to 0.961$ and $a_2 \to -0.232$} \,, \\
    \sigma_B\big|_4^{\rm max} & = 0.217 & & \mbox{for $\xi \to 0.905$ and $a_2 \to -0.030$} \,.
\end{align}
This interval is compatible with the estimate obtained by using only $p_1$ and $p_2$.
However, its size is reduced by more than $60\,\%$.
The value $\sigma_B = 0.315$ in the Lee/Neubert model is not contained in this estimate.

We show the resulting momentum-space LCDA and Laplace transform in \reffig{pheno2:phi} and  \reffig{pheno2:ell},
respectively, where we find that the parameters are restricted as $0.745 < \xi < 1.577$, $-0.319 < a_4 < 0.201$,
and $0.743 < \xi < 1.522$, $-0.312 < a_4 < 0.120$.
In \reffig{pheno2:pararms-a-combined},
we compare the allowed regions for the coefficients $a_0$ through $a_2$ obtained here with the ones as obtained in \refsec{pseudo-pheno}.
We find that the regions largely overlap, while shrinking significantly for $a_0$ and $a_2$ and staying approximately constant for $a_1$.
We further show the regions for the additional coefficients $a_3$ and $a_4$
in \reffig{pheno2:params-b}.
\\

Once more, we caution that the plots and numerical results here illustrate the applicability of our method.
However, they cannot be interpreted as predictions, which would require a more careful treatment of uncertainties 
on the basis of experimental data.

\section{Conclusion and Outlook}
\label{sec:conclusion}

We have proposed a novel systematic parametrization 
of the leading-twist $B$-meson light-cone
distribution amplitude (LCDA) in position space. 
At the center of our derivation is the Taylor expansion of the
LCDA in a conveniently chosen variable $y$, which arises from the conformal transformation
in \refeq{tauymap}.
The coefficients of that expansion obey an integral bound \refeq{akbound}, 
which provides qualitative control of the truncation error of the expansion, with the numerical value of the
bound presently unknown.
Our parametrization yields simple expressions for a variety of quantities connected
to the LCDA, including its logarithmic moments and a set of ``pseudo-observables''
describing the low-momentum behavior.
For convenience we summarize the resulting formulas for the most important functions
in Table~\ref{tab:para}.
We have also discussed three different approaches to implement the renormalization-group (RG)
evolution of the LCDA and its derived quantities within our framework. We have identified one approach that allows a computationally efficient implementation in 
future phenomenological analyses.  
\\

We have performed detailed numerical studies to show that our parametrization can successfully reproduce different benchmark models,
including non-trivial features like the ''radiative tail'' at large light-cone momentum.
Furthermore, we have illustrated the power of our approach to combine different types of phenomenological
and theoretical constraints. This is achieved through matching our parametrization to hypothetical values of two ``pseudo-observables'' in \refeq{pseudo} that are expected to be constrained by future experimental data on the photo-leptonic $B \to \gamma\ell\nu$ decay.
Moreover, we have shown that theoretical constraints on the expansion parameters from the local operator product expansion (OPE) can be implemented at small but finite light-cone separation in a natural and straight-forward manner. We have used this  to define a new renormalization 
scheme \refeq{Lama1} for the mass parameter 
in heavy-quark effective theory (HQET), which resembles the so-called ``DA-scheme'' that has been introduced  by Lee and Neubert from the consideration of ``cut-off'' moments.
\\

Our framework is general enough to allow theoretical refinements in the future. First, it can be applied to higher-twist LCDA of the $B$-meson, as we have briefly discussed for the Wandzura-Wilczek part of the twist-three LCDA $\phi_-$. Second, the available two-loop RG evolution can be implemented on the level of our truncated expansion.
Third, the OPE constraints from dimension-five HQET operators can be included as well.
Finally, on the phenomenological side, 
a future determination of the very value of the integral bound, e.g.\ from lattice QCD studies, would allow us
to quantify the truncation errors.

\begin{table}[t]
	\begin{tabular}{p{4cm} r c l}
	    \hline
		position-space LCDA
		    & $\displaystyle \tilde \phi_+(\tau,\mu_0)$
		    & $\displaystyle =$
		    & $\displaystyle \frac{1}{(1+i\omega_0\tau)^2} \, \sum\limits_{k=0}^K a_k(\mu_0) \left(
		        \frac{i\omega_0 \tau -1}{i\omega_0\tau +1}
		       \right)^k$
		\\
		momentum-space LCDA
		    & $\displaystyle \phi_+(\omega,\mu_0)$
		    & $\displaystyle =$
		    & $\displaystyle \frac{\omega \, e^{-\omega/\omega_0}}{\omega_0^2} \, \sum\limits_{k=0}^K a_k(\mu_0) \, \frac{1}{1+k} \, L_k^{(1)}(2\omega/\omega_0)$
		\\
		dual-space LCDA
		    & $\displaystyle \eta_+(s,\mu_0)$
		    & $\displaystyle =$
		    & $\displaystyle e^{-s\omega_0} \, \sum\limits_{k=0}^K a_k(\mu_0) \, \frac{(-1)^k}{1+k} \, L_k^{(1)}(2\omega_0 s)$
		\\ 
		generating function
		    & $\displaystyle F_{[\eta_+]}(t;\mu_0,\mu_m)$
		    & $\displaystyle =$
		    & $\displaystyle \frac{\Gamma(1-t)}{\omega_0} \, \left(\frac{\hat\mu_m}{\omega_0} \right)^{-t}  \,
		            \sum\limits_{k=0}^K a_k(\mu_0) \, {}_2F_1(-k,1+t,2,2)$
		\\
		\hline
		inverse moment
		    & $\displaystyle \lambda_B^{-1}(\mu_0)$
		    & $\displaystyle =$
		    & $\displaystyle \frac{1}{\omega_0} \, \sum\limits_{k=0}^K a_k(\mu_0) \, \frac{1+(-1)^{k}}{2 \, (1+k)}$ \qquad (only even $k$)\\
		logarithmic moment
		    & $\displaystyle \sigma_B(\mu_0)$
		    & $\displaystyle =$
		    & $\displaystyle 
		    - \ln\xi 
        - \frac{1}{\xi} \, \sum_{k=0}^K a_k(\mu_0) \left[\dv{t} \, {}_2F_1(-k, 1+t; 2; 2)\right]_{t = 0} $ \qquad (only odd $k$)
		\\
		derivative at $\omega=0$
		    & $\displaystyle \phi_+'(0,\mu_0)$
		    & $\displaystyle =$
		    & $\displaystyle \frac{1}{\omega_0^2} \, \sum\limits_{k=0}^K a_k(\mu_0)$
		\\
		\hline
	\end{tabular}	
	\caption{%
	    \label{tab:para}
	    Summary of representations and pseudo observables connected to the leading-twist $B$-meson LCDA
	    within our proposed parametrization at the low-energy reference scale $\mu_0$.
	    Here $L_k^{(1)}$ are associated Laguerre polynomials, and $\xi = \omega_0/\lambda_B$.
	}
\end{table}

\section*{Note Added}

During the final phase of this work, Ref.~\cite{Galda:2022dhp} appeared.
Among others, it discusses a complementary approach to parametrizing the leading $B$-meson LCDA,
where the generating function \refeq{eta:genfunc} is expanded
in $t$. The logarithmic moments appear as the expansion coefficients.
In contrast to our work, this expansion is not controlled by an integral bound.

\acknowledgments

We thank Guido Bell, Martin Beneke, and Bj\"orn O.~Lange for helpful discussions.
T.F.\ would like to thank Roman Zwicky for discussions about the $B$-meson LCDA which initiated some of the ideas presented in this work.
The research of T.F.\ is supported by the Deutsche Forschungsgemeinschaft (DFG, German Research Foundation) under grant 396021762 -- TRR 257. 
The work of D.v.D.\ is supported by the DFG within the Emmy Noether Programme under grant DY-130/1-1
and the Sino-German Collaborative Research Center TRR110 ``Symmetries and the Emergence of Structure in QCD''
(DFG Project-ID 196253076, NSFC Grant No.~12070131001, TRR 110).
P.L.\ and D.v.D.\ were supported in the final phase of this work by the Munich Institute for Astro- and Particle Physics (MIAPP),
which is funded by the DFG under Germany's Excellence Strategy -- EXC-2094 -- 390783311.

\appendix

\parindent0pt

\section{Useful Definitions and Formulas}

Our definition of the RG functions $V(\mu;\mu_0)$ and $g(\mu;\mu_0)$ reads (see e.g.\ Ref.~\cite{Neubert:2004dd}),
\begin{align}
    \label{eq:def:V}
    V(\mu, \mu_0)
        & = -\int_{\alpha_s(\mu_0)}^{\alpha_s(\mu)}
            \frac{\mathrm{d}\alpha}{\beta(\alpha)}
            \left[
                \gamma_+(\alpha)
                + \GammaCusp(\alpha)\, \int_{\alpha_s(\mu_0)}^\alpha \frac{\mathrm{d}\alpha}{\beta(\alpha)}
            \right] \,, \\
    \label{eq:def:g}
    g(\mu;\mu_0)
        & = \int_{\alpha_s(\mu_0)}^{\alpha_s(\mu)} 
            \frac{d\alpha}{\beta(\alpha)} \, \GammaCusp(\alpha) \,.
\end{align}

A useful relation for Bessel functions reads (see e.g.\ Ref.~\cite{Braun:2017liq})
\begin{equation}
    \label{ftrafo-bessel}
    \int_0^\infty \dd \omega e^{-i \omega z}\, \left(\frac{\omega}{s}\right)^{j - 1/2}\, J_{2j - 1}(2 \sqrt{s \omega})
        = e^{-i\pi j}\, \frac{e^{i s/z}}{z^{2j}}\,.
\end{equation}

Furthermore, Bessel functions can be expanded as 
an infinite series of associated Laguerre polynomials,
\begin{align}
 J_\alpha(x) &=
     \left( \frac{x}{2} \right)^\alpha \, \frac{e^{-t}}{\Gamma(1+\alpha)} 
     \, \sum_{k=0}^\infty \frac{\alpha!}{(k+\alpha)!} \, L_k^{(\alpha)}\left(\frac{x^2}{4t}\right) \, t^k \,,
\end{align}
with an arbitrary parameter $t$. Especially, using $t = \omega/\omega_0$, one gets 
\begin{align}
 J_1(2 \sqrt{\omega s}) &=
     \sqrt{\omega s} \, e^{-\omega/\omega_0} 
     \, \sum_{k=0}^\infty \frac{1}{(1+k)!} \, L_k^{(1)}(s\omega_0) \, \left(\frac{\omega}{\omega_0} \right)^k \,.
\end{align}

The associated Laguerre polynomials can be written in closed form,
\begin{align}
    L_n^{(\alpha)}(x) &= \sum_{i=0}^n (-1)^i \, \binom{n+\alpha}{n-i} \, \frac{x^i}{i!} \,.
\end{align}

\section{Solutions for Expansion Coefficients $a_k$ from Pseudo-phenomenology and OPE}

\label{app:aksolutions}

The solutions for the expansion coefficients $a_{0-3}$ with the constraints from the pseudo-observables $p_1$ and $p_2$ and the two theory inputs $t_1$ and $t_2$ (see \refsec{pseudo-pheno}) for arbitrary values of $\xi$, $x_0$, $n_0$ and $a_4$ are given by
\begin{align}
a_0 &= 
     Z \left( -\frac{(5\xi-1)(1+30\xi+25\xi^2)}{100\xi^2(5\xi-3)}
     + \frac{5\xi-1}{15 \xi^2 (5\xi-3)} \, \bar \Lambda_a \, p_1 \right) 
     \\ & \quad {}
     + \frac{3\xi(5\xi-1)}{2 (5\xi-3)}- \frac{(5\xi+1)^5}{2500\xi^2(5\xi-3)} \, p_2
     - \frac{(9-5\xi)(5\xi-1)}{5(5\xi-3)(5\xi+1)} \, a_4
     \cr & \quad {}+
      \frac{( 5\xi-1)(1+30\xi+25\xi^2) \, \alpha_s C_F}{75\pi \xi^3 (5\xi-3) } \, \bar\Lambda_a \, n_0 \, p_1 \, (1+\ln x_0) 
      \,, \cr
a_1 &= 
      Z \left( - 2 + \frac{1}{3 \xi} \, \bar\Lambda_a \, p_1\right)    + \frac{3 \xi}{2}+
      \frac{8 \alpha_s C_F}{3\pi} \, \frac{\bar\Lambda_a n_0}{\xi}  \, (1+\ln x_0) \, p_1  - \frac{4a_4}{5}
    \,, \\
a_2 &=
    Z \left( \frac {3 \, (5\xi-1)(1+30\xi+25\xi^2)}{100\xi^2(5\xi-3)} 
    - \frac{5\xi-1}{5\xi^2(5\xi-3)} \, \bar\Lambda_a \, p_1 \right)
    \\ 
    & \quad {} 
    - \frac{3\xi(5\xi+3)}{2(5\xi-3)}  + \frac{3(5\xi+1)^5}{2500\xi^2(5\xi-3)} \, p_2 
    + \frac{6 (-3+30\xi-25\xi^2)}{5(5\xi-3)(5\xi+1)} \, a_4
    \cr & \quad {} - 
      \frac{(5\xi-1)(1+30\xi+25\xi^2) \alpha_s C_F}{25\pi \xi^3(5\xi-3)}  \, \bar\Lambda_a \, n_0 \, p_1 \, (1+\ln x_0) 
    \,, \cr
a_3 &=
    Z \left( \frac{(5\xi+1)(-1-20\xi+75\xi^2)}{50\xi^2(5\xi-3)} 
    - \frac{(5\xi-2)(5\xi+1)}{15\xi^2(5\xi-3)} \, \bar\Lambda_a \, p_1\right) 
    \\ & \quad {}
    - \frac{3\xi(5\xi+1)}{2(5\xi-3)} + \frac{(5\xi+1)^{5} }{1250 \xi^2(5\xi-3)}\, p_2 
    + \frac{4(-9+10\xi+25\xi^2)}{5(5\xi-3)(5\xi+1)} \,  a_4
    \cr & \quad {} -
    \frac{2 (5\xi+1)(-1-20\xi+75\xi^2)\, \alpha_s C_F}{75\pi \xi^3(5\xi-3)}  \, \bar\Lambda_a \, n_0 \, p_1 \, (1+\ln x_0) \,. \nonumber
\end{align}

\clearpage 
\bibliography{blcdas}

\begin{thebibliography}{33}%
\makeatletter
\providecommand \@ifxundefined [1]{%
 \@ifx{#1\undefined}
}%
\providecommand \@ifnum [1]{%
 \ifnum #1\expandafter \@firstoftwo
 \else \expandafter \@secondoftwo
 \fi
}%
\providecommand \@ifx [1]{%
 \ifx #1\expandafter \@firstoftwo
 \else \expandafter \@secondoftwo
 \fi
}%
\providecommand \natexlab [1]{#1}%
\providecommand \enquote  [1]{``#1''}%
\providecommand \bibnamefont  [1]{#1}%
\providecommand \bibfnamefont [1]{#1}%
\providecommand \citenamefont [1]{#1}%
\providecommand \href@noop [0]{\@secondoftwo}%
\providecommand \href [0]{\begingroup \@sanitize@url \@href}%
\providecommand \@href[1]{\@@startlink{#1}\@@href}%
\providecommand \@@href[1]{\endgroup#1\@@endlink}%
\providecommand \@sanitize@url [0]{\catcode `\\12\catcode `\$12\catcode
  `\&12\catcode `\#12\catcode `\^12\catcode `\_12\catcode `\%12\relax}%
\providecommand \@@startlink[1]{}%
\providecommand \@@endlink[0]{}%
\providecommand \url  [0]{\begingroup\@sanitize@url \@url }%
\providecommand \@url [1]{\endgroup\@href {#1}{\urlprefix }}%
\providecommand \urlprefix  [0]{URL }%
\providecommand \Eprint [0]{\href }%
\providecommand \doibase [0]{http://dx.doi.org/}%
\providecommand \selectlanguage [0]{\@gobble}%
\providecommand \bibinfo  [0]{\@secondoftwo}%
\providecommand \bibfield  [0]{\@secondoftwo}%
\providecommand \translation [1]{[#1]}%
\providecommand \BibitemOpen [0]{}%
\providecommand \bibitemStop [0]{}%
\providecommand \bibitemNoStop [0]{.\EOS\space}%
\providecommand \EOS [0]{\spacefactor3000\relax}%
\providecommand \BibitemShut  [1]{\csname bibitem#1\endcsname}%
\let\auto@bib@innerbib\@empty
\bibitem [{\citenamefont {Beneke}\ \emph {et~al.}(1999)\citenamefont {Beneke},
  \citenamefont {Buchalla}, \citenamefont {Neubert},\ and\ \citenamefont
  {Sachrajda}}]{Beneke:1999br}%
  \BibitemOpen
  \bibfield  {author} {\bibinfo {author} {\bibfnamefont {M.}~\bibnamefont
  {Beneke}}, \bibinfo {author} {\bibfnamefont {G.}~\bibnamefont {Buchalla}},
  \bibinfo {author} {\bibfnamefont {M.}~\bibnamefont {Neubert}}, \ and\
  \bibinfo {author} {\bibfnamefont {C.~T.}\ \bibnamefont {Sachrajda}},\ }\href
  {\doibase 10.1103/PhysRevLett.83.1914} {\bibfield  {journal} {\bibinfo
  {journal} {Phys. Rev. Lett.}\ }\textbf {\bibinfo {volume} {83}},\ \bibinfo
  {pages} {1914} (\bibinfo {year} {1999})},\ \Eprint
  {http://arxiv.org/abs/hep-ph/9905312} {arXiv:hep-ph/9905312 [hep-ph]}
  \BibitemShut {NoStop}%
\bibitem [{\citenamefont {Beneke}\ \emph {et~al.}(2001)\citenamefont {Beneke},
  \citenamefont {Buchalla}, \citenamefont {Neubert},\ and\ \citenamefont
  {Sachrajda}}]{Beneke:2001ev}%
  \BibitemOpen
  \bibfield  {author} {\bibinfo {author} {\bibfnamefont {M.}~\bibnamefont
  {Beneke}}, \bibinfo {author} {\bibfnamefont {G.}~\bibnamefont {Buchalla}},
  \bibinfo {author} {\bibfnamefont {M.}~\bibnamefont {Neubert}}, \ and\
  \bibinfo {author} {\bibfnamefont {C.~T.}\ \bibnamefont {Sachrajda}},\ }\href
  {\doibase 10.1016/S0550-3213(01)00251-6} {\bibfield  {journal} {\bibinfo
  {journal} {Nucl. Phys.}\ }\textbf {\bibinfo {volume} {B606}},\ \bibinfo
  {pages} {245} (\bibinfo {year} {2001})},\ \Eprint
  {http://arxiv.org/abs/hep-ph/0104110} {arXiv:hep-ph/0104110 [hep-ph]}
  \BibitemShut {NoStop}%
\bibitem [{\citenamefont {Altmannshofer}\ \emph {et~al.}(2019)\citenamefont
  {Altmannshofer} \emph {et~al.}}]{Belle-II:2018jsg}%
  \BibitemOpen
  \bibfield  {author} {\bibinfo {author} {\bibfnamefont {W.}~\bibnamefont
  {Altmannshofer}} \emph {et~al.} (\bibinfo {collaboration} {Belle-II}),\
  }\href {\doibase 10.1093/ptep/ptz106} {\bibfield  {journal} {\bibinfo
  {journal} {PTEP}\ }\textbf {\bibinfo {volume} {2019}},\ \bibinfo {pages}
  {123C01} (\bibinfo {year} {2019})},\ \bibinfo {note} {[Erratum: PTEP 2020,
  029201 (2020)]},\ \Eprint {http://arxiv.org/abs/1808.10567} {arXiv:1808.10567
  [hep-ex]} \BibitemShut {NoStop}%
\bibitem [{\citenamefont {Khodjamirian}\ \emph {et~al.}(2005)\citenamefont
  {Khodjamirian}, \citenamefont {Mannel},\ and\ \citenamefont
  {Offen}}]{Khodjamirian:2005ea}%
  \BibitemOpen
  \bibfield  {author} {\bibinfo {author} {\bibfnamefont {A.}~\bibnamefont
  {Khodjamirian}}, \bibinfo {author} {\bibfnamefont {T.}~\bibnamefont
  {Mannel}}, \ and\ \bibinfo {author} {\bibfnamefont {N.}~\bibnamefont
  {Offen}},\ }\href {\doibase 10.1016/j.physletb.2005.06.021} {\bibfield
  {journal} {\bibinfo  {journal} {Phys. Lett.}\ }\textbf {\bibinfo {volume}
  {B620}},\ \bibinfo {pages} {52} (\bibinfo {year} {2005})},\ \Eprint
  {http://arxiv.org/abs/hep-ph/0504091} {arXiv:hep-ph/0504091 [hep-ph]}
  \BibitemShut {NoStop}%
\bibitem [{\citenamefont {De~Fazio}\ \emph {et~al.}(2006)\citenamefont
  {De~Fazio}, \citenamefont {Feldmann},\ and\ \citenamefont
  {Hurth}}]{DeFazio:2005dx}%
  \BibitemOpen
  \bibfield  {author} {\bibinfo {author} {\bibfnamefont {F.}~\bibnamefont
  {De~Fazio}}, \bibinfo {author} {\bibfnamefont {T.}~\bibnamefont {Feldmann}},
  \ and\ \bibinfo {author} {\bibfnamefont {T.}~\bibnamefont {Hurth}},\ }\href
  {\doibase 10.1016/j.nuclphysb.2008.03.022, 10.1016/j.nuclphysb.2005.09.047}
  {\bibfield  {journal} {\bibinfo  {journal} {Nucl. Phys.}\ }\textbf {\bibinfo
  {volume} {B733}},\ \bibinfo {pages} {1} (\bibinfo {year} {2006})},\ \bibinfo
  {note} {[Erratum: Nucl. Phys.B800,405(2008)]},\ \Eprint
  {http://arxiv.org/abs/hep-ph/0504088} {arXiv:hep-ph/0504088 [hep-ph]}
  \BibitemShut {NoStop}%
\bibitem [{\citenamefont {Khodjamirian}\ \emph {et~al.}(2007)\citenamefont
  {Khodjamirian}, \citenamefont {Mannel},\ and\ \citenamefont
  {Offen}}]{Khodjamirian:2006st}%
  \BibitemOpen
  \bibfield  {author} {\bibinfo {author} {\bibfnamefont {A.}~\bibnamefont
  {Khodjamirian}}, \bibinfo {author} {\bibfnamefont {T.}~\bibnamefont
  {Mannel}}, \ and\ \bibinfo {author} {\bibfnamefont {N.}~\bibnamefont
  {Offen}},\ }\href {\doibase 10.1103/PhysRevD.75.054013} {\bibfield  {journal}
  {\bibinfo  {journal} {Phys. Rev.}\ }\textbf {\bibinfo {volume} {D75}},\
  \bibinfo {pages} {054013} (\bibinfo {year} {2007})},\ \Eprint
  {http://arxiv.org/abs/hep-ph/0611193} {arXiv:hep-ph/0611193 [hep-ph]}
  \BibitemShut {NoStop}%
\bibitem [{\citenamefont {De~Fazio}\ \emph {et~al.}(2008)\citenamefont
  {De~Fazio}, \citenamefont {Feldmann},\ and\ \citenamefont
  {Hurth}}]{DeFazio:2007hw}%
  \BibitemOpen
  \bibfield  {author} {\bibinfo {author} {\bibfnamefont {F.}~\bibnamefont
  {De~Fazio}}, \bibinfo {author} {\bibfnamefont {T.}~\bibnamefont {Feldmann}},
  \ and\ \bibinfo {author} {\bibfnamefont {T.}~\bibnamefont {Hurth}},\ }\href
  {\doibase 10.1088/1126-6708/2008/02/031} {\bibfield  {journal} {\bibinfo
  {journal} {JHEP}\ }\textbf {\bibinfo {volume} {02}},\ \bibinfo {pages} {031}
  (\bibinfo {year} {2008})},\ \Eprint {http://arxiv.org/abs/0711.3999}
  {arXiv:0711.3999 [hep-ph]} \BibitemShut {NoStop}%
\bibitem [{\citenamefont {Beneke}\ and\ \citenamefont
  {Rohrwild}(2011)}]{Beneke:2011nf}%
  \BibitemOpen
  \bibfield  {author} {\bibinfo {author} {\bibfnamefont {M.}~\bibnamefont
  {Beneke}}\ and\ \bibinfo {author} {\bibfnamefont {J.}~\bibnamefont
  {Rohrwild}},\ }\href {\doibase 10.1140/epjc/s10052-011-1818-8} {\bibfield
  {journal} {\bibinfo  {journal} {Eur. Phys. J.}\ }\textbf {\bibinfo {volume}
  {C71}},\ \bibinfo {pages} {1818} (\bibinfo {year} {2011})},\ \Eprint
  {http://arxiv.org/abs/1110.3228} {arXiv:1110.3228 [hep-ph]} \BibitemShut
  {NoStop}%
\bibitem [{\citenamefont {Braun}\ and\ \citenamefont
  {Khodjamirian}(2013)}]{Braun:2012kp}%
  \BibitemOpen
  \bibfield  {author} {\bibinfo {author} {\bibfnamefont {V.~M.}\ \bibnamefont
  {Braun}}\ and\ \bibinfo {author} {\bibfnamefont {A.}~\bibnamefont
  {Khodjamirian}},\ }\href {\doibase 10.1016/j.physletb.2012.11.047} {\bibfield
   {journal} {\bibinfo  {journal} {Phys. Lett.}\ }\textbf {\bibinfo {volume}
  {B718}},\ \bibinfo {pages} {1014} (\bibinfo {year} {2013})},\ \Eprint
  {http://arxiv.org/abs/1210.4453} {arXiv:1210.4453 [hep-ph]} \BibitemShut
  {NoStop}%
\bibitem [{\citenamefont {Beneke}\ \emph {et~al.}(2018)\citenamefont {Beneke},
  \citenamefont {Braun}, \citenamefont {Ji},\ and\ \citenamefont
  {Wei}}]{Beneke:2018wjp}%
  \BibitemOpen
  \bibfield  {author} {\bibinfo {author} {\bibfnamefont {M.}~\bibnamefont
  {Beneke}}, \bibinfo {author} {\bibfnamefont {V.~M.}\ \bibnamefont {Braun}},
  \bibinfo {author} {\bibfnamefont {Y.}~\bibnamefont {Ji}}, \ and\ \bibinfo
  {author} {\bibfnamefont {Y.-B.}\ \bibnamefont {Wei}},\ }\href {\doibase
  10.1007/JHEP07(2018)154} {\bibfield  {journal} {\bibinfo  {journal} {JHEP}\
  }\textbf {\bibinfo {volume} {07}},\ \bibinfo {pages} {154} (\bibinfo {year}
  {2018})},\ \Eprint {http://arxiv.org/abs/1804.04962} {arXiv:1804.04962
  [hep-ph]} \BibitemShut {NoStop}%
\bibitem [{\citenamefont {Grozin}\ and\ \citenamefont
  {Neubert}(1997)}]{Grozin:1996pq}%
  \BibitemOpen
  \bibfield  {author} {\bibinfo {author} {\bibfnamefont {A.~G.}\ \bibnamefont
  {Grozin}}\ and\ \bibinfo {author} {\bibfnamefont {M.}~\bibnamefont
  {Neubert}},\ }\href {\doibase 10.1103/PhysRevD.55.272} {\bibfield  {journal}
  {\bibinfo  {journal} {Phys. Rev.}\ }\textbf {\bibinfo {volume} {D55}},\
  \bibinfo {pages} {272} (\bibinfo {year} {1997})},\ \Eprint
  {http://arxiv.org/abs/hep-ph/9607366} {arXiv:hep-ph/9607366 [hep-ph]}
  \BibitemShut {NoStop}%
\bibitem [{\citenamefont {Lee}\ and\ \citenamefont
  {Neubert}(2005)}]{Lee:2005gza}%
  \BibitemOpen
  \bibfield  {author} {\bibinfo {author} {\bibfnamefont {S.~J.}\ \bibnamefont
  {Lee}}\ and\ \bibinfo {author} {\bibfnamefont {M.}~\bibnamefont {Neubert}},\
  }\href {\doibase 10.1103/PhysRevD.72.094028} {\bibfield  {journal} {\bibinfo
  {journal} {Phys. Rev.}\ }\textbf {\bibinfo {volume} {D72}},\ \bibinfo {pages}
  {094028} (\bibinfo {year} {2005})},\ \Eprint
  {http://arxiv.org/abs/hep-ph/0509350} {arXiv:hep-ph/0509350 [hep-ph]}
  \BibitemShut {NoStop}%
\bibitem [{\citenamefont {Lange}\ and\ \citenamefont
  {Neubert}(2003)}]{Lange:2003ff}%
  \BibitemOpen
  \bibfield  {author} {\bibinfo {author} {\bibfnamefont {B.~O.}\ \bibnamefont
  {Lange}}\ and\ \bibinfo {author} {\bibfnamefont {M.}~\bibnamefont
  {Neubert}},\ }\href {\doibase 10.1103/PhysRevLett.91.102001} {\bibfield
  {journal} {\bibinfo  {journal} {Phys. Rev. Lett.}\ }\textbf {\bibinfo
  {volume} {91}},\ \bibinfo {pages} {102001} (\bibinfo {year} {2003})},\
  \Eprint {http://arxiv.org/abs/hep-ph/0303082} {arXiv:hep-ph/0303082 [hep-ph]}
  \BibitemShut {NoStop}%
\bibitem [{\citenamefont {Bell}\ \emph {et~al.}(2013)\citenamefont {Bell},
  \citenamefont {Feldmann}, \citenamefont {Wang},\ and\ \citenamefont
  {Yip}}]{Bell:2013tfa}%
  \BibitemOpen
  \bibfield  {author} {\bibinfo {author} {\bibfnamefont {G.}~\bibnamefont
  {Bell}}, \bibinfo {author} {\bibfnamefont {T.}~\bibnamefont {Feldmann}},
  \bibinfo {author} {\bibfnamefont {Y.-M.}\ \bibnamefont {Wang}}, \ and\
  \bibinfo {author} {\bibfnamefont {M.~W.~Y.}\ \bibnamefont {Yip}},\ }\href
  {\doibase 10.1007/JHEP11(2013)191} {\bibfield  {journal} {\bibinfo  {journal}
  {JHEP}\ }\textbf {\bibinfo {volume} {11}},\ \bibinfo {pages} {191} (\bibinfo
  {year} {2013})},\ \Eprint {http://arxiv.org/abs/1308.6114} {arXiv:1308.6114
  [hep-ph]} \BibitemShut {NoStop}%
\bibitem [{\citenamefont {Braun}\ and\ \citenamefont
  {Manashov}(2014)}]{Braun:2014owa}%
  \BibitemOpen
  \bibfield  {author} {\bibinfo {author} {\bibfnamefont {V.~M.}\ \bibnamefont
  {Braun}}\ and\ \bibinfo {author} {\bibfnamefont {A.~N.}\ \bibnamefont
  {Manashov}},\ }\href {\doibase 10.1016/j.physletb.2014.02.051} {\bibfield
  {journal} {\bibinfo  {journal} {Phys. Lett.}\ }\textbf {\bibinfo {volume}
  {B731}},\ \bibinfo {pages} {316} (\bibinfo {year} {2014})},\ \Eprint
  {http://arxiv.org/abs/1402.5822} {arXiv:1402.5822 [hep-ph]} \BibitemShut
  {NoStop}%
\bibitem [{\citenamefont {Braun}\ \emph {et~al.}(2019)\citenamefont {Braun},
  \citenamefont {Ji},\ and\ \citenamefont {Manashov}}]{Braun:2019wyx}%
  \BibitemOpen
  \bibfield  {author} {\bibinfo {author} {\bibfnamefont {V.~M.}\ \bibnamefont
  {Braun}}, \bibinfo {author} {\bibfnamefont {Y.}~\bibnamefont {Ji}}, \ and\
  \bibinfo {author} {\bibfnamefont {A.~N.}\ \bibnamefont {Manashov}},\ }\href
  {\doibase 10.3204/PUBDB-2019-02451} {\bibfield  {journal} {\bibinfo
  {journal} {Phys. Rev. D}\ }\textbf {\bibinfo {volume} {100}},\ \bibinfo
  {pages} {014023} (\bibinfo {year} {2019})},\ \Eprint
  {http://arxiv.org/abs/1905.04498} {arXiv:1905.04498 [hep-ph]} \BibitemShut
  {NoStop}%
\bibitem [{\citenamefont {Liu}\ \emph {et~al.}(2020)\citenamefont {Liu},
  \citenamefont {Mecaj}, \citenamefont {Neubert}, \citenamefont {Wang},\ and\
  \citenamefont {Fleming}}]{Liu:2020eqe}%
  \BibitemOpen
  \bibfield  {author} {\bibinfo {author} {\bibfnamefont {Z.~L.}\ \bibnamefont
  {Liu}}, \bibinfo {author} {\bibfnamefont {B.}~\bibnamefont {Mecaj}}, \bibinfo
  {author} {\bibfnamefont {M.}~\bibnamefont {Neubert}}, \bibinfo {author}
  {\bibfnamefont {X.}~\bibnamefont {Wang}}, \ and\ \bibinfo {author}
  {\bibfnamefont {S.}~\bibnamefont {Fleming}},\ }\href {\doibase
  10.1007/JHEP07(2020)104} {\bibfield  {journal} {\bibinfo  {journal} {JHEP}\
  }\textbf {\bibinfo {volume} {07}},\ \bibinfo {pages} {104} (\bibinfo {year}
  {2020})},\ \Eprint {http://arxiv.org/abs/2005.03013} {arXiv:2005.03013
  [hep-ph]} \BibitemShut {NoStop}%
\bibitem [{\citenamefont {Galda}\ and\ \citenamefont
  {Neubert}(2020)}]{Galda:2020epp}%
  \BibitemOpen
  \bibfield  {author} {\bibinfo {author} {\bibfnamefont {A.~M.}\ \bibnamefont
  {Galda}}\ and\ \bibinfo {author} {\bibfnamefont {M.}~\bibnamefont
  {Neubert}},\ }\href {\doibase 10.1103/PhysRevD.102.071501} {\bibfield
  {journal} {\bibinfo  {journal} {Phys. Rev. D}\ }\textbf {\bibinfo {volume}
  {102}},\ \bibinfo {pages} {071501} (\bibinfo {year} {2020})},\ \Eprint
  {http://arxiv.org/abs/2006.05428} {arXiv:2006.05428 [hep-ph]} \BibitemShut
  {NoStop}%
\bibitem [{\citenamefont {Braun}\ \emph {et~al.}(2004)\citenamefont {Braun},
  \citenamefont {Ivanov},\ and\ \citenamefont {Korchemsky}}]{Braun:2003wx}%
  \BibitemOpen
  \bibfield  {author} {\bibinfo {author} {\bibfnamefont {V.~M.}\ \bibnamefont
  {Braun}}, \bibinfo {author} {\bibfnamefont {D.~{\relax Yu}.}\ \bibnamefont
  {Ivanov}}, \ and\ \bibinfo {author} {\bibfnamefont {G.~P.}\ \bibnamefont
  {Korchemsky}},\ }\href {\doibase 10.1103/PhysRevD.69.034014} {\bibfield
  {journal} {\bibinfo  {journal} {Phys. Rev.}\ }\textbf {\bibinfo {volume}
  {D69}},\ \bibinfo {pages} {034014} (\bibinfo {year} {2004})},\ \Eprint
  {http://arxiv.org/abs/hep-ph/0309330} {arXiv:hep-ph/0309330 [hep-ph]}
  \BibitemShut {NoStop}%
\bibitem [{\citenamefont {Grozin}\ and\ \citenamefont
  {Korchemsky}(1996)}]{Grozin:1994ni}%
  \BibitemOpen
  \bibfield  {author} {\bibinfo {author} {\bibfnamefont {A.~G.}\ \bibnamefont
  {Grozin}}\ and\ \bibinfo {author} {\bibfnamefont {G.~P.}\ \bibnamefont
  {Korchemsky}},\ }\href {\doibase 10.1103/PhysRevD.53.1378} {\bibfield
  {journal} {\bibinfo  {journal} {Phys. Rev. D}\ }\textbf {\bibinfo {volume}
  {53}},\ \bibinfo {pages} {1378} (\bibinfo {year} {1996})},\ \Eprint
  {http://arxiv.org/abs/hep-ph/9411323} {arXiv:hep-ph/9411323} \BibitemShut
  {NoStop}%
\bibitem [{\citenamefont {Strichartz}(2003)}]{Strichartz:2003}%
  \BibitemOpen
  \bibfield  {author} {\bibinfo {author} {\bibfnamefont {R.~S.}\ \bibnamefont
  {Strichartz}},\ }\href@noop {} {\emph {\bibinfo {title} {A Guide To
  Distribution Theory And Fourier Transforms}}}\ (\bibinfo  {publisher} {CRC
  Press},\ \bibinfo {year} {2003})\BibitemShut {NoStop}%
\bibitem [{\citenamefont {Braun}\ \emph {et~al.}(2017)\citenamefont {Braun},
  \citenamefont {Ji},\ and\ \citenamefont {Manashov}}]{Braun:2017liq}%
  \BibitemOpen
  \bibfield  {author} {\bibinfo {author} {\bibfnamefont {V.~M.}\ \bibnamefont
  {Braun}}, \bibinfo {author} {\bibfnamefont {Y.}~\bibnamefont {Ji}}, \ and\
  \bibinfo {author} {\bibfnamefont {A.~N.}\ \bibnamefont {Manashov}},\ }\href
  {\doibase 10.1007/JHEP05(2017)022} {\bibfield  {journal} {\bibinfo  {journal}
  {JHEP}\ }\textbf {\bibinfo {volume} {05}},\ \bibinfo {pages} {022} (\bibinfo
  {year} {2017})},\ \Eprint {http://arxiv.org/abs/1703.02446} {arXiv:1703.02446
  [hep-ph]} \BibitemShut {NoStop}%
\bibitem [{\citenamefont {Feldmann}\ \emph {et~al.}(2014)\citenamefont
  {Feldmann}, \citenamefont {Lange},\ and\ \citenamefont
  {Wang}}]{Feldmann:2014ika}%
  \BibitemOpen
  \bibfield  {author} {\bibinfo {author} {\bibfnamefont {T.}~\bibnamefont
  {Feldmann}}, \bibinfo {author} {\bibfnamefont {B.~O.}\ \bibnamefont {Lange}},
  \ and\ \bibinfo {author} {\bibfnamefont {Y.-M.}\ \bibnamefont {Wang}},\
  }\href {\doibase 10.1103/PhysRevD.89.114001} {\bibfield  {journal} {\bibinfo
  {journal} {Phys. Rev.}\ }\textbf {\bibinfo {volume} {D89}},\ \bibinfo {pages}
  {114001} (\bibinfo {year} {2014})},\ \Eprint {http://arxiv.org/abs/1404.1343}
  {arXiv:1404.1343 [hep-ph]} \BibitemShut {NoStop}%
\bibitem [{\citenamefont {Caprini}(2019)}]{Caprini:2019osi}%
  \BibitemOpen
  \bibfield  {author} {\bibinfo {author} {\bibfnamefont {I.}~\bibnamefont
  {Caprini}},\ }\href {\doibase 10.1007/978-3-030-18948-8} {\emph {\bibinfo
  {title} {{Functional Analysis and Optimization Methods in Hadron
  Physics}}}},\ SpringerBriefs in Physics\ (\bibinfo  {publisher} {Springer},\
  \bibinfo {year} {2019})\BibitemShut {NoStop}%
\bibitem [{\citenamefont {Kawamura}\ and\ \citenamefont
  {Tanaka}(2009)}]{Kawamura:2008vq}%
  \BibitemOpen
  \bibfield  {author} {\bibinfo {author} {\bibfnamefont {H.}~\bibnamefont
  {Kawamura}}\ and\ \bibinfo {author} {\bibfnamefont {K.}~\bibnamefont
  {Tanaka}},\ }\href {\doibase 10.1016/j.physletb.2009.02.028} {\bibfield
  {journal} {\bibinfo  {journal} {Phys. Lett.}\ }\textbf {\bibinfo {volume}
  {B673}},\ \bibinfo {pages} {201} (\bibinfo {year} {2009})},\ \Eprint
  {http://arxiv.org/abs/0810.5628} {arXiv:0810.5628 [hep-ph]} \BibitemShut
  {NoStop}%
\bibitem [{\citenamefont {Kawamura}\ \emph {et~al.}(2001)\citenamefont
  {Kawamura}, \citenamefont {Kodaira}, \citenamefont {Qiao},\ and\
  \citenamefont {Tanaka}}]{Kawamura:2001jm}%
  \BibitemOpen
  \bibfield  {author} {\bibinfo {author} {\bibfnamefont {H.}~\bibnamefont
  {Kawamura}}, \bibinfo {author} {\bibfnamefont {J.}~\bibnamefont {Kodaira}},
  \bibinfo {author} {\bibfnamefont {C.-F.}\ \bibnamefont {Qiao}}, \ and\
  \bibinfo {author} {\bibfnamefont {K.}~\bibnamefont {Tanaka}},\ }\href
  {\doibase 10.1016/S0370-2693(01)01299-0, 10.1016/S0370-2693(02)01866-X}
  {\bibfield  {journal} {\bibinfo  {journal} {Phys. Lett.}\ }\textbf {\bibinfo
  {volume} {B523}},\ \bibinfo {pages} {111} (\bibinfo {year} {2001})},\
  \bibinfo {note} {[Erratum: Phys. Lett.B536,344(2002)]},\ \Eprint
  {http://arxiv.org/abs/hep-ph/0109181} {arXiv:hep-ph/0109181 [hep-ph]}
  \BibitemShut {NoStop}%
\bibitem [{\citenamefont {Descotes-Genon}\ and\ \citenamefont
  {Offen}(2009)}]{Descotes-Genon:2009jif}%
  \BibitemOpen
  \bibfield  {author} {\bibinfo {author} {\bibfnamefont {S.}~\bibnamefont
  {Descotes-Genon}}\ and\ \bibinfo {author} {\bibfnamefont {N.}~\bibnamefont
  {Offen}},\ }\href {\doibase 10.1088/1126-6708/2009/05/091} {\bibfield
  {journal} {\bibinfo  {journal} {JHEP}\ }\textbf {\bibinfo {volume} {05}},\
  \bibinfo {pages} {091} (\bibinfo {year} {2009})},\ \Eprint
  {http://arxiv.org/abs/0903.0790} {arXiv:0903.0790 [hep-ph]} \BibitemShut
  {NoStop}%
\bibitem [{\citenamefont {Bell}\ and\ \citenamefont
  {Feldmann}(2008)}]{Bell:2008er}%
  \BibitemOpen
  \bibfield  {author} {\bibinfo {author} {\bibfnamefont {G.}~\bibnamefont
  {Bell}}\ and\ \bibinfo {author} {\bibfnamefont {T.}~\bibnamefont
  {Feldmann}},\ }\href {\doibase 10.1088/1126-6708/2008/04/061} {\bibfield
  {journal} {\bibinfo  {journal} {JHEP}\ }\textbf {\bibinfo {volume} {04}},\
  \bibinfo {pages} {061} (\bibinfo {year} {2008})},\ \Eprint
  {http://arxiv.org/abs/0802.2221} {arXiv:0802.2221 [hep-ph]} \BibitemShut
  {NoStop}%
\bibitem [{\citenamefont {Beneke}\ and\ \citenamefont
  {Feldmann}(2001)}]{Beneke:2000wa}%
  \BibitemOpen
  \bibfield  {author} {\bibinfo {author} {\bibfnamefont {M.}~\bibnamefont
  {Beneke}}\ and\ \bibinfo {author} {\bibfnamefont {T.}~\bibnamefont
  {Feldmann}},\ }\href {\doibase 10.1016/S0550-3213(00)00585-X} {\bibfield
  {journal} {\bibinfo  {journal} {Nucl. Phys. B}\ }\textbf {\bibinfo {volume}
  {592}},\ \bibinfo {pages} {3} (\bibinfo {year} {2001})},\ \Eprint
  {http://arxiv.org/abs/hep-ph/0008255} {arXiv:hep-ph/0008255} \BibitemShut
  {NoStop}%
\bibitem [{\citenamefont {Heller}\ \emph {et~al.}(2015)\citenamefont {Heller}
  \emph {et~al.}}]{Belle:2015mpp}%
  \BibitemOpen
  \bibfield  {author} {\bibinfo {author} {\bibfnamefont {A.}~\bibnamefont
  {Heller}} \emph {et~al.} (\bibinfo {collaboration} {Belle}),\ }\href
  {\doibase 10.1103/PhysRevD.91.112009} {\bibfield  {journal} {\bibinfo
  {journal} {Phys. Rev. D}\ }\textbf {\bibinfo {volume} {91}},\ \bibinfo
  {pages} {112009} (\bibinfo {year} {2015})},\ \Eprint
  {http://arxiv.org/abs/1504.05831} {arXiv:1504.05831 [hep-ex]} \BibitemShut
  {NoStop}%
\bibitem [{\citenamefont {Gelb}\ \emph {et~al.}(2018)\citenamefont {Gelb} \emph
  {et~al.}}]{Belle:2018jqd}%
  \BibitemOpen
  \bibfield  {author} {\bibinfo {author} {\bibfnamefont {M.}~\bibnamefont
  {Gelb}} \emph {et~al.} (\bibinfo {collaboration} {Belle}),\ }\href {\doibase
  10.1103/PhysRevD.98.112016} {\bibfield  {journal} {\bibinfo  {journal} {Phys.
  Rev. D}\ }\textbf {\bibinfo {volume} {98}},\ \bibinfo {pages} {112016}
  (\bibinfo {year} {2018})},\ \Eprint {http://arxiv.org/abs/1810.12976}
  {arXiv:1810.12976 [hep-ex]} \BibitemShut {NoStop}%
\bibitem [{\citenamefont {Galda}\ \emph {et~al.}(2022)\citenamefont {Galda},
  \citenamefont {Neubert},\ and\ \citenamefont {Wang}}]{Galda:2022dhp}%
  \BibitemOpen
  \bibfield  {author} {\bibinfo {author} {\bibfnamefont {A.~M.}\ \bibnamefont
  {Galda}}, \bibinfo {author} {\bibfnamefont {M.}~\bibnamefont {Neubert}}, \
  and\ \bibinfo {author} {\bibfnamefont {X.}~\bibnamefont {Wang}},\ }\href@noop
  {} {\  (\bibinfo {year} {2022})},\ \Eprint {http://arxiv.org/abs/2203.08202}
  {arXiv:2203.08202 [hep-ph]} \BibitemShut {NoStop}%
\bibitem [{\citenamefont {Neubert}(2005)}]{Neubert:2004dd}%
  \BibitemOpen
  \bibfield  {author} {\bibinfo {author} {\bibfnamefont {M.}~\bibnamefont
  {Neubert}},\ }\href {\doibase 10.1140/epjc/s2005-02141-1} {\bibfield
  {journal} {\bibinfo  {journal} {Eur. Phys. J. C}\ }\textbf {\bibinfo {volume}
  {40}},\ \bibinfo {pages} {165} (\bibinfo {year} {2005})},\ \Eprint
  {http://arxiv.org/abs/hep-ph/0408179} {arXiv:hep-ph/0408179} \BibitemShut
  {NoStop}%
\end{thebibliography}%

\end{document}